%% file: main.tex
\documentclass[fleqn,usenatbib]{mnras}

\usepackage{newtxtext,newtxmath}

\usepackage[T1]{fontenc}

\DeclareRobustCommand{\VAN}[3]{#2}
\let\VANthebibliography\thebibliography
\def\thebibliography{\DeclareRobustCommand{\VAN}[3]{##3}\VANthebibliography}

\usepackage{scalerel}
\usepackage{tikz}
\usepackage{subfiles}
\usepackage{booktabs}
\usepackage{multirow}
\usetikzlibrary{svg.path}
\usepackage{siunitx}
\definecolor{orcidlogocol}{HTML}{A6CE39}
\tikzset{
  orcidlogo/.pic={
    \fill[orcidlogocol] svg{M256,128c0,70.7-57.3,128-128,128C57.3,256,0,198.7,0,128C0,57.3,57.3,0,128,0C198.7,0,256,57.3,256,128z};
    \fill[white] svg{M86.3,186.2H70.9V79.1h15.4v48.4V186.2z}
                 svg{M108.9,79.1h41.6c39.6,0,57,28.3,57,53.6c0,27.5-21.5,53.6-56.8,53.6h-41.8V79.1z M124.3,172.4h24.5c34.9,0,42.9-26.5,42.9-39.7c0-21.5-13.7-39.7-43.7-39.7h-23.7V172.4z}
                 svg{M88.7,56.8c0,5.5-4.5,10.1-10.1,10.1c-5.6,0-10.1-4.6-10.1-10.1c0-5.6,4.5-10.1,10.1-10.1C84.2,46.7,88.7,51.3,88.7,56.8z};
  }
}

\newcommand\orcidicon[1]{\href{https://orcid.org/#1}{\mbox{\scalerel*{
\begin{tikzpicture}[yscale=-1,transform shape]
\pic{orcidlogo};
\end{tikzpicture}
}{|}}}}

\usepackage{graphicx}	
\usepackage{amsmath}	

\usepackage{hyperref}
\usepackage{lscape} 

\def\aap{A\& A}
\def\apjl{ApJL}
\def\apjs{ApJS}

\newcommand{\ism}{ISM}
\newcommand{\ifu}{IFU}
\newcommand{\vlt}{VLT}
\newcommand{\bpt}{BPT}
\newcommand{\agn}{AGN}
\newcommand{\cmf}{CMF}

\newcommand{\sparc}{{\sc Sparcs}}
\newcommand{\cloudy}{{\sc Cloudy}}
\newcommand{\radmc}{{\sc Radmc-3d}}
\newcommand{\hylight}{{\sc HyLight}}

\newcommand{\swift}{{\sc Swift}}

\newcommand{\chimes}{{\sc Chimes}}

\title[Hydrogen recombination line modelling]{The {\sc HyLight} model for hydrogen emission lines in simulated nebulae}

\author[Y. Liu et al.]{Yuankang Liu$^{\orcidicon{0000-0003-0771-763X}}$,$^{1,2}$\thanks{E-mail: yuankang.liu@durham.ac.uk (YKL)}
Tom Theuns$^{\orcidicon{0000-0002-3790-9520}}$,$^{1}$
Tsang Keung Chan$^{\orcidicon{0000-0003-2544-054X}}$,$^{3,4}$
Alexander J. Richings$^{\orcidicon{0000-0003-0502-9235}}$$^{5,6}$
\newauthor
and Anna F. McLeod$^{\orcidicon{0000-0002-5456-523X}2,1}$
\\
$^{1}$Institute for Computational Cosmology, Department of Physics, Durham University, South Road, Durham DH1 3LE, UK\\
$^{2}$Centre for Extragalactic Astronomy, Department of Physics, Durham University, South Road, Durham DH1 3LE, UK\\
$^{3}$Department of Physics, The Chinese University of Hong Kong, Shatin, Hong Kong, China\\
$^{4}$Department of Astronomy and Astrophysics, the University of Chicago, Chicago, IL 60637, USA\\
$^{5}$Centre for Data Science, Artificial Intelligence and Modelling, University of Hull, Cottingham Road, Hull, HU6 7RX, UK\\
$^{6}$E.A. Milne Centre for Astrophysics, University of Hull, Cottingham Road, Hull HU6 7RX, UK
}

\date{Accepted XXX. Received YYY; in original form ZZZ}

\pubyear{2025}

\begin{document}
\label{firstpage}
\pagerange{\pageref{firstpage}--\pageref{lastpage}}
\maketitle

\begin{abstract}
Hydrogen recombination lines provide key diagnostics of ionized gas in galaxies, yet most hydrodynamical simulations estimate hydrogen level populations using interpolated emissivity tables rather than computing them directly from local physical conditions. We present \hylight{}, a Python-based atomic model that calculates hydrogen level populations and line emissivities from the gas density, temperature, and ionization state, enabling accurate predictions in both equilibrium and non-equilibrium environments.
Benchmark comparisons show that \hylight{} reproduces \cloudy{} predictions for Balmer, Paschen, and Brackett emissivities to within 1 per cent under typical photoionized nebular conditions, while discrepancies of several tens of per cent arise relative to other published calculations. As an illustrative application, we use \hylight{} to compute photoionization-to-line intensity ratios in an \ion{H}{ii} nebula and generate synthetic hydrogen emission maps from a radiation-hydrodynamical simulation that includes non-equilibrium thermochemistry.
Combining physical consistency with flexibility, \hylight{} provides a robust framework for connecting hydrodynamical simulations with observational diagnostics of photoionized regions, and enhances our ability to interpret hydrogen emission in complex, non-equilibrium astrophysical environments.
\end{abstract}




\subfile{sections/introduction}
\subfile{sections/motivation}

\subfile{sections/model}

\subfile{sections/application}
\subfile{sections/conclusion}

\section*{Acknowledgements}
We thank the anonymous reviewer for their constructive feedback. 
We also thank Peter van Hoof and the \cloudy{} discussion forum for many helpful discussions.

This work is co-funded by the European Union (Widening Participation, ExGal-Twin, GA 101158446). Views and opinions expressed are however those of the author(s) only and do not necessarily reflect those of the European Union. Neither the European Union nor the granting authority can be held responsible for them. This work received funding from the Horizon Europe guarantee scheme of UK Research and Innovation (UKRI). 

We acknowledge the support of the DiRAC@Durham facility managed by the Institute for Computational Cosmology on behalf of the STFC DiRAC HPC Facility (www.dirac.ac.uk). The equipment was funded by BEIS capital funding via STFC capital grants ST/K00042X/1, ST/P002293/1, ST/R002371/1 and ST/S002502/1, Durham University and STFC operations grant ST/R000832/1. DiRAC is part of the National e-Infrastructure. 

YKL acknowledges the support from the Durham Doctoral Studentship (DDS). 

TKC is supported by the `Improvement on Competitiveness in Hiring New Faculties' Funding Scheme (4937210-4937211-4937212), the Direct Grant project (4053662,4443786,4053719) from the Chinese University of Hong Kong, and the RGC Early Career Scheme (24301524). TKC was supported by the E. Margaret Burbidge Prize Postdoctoral Fellowship from the Brinson Foundation at the Departments of Astronomy and Astrophysics at the University of Chicago. 

The research in this paper made use of the \swift{} open-source simulation code
(\href{http://www.swiftsim.com}{http://www.swiftsim.com}, \citealt{Schaller2024}). The analysis made use of {\sc Matplotlib}~\citep{Hunter2007}, {\sc NumPy}~\citep{Harris2020}, {\sc Scipy}~\citep{Jones2001, Virtanen2020}, {\sc Swiftsimio}~\citep{Borrow2020}, {\sc Unyt}~\citep{Goldbaum2018}, {\sc H5py}~\citep{Collette2022} and {\sc Pandas}~\citep{McKinney2010, Reback2020}. We also made use of the arXiv preprint service and NASA’s Astrophysics Data System. 
\section*{Data Availability}
The \hylight{} package is publicly available at \href{https://github.com/YuankangLiu/HyLight}{https://github.com/YuankangLiu/HyLight}. The \cloudy{} code is available at \href{https://gitlab.nublado.org/cloudy/cloudy/-/wikis/home}{https://gitlab.nublado.org/cloudy/cloudy/-/wikis/home}. The {\sc Fortran 77} code to calculate Einstein $A$ values can be found at \href{https://data.mendeley.com/datasets/3drgznwty8/1}{https://data.mendeley.com/datasets/3drgznwty8/1}. The public version of the {\sc Swift} code is available at \href{http://www.swiftsim.com}{http://www.swiftsim.com}, which includes the {\sc Sph-m1rt} radiative transfer module coupled with a simple hydrogen and helium thermochemistry network. The \chimes{} code and the associated package are publicly available at \href{https://richings.bitbucket.io/chimes/home.html}{https://richings.bitbucket.io/chimes/home.html}. Currently, \sparc{} is a \swift{} module located in a private \swift{} branch. \sparc{} will be released to the public in the future. \radmc{} is publicly available at \href{https://www.ita.uni-heidelberg.de/~dullemond/software/radmc-3d/}{https://www.ita.uni-heidelberg.de/$\sim$dullemond/software/radmc-3d/}. The input scripts for the \cloudy{} models and the analysis code used in this paper can be accessed at \href{https://doi.org/10.5281/zenodo.16911175}{https://doi.org/10.5281/zenodo.16911175}.



\bibliographystyle{mnras}
\bibliography{references}



\appendix
\subfile{sections/appendices}


\bsp	
\label{lastpage}

\end{document}


%% file: sections/introduction.tex
\section{Introduction}
Nebular emission lines play a crucial role in determining the physical conditions in star-forming regions and the interstellar medium (\ism{}) of galaxies~\citep[see, for example,  the review by][]{Kewley2019}. Physical properties inferred from such lines include the composition (\lq metallicity\rq) of the gas~\citep[e.g.][]{Kewley2002, Marino2013, Rerez-Montero2014, Perez-Montero2017}, or its electron density, temperature, and pressure~\citep[e.g.][]{Seaton1957, Peimbert1967, Rubin1989}. These studies are often based on the \lq Baldwin-Phillips-Terlevich diagram\rq\ (the \bpt\ diagram, \citealt[][]{Baldwin1981}), which uses the ratios of [\ion{O}{iii}] $\lambdaup$5007/H$\beta$ vs [\ion{N}{ii}] $\lambdaup$6583/H$\alpha$ emission lines to distinguish star-forming galaxies from \agn-dominated galaxies (see e.g.~\citealt{Veilleux1987, Kewley2001, Kauffmann2003}). Other recent examples of using emission line diagnostics include \cite{McLeod2015}, who used the \bpt\ diagram to analyse MUSE\footnote{\href{https://www.eso.org/sci/facilities/develop/instruments/muse.html}{https://www.eso.org/sci/facilities/develop/instruments/muse.html}} integral field unit (\ifu) data from \vlt\ of the \lq Pillars of Creation\rq. \cite{McLeod2019} conducted spatially resolved ionized gas studies in the Large Magellanic Cloud (LMC) using emission line ratios. \cite{Mcleod2021} used emission line ratios to distinguish between \ion{H}{ii} regions, supernova remnants, and planetary nebulae. \citet{Isobe2025} and \citet{Shapley2025} analysed line-ratio diagrams to study the nature of star formation in galaxies at redshifts $3-7$ observed by the James Webb Space Telescope. \cite{Groves2023} analysed thousands of \ion{H}{ii} regions in 19 nearby main-sequence galaxies with the help of the bright emission lines. \cite{Kennicutt1998} reviews the use of emission lines to infer star formation rates.

Connecting the physical properties of the nebular gas to the luminosity of its emission lines requires detailed modelling of a range of physical processes, from computing how photons travel through the gas and dust (radiative transfer, henceforth RT), accounting for the impact of collisions on the ionization and excitation state of ions, and finally relating excitation states to emissivity. Two examples of spectral synthesis codes that model these processes include {\sc Mappings-v}~\citep[][detailed in \citealt{Dopita1996}]{Sutherland2018} and \cloudy{}~\citep[][recently described in \citealt{Cloudy2017, Cloudy2025}]{Cloudy1999}. Both include very detailed atomic/molecular physics models and RT, and can be used with relatively simple nebular geometries and situations in which the gas is either in thermal equilibrium or in photoionization equilibrium. 

Nebulae, such as, for example, \ion{H}{ii} regions, have been described with the spherical model of \cite{Osterbrock1974}~\citep[e.g.][]{Sankrit2000}, yet in reality most have a more complex geometry~\citep[][]{Perez2001, Simpson2004, McLeod2016}. \citet{Morisset2013} combines spherically symmetric \cloudy{} models to account for this more complex geometry. \citet{Jin2022} adopted a fully self-consistent Monte Carlo radiative transfer (MCRT) technique in {\sc Mappings-v} that can handle such complex geometries. Both models solve for the equilibrium solution and ignore the possible impact of the non-equilibrium effects on emission line diagnostics, and, more importantly, assume that the density structure of the nebula is relatively simple and known a priori.

Nearby star-forming regions have been mapped in detail using integral field spectroscopy~\citep[e.g.][]{Rousseau-nepton2019, Groves2023, Drory2024}, revealing filaments of high-density neutral or molecular gas delineating ionized regions of lower density~\citep[e.g.][]{Mcleod2021}. It is unclear whether the emission lines emitted by these intricate structures can be well described or understood with current relatively simple models of such \ion{H}{ii} regions which assume simple geometries (plane-parallel or spherical) in photoionization equilibrium when calculating species abundances. This can be tested by performing more realistic radiation-hydrodynamic simulations of nebular regions and generating mock IFU data cubes from them. Such radiation-hydrodynamical simulations model in detail how a star-forming cloud or the \ism{} of a galaxy is shaped by radiation, supernovae, winds and jets (see, for example, the SILCC\footnote{\href{https://hera.ph1.uni-koeln.de/~silcc/}{https://hera.ph1.uni-koeln.de/~silcc/}}~\citep{Walch2015} and STARFORGE\footnote{\href{https://starforge.space/}{https://starforge.space/}} \citep{Grudic2021} projects). These simulations perform RT for a small set of photon energies with the aim of computing the ionization states of important gas coolants, i.e., typically \ion{H}{i}, \ion{He}{i} and \ion{He}{ii}-ionizing photons. The emissivity of this gas in the lines of interest is computed in post-processing~\citep[][]{Hirschmann2017, Hirschmann2019, Hirschmann2023}. 

Unlike lines such as [\ion{O}{iii}] $\lambdaup$5007 or [\ion{N}{ii}] $\lambdaup$6583, which are excited by collisions, excited states of hydrogen result predominantly from recombinations in a photoionized gas, hence they are referred to as hydrogen {\em recombination} lines. Hydrogen recombination lines are typically amongst the most luminous nebular emission lines and as such have often been the main focus in astrophysical studies. For example, the H$\alpha$ ($n=3\to 2$) line is widely used as an indicator for star formation, since, in photoionization equilibrium, the recombination rate equals the photoionization rate, and the latter is a measure of the rate at which young, massive stars are formed~\citep[][]{Kennicutt1998}. The emissivity of gas of a given density, ionization state, and temperature in hydrogen recombination lines was calculated in a series of papers in the 1930s~\citep[][]{Menzel1937a, Menzel1937b, Baker1938}. These papers showed that recombination, rather than collisional excitation, is the main channel that populates the hydrogen atom's excited states under typical nebular conditions. The excited atom emits a series of hydrogen lines as it cascades to the ground state, making the nebula glow at discrete wavelengths.

These early papers discussed two limiting cases to describe hydrogen recombination in a nebula. In \lq Case~A\rq, the nebula is optically thin in the Lyman series lines (transitions to $n=1$, where $n$ is the usual principal quantum number), whereas in \lq Case~B\rq, the nebula is optically thick in those lines. We note that these are just limiting cases, and in practice, neither approximation is accurate in all situations. \citet{Baker1938} examined an $n$-resolved atomic model, assuming states with different $l$ are populated according to their statistical weight, $2l+1$ (where $l$ is the usual orbital angular momentum  quantum number). They argued that Case~B is the better approximation under typical nebular conditions. \citet{Brocklehurst1971} revisited the problem and discussed an $nl$-resolved atomic model, where the $l$-states at a given $n$ are generally {\em not} distributed according to their statistical weights, leading to different predictions for the emissivity of recombination lines. Later, tables of recombination line emissivities for a range of gas densities and temperatures were published by~\citet{Hummer1987} and~\citet{Storey1988, Storey1995}. The lowest density in the table is 10$^2$ cm$^{-3}$, which may be insufficient for certain applications. \citet{Martin1988} extended these calculations to lower densities. 

Simulations can be post-processed to compute hydrogen recombination line emissivities by interpolating from published tables \citep[e.g.][]{Katz2022, McClymont2025}, or by constructing tailor-made interpolation tables using, for example, \cloudy{}~\cite[e.g.][]{Hirschmann2023b}. Alternatively, \citet{Raga2015} suggested computing the emissivities directly, given the temperature, ionization state and density of the gas, by computing the rate at which levels are populated by recombinations and depopulated by radiative transitions under the assumption that $l$-states within an $n$-level are populated according to their statistical weights, which we will show in this paper is not accurate for photoionized nebulae. The method of \citealt{Raga2015} is widely used in ISM emission line studies. For example, \citet{Silva2018} used it to predict intensity maps and power spectra of H$\alpha$, \citet{Rodriguez-gonzalez2023} used it to calculate H$\alpha$ emission in planetary nebulae. The model was also used to calculate H$\alpha$ emission in post-processing of simulations of AGN outflows~\citep{Richings2021} and simulations of \ion{H}{ii} regions and the ISM of isolated disc galaxies~\citep{Richings2022}. 

In this paper we examine whether these different predictions give consistent results. We also present a method called \hylight{} for computing hydrogen population levels and hence line emissivities for hydrogen recombination lines directly, given the temperature, density, and ionization state of the emitting gas. This allows one to predict hydrogen line emissivities for hydrodynamical simulations that include radiative transfer, and enables such simulations to be compared to observations. An advantage of \hylight{} over using interpolation from a \cloudy{} table is that our method uses the ionization state computed by the simulation directly. This ionization state may include non-equilibrium effects, which cannot be handled by \cloudy{}, which assumes photoionization equilibrium. In addition, \hylight{} is not restricted to any particular geometry.

The paper is organised as follows. In \S~\ref{sec:motivation}, we compare predictions for hydrogen line emissivities between different published models, showing that there are significant differences between them.  We introduce our own model in \S~\ref{sec:model}. The code is implemented as a publicly available {\sc Python} package\footnote{\href{https://github.com/YuankangLiu/HyLight}{https://github.com/YuankangLiu/HyLight}}. \hylight{} directly computes the level population of hydrogen atoms, taking into account radiative recombination and collisional excitation from the ground state. We compare the resulting level population and line emissivities to \cloudy{}. As an example application, we post-process a snapshot of a non-equilibrium radiation-hydrodynamical simulation of an \ion{H}{ii} region, performed with \sparc{}~\citep{Chan2026}. We use \hylight{} to compute level populations and process the result with \radmc{}~\citep{Dullemond2012} to obtain a full mock IFU data cube. This enables direct comparison between simulated and observed \ion{H}{ii} regions, which we will present elsewhere.
\end{document}

%% file: sections/motivation.tex
\section{Hydrogen recombination lines: comparison of model predictions}
\label{sec:motivation}
Consider a hydrogen atom in the $n=3$ state. When the electron transitions to the $n=2$ state, an H$\alpha$ photon is emitted. The H$\alpha$ volume emissivity, $\epsilon_{3,2}$, is set by the level population of the $n=3$ state, 
\begin{align}
    \epsilon_{3,2} &= \left(n_{3s} A_{3s, 2p} + n_{3p} A_{3p, 2s} + n_{3d} A_{3d, 2p}\right) h\nu_{3,2},
    \label{eq:Ha_emis_calc}
\end{align}
where $n_{nl}$ is the number density of hydrogen atoms in state $nl$, $A_{nl, n'l'}$ is the spontaneous transition rate from the $nl\to n'l'$ state (the Einstein $A$ coefficient), and $h\nu_{3,2}$ is the photon energy corresponding to the transition; $l$ indicates the angular momentum quantum number, with the usual convention that the $l=0$, 1 and 2 states are denoted by $s$, $p$ and $d$.
This equation accounts for the $\Delta \, l=\pm 1$ selection rule for dipole-allowed transitions but neglects stimulated emissions \cite[e.g.][]{Rybicki1979}. The key to computing the emissivities of hydrogen lines is therefore computing the level populations $n_{nl}$ ($A_{nl, n'l'}$ is an integral involving the products of the $nl$ and $n'l'$ electron orbitals). Calculating the $n_{nl}$ is not trivial. Various authors use different approximations, and we begin by quantifying the resulting differences in the predicted emissivity for common recombination lines in two simple setups:
(1) a spherical cloud of uniform-density hydrogen gas at constant temperature, with a central ionizing source, and (2) a thin, spherical shell of uniform-density hydrogen gas at constant temperature, with a central ionizing source.
\begin{table*}
    \caption{List of setups used in this paper with the detailed parameters. We vary the properties of the source of ionizing photons (blackbody spectrum or a narrow laser beam), the geometry of the gas (spherical or shell), and whether we specify the nature of recombinations (Case A, Case B, or unspecified). The table also lists the number of $l$-resolved levels (for $n=1 \to n_{\rm resolved}$ and the number of collapsed levels in the \cloudy{} modelling.
    In all models, the inner radius $r_{i}$ of all models is kept at 0.01 pc, the hydrogen density is 100 cm$^{-3}$ and the gas temperature is $10^4$ K (typical values for \ion{H}{ii} regions). The ionizing photon rate of the source is $Q({\rm H})=10^{48}$ s$^{-1}$ to mimic a hot star. For simplicity, only hydrogen is taken into account in these models.}
    \label{tab:cloudy_models}
    \begin{tabular}{cccccc}
    \hline
    Label & Source spectrum & Cloud geometry & Recombination method & $n_\mathrm{resolved}$ & $n_\mathrm{collapsed}$\\
    \hline
    Ref - LSph & Laser at 1.1 Ryd & Sphere & Not specified & 100 & 1\\
    Ref - Sph & Blackbody of 10$^{4.6}$ K & Sphere & Not specified & 100 & 1\\
    Ref - ASh & Blackbody of 10$^{4.6}$ K & Shell & Case A & 100 & 1\\
    Def - ASh & Blackbody of 10$^{4.6}$ K & Shell & Case A & 10 & 15\\
    Ref - BSh & Blackbody of 10$^{4.6}$ K & Shell & Case B & 100 & 1\\
    Def - BSh & Blackbody of 10$^{4.6}$ K & Shell & Case B & 10 & 15\\
    Ref - ALSh & Laser at 1.1 Ryd & Shell & Case A & 100 & 1\\
    Def - ALSh & Laser at 1.1 Ryd & Shell & Case A & 10 & 15\\
    \hline
    \end{tabular}
\end{table*}

\subsection{Idealized test setups for model comparisons}
We use two simple geometries to contrast different models for calculating hydrogen line emissivities. In both cases, the gas is static, consists of pure hydrogen gas with a density of $100~\unit{cm^{-3}}$, does not contain dust or molecules, and its temperature is kept constant at $T=10^4~\unit{K}$. The photon source emits ionizing photons at a rate of $10^{48}~\unit{s^{-1}}$ with a blackbody spectrum of temperature 10$^{4.6}$~K. These values are typical for \ion{H}{ii} regions~\citep{Osterbrock2006}. The two example setups are as follows:
\begin{enumerate}
\item {\em Idealised spherical model.}
Spherical models of \ion{H}{ii} regions are widely used historically~\citep[see][for examples]{Osterbrock1974, Pellegrini2012}. In these models, a central hot star ionizes the surrounding gas, with the ionization front stalling at the Str{\"o}mgren radius. In our setup, the inner edge of the gas cloud is positioned 0.01 pc away from the ionizing source. These spherical setups are labelled {\lq Sph\rq} in Table~\ref{tab:cloudy_models}, which also lists all the other setups we have used in this paper along with their specific parameters.
\item {\em Idealised shell models.} Shell models are widely used to explore the ionization and emission properties of gas for a particular set of physical conditions and for a particular incident radiation field, without taking into account the effects of radiative transfer on the incident spectrum~\cite[e.g.][]{Dagostino2019}. In our setup, the shell of gas is positioned $0.01~\unit{pc}$ from the same source as in (i), and the shell is $1~\unit{cm}$ thick to keep any radiative transfer effects to a minimum. Such fiducial setups are labelled as {\lq Sh\rq} in Table~\ref{tab:cloudy_models} depending on the method to compute the recombination rate.
\end{enumerate}

We compute the ionization state of the gas in photoionization equilibrium using \cloudy{} version C25\footnote{\href{https://gitlab.nublado.org/cloudy/cloudy/-/wikis/home}{https://gitlab.nublado.org/cloudy/cloudy/-/wikis/home}.}~\citep[][recently described in \citealt{Cloudy2017, Cloudy2023, Cloudy2025}]{Cloudy1999}. We extract the electron density, $n_e$, proton density, $n_p$, and neutral hydrogen density, $n_{\rm HI}$, as a function of distance $r$ from the source; these serve as input for calculating the emissivities in all models. 

\subsection{Atomic models}
\label{sec:atomic_models}
Given the two types of setups, we compute and compare the level population and hence emissivity in hydrogen recombination lines for several atomic models:
\begin{enumerate}
    \item {\em The \cloudy{} model}. \cloudy{} calculates the emissivity using a built-in hydrogen model. The level population is computed accounting for collisional excitation and de-exicitation (including collisions that change $n$ and those that change $l$ at given $n$), level-resolved recombinations ({\em i.e.} accounting for the fact that the recombination rates depend on both $n$ and $l$), stimulated emission, and absorption of lines from the source and from the nebula. By default, \cloudy{} uses 10 $l$-resolved energy levels, (i.e. principal quantum numbers $n=1\to 10$, with $l=0\to n-1$ for each $n$). The code accounts for levels $n>10$ using \lq collapsed\rq\ (not $l$-resolved) levels, by default using 10 of these.  We investigate the convergence of the \cloudy{} results when increasing the number of level-resolved levels in \S~\ref{sec:cmf_validation} and Appendix~\ref{app:cloudy_model_convergence} (see also \citealt{Guzman2025}). Based on our findings, we generate a \lq Reference\rq\ \cloudy{} model which employs 100 $l$-resolved levels and 1 collapsed level. This reference atomic model is labelled {\lq Ref\rq} throughout this paper, see also Table~\ref{tab:cloudy_models}. 
    \item {\em The Storey \& Hummer tables.} \cite{Storey1995} published interpolation tables of level populations and emissivities as a function of density and temperature of emitting gas for hydrogenic ions. Their model includes an adaptive number of $nl$-resolved levels and is described by~\citealt{Hummer1987, Storey1988} and~\citealt{Storey1995}.
    \item {\em The Raga et al. model.} \cite{Raga2015} computed the level populations of a 5-level hydrogen atom. The model includes collisional excitations, but recombinations and collisions are not $l$-resolved. The model assumes that the level population for a given $n$ are proportional to their statistical weights, $2l+1$.  
    \item {\em Our model.} The \hylight{} model, presented in this paper, calculates the level population for a hydrogen atom where the user can select the maximum number of levels used, $n_{\rm max}$, and accounts for all $l$-states for levels $n=1\to n_{\rm max}$. Details of the model are presented in the next section. In this section, we set $n_{\rm max}=100$. The model is similar to \cloudy{}'s built-in methods but omits certain processes (e.g. Lyman pumping, $l$-mixing collisions, collisional excitation and de-excitation above the ground state). These approximations are good under typical nebular conditions, as we show in \S~\ref{sec:model}.
    \item {\em The {\sc LTE} model}. We also compute the level population assuming local thermodynamic equilibrium ({\sc LTE}) for reference. Details of the LTE calculation are in Appendix~\ref{sec:appendix_depar_coeff}.
\end{enumerate}
The Einstein $A$ coefficients used in \hylight{} are computed using the code described by \citealt{Hoang-Binh1990} and \citealt{Hoang-Binh2005}. We verified that they are identical to those used in \cloudy{}.
\subsection{Model comparison}
\begin{figure*}
    \includegraphics[width=\textwidth]{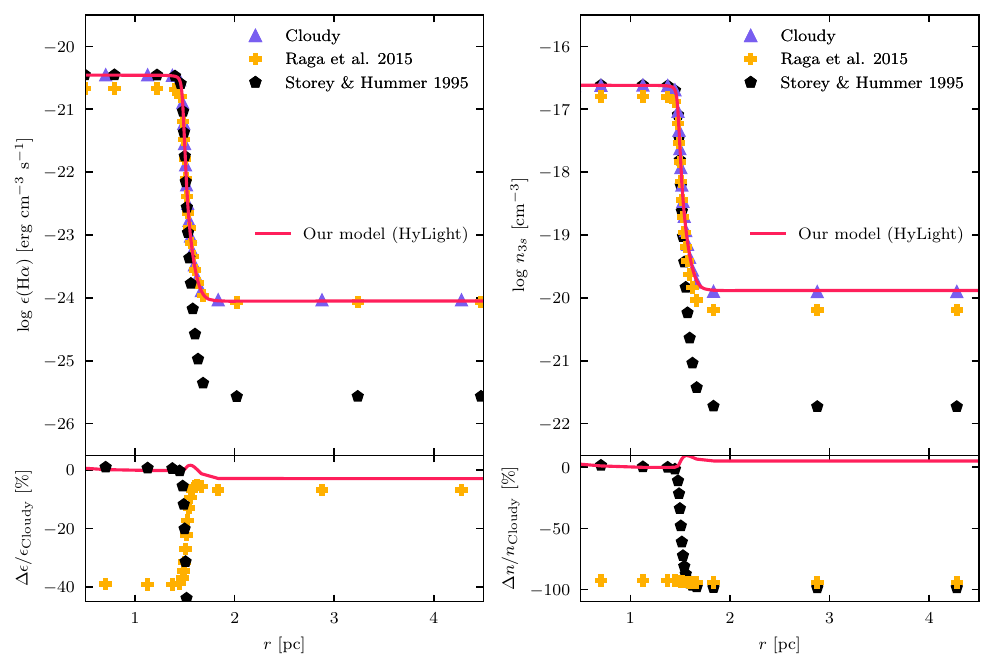}
    \caption{{\em Left panel:} Profile of H$\alpha$ emissivity, $\epsilon({\rm H \alpha}) \equiv \epsilon_{3,2}$, of the idealised spherical setup for different models: the reference \cloudy{} model \lq Ref - Sph\rq\ ({\em purple triangles}), the \protect\cite{Raga2015} model ({\em yellow crosses}) and the tabulated values from \protect\cite{Storey1995} ({\em black diamonds}).
    The {\em top panel} shows the emissivity; the {\em lower panel} is the relative difference in emissivity compared to \lq Ref-Sph\rq\ in per cent. In all models, the emissivity is approximately constant inside the Str\"omgren radius, $r\approx 1.5~\unit{pc}$, then drops rapidly. Different models differ by up to 50 per cent from the \cloudy{} prediction. In particular, \citet{Storey1995} values at large radii ($\gtrsim$ 2 pc) differ by over 40~per cent in terms of emissivity. 
    {\em Right panel:} Same as left panel, but for the density of hydrogen atoms in the $3s$ state. Different models now differ by up to 100~per cent from the \cloudy{} prediction.
    In all panels, the {\em red line} is the prediction from the model described in this paper. Our model agrees with the \cloudy{} prediction within 1 per cent in terms of the total H$\alpha$ luminosity.}
    \label{fig:sph_model_comp}
\end{figure*}

The H$\alpha$ emissivity, $\epsilon({\rm H \alpha}) \equiv \epsilon_{3,2}$, is compared for the different models and the \lq spherical cloud\rq\ setup in Fig.~\ref{fig:sph_model_comp} (left panel). In all models, $\epsilon_{3,2}$ is nearly constant out to the Str{\"o}mgren radius, $r\approx 1.5~\unit{pc}$, then drops rapidly. \cloudy{} (purple symbols) and the tabulated values from \cite{Storey1995} (black diamonds) agree to better than a per cent in the inner parts of the spherical cloud, but start to deviate at larger $r$. In contrast, the \cite{Raga2015} results (yellow crosses) are 40~per cent lower than \cloudy{} in the inner parts of the cloud, but agree to within a few per cent in the outskirts. The level of (dis)agreement between the models is mirrored in the level population of the $3s$ level (Fig.~\ref{fig:sph_model_comp}, right panel). Interestingly, here \cite{Raga2015} differs significantly from the \cloudy{} values, including in the outer (neutral) parts of the cloud where the emissivity agrees. The reason behind the discrepancy between \cite{Raga2015} and \cloudy{} is that the model by \cite{Raga2015} only takes into account atomic levels up to $n=5$, and $l$-levels are populated according to their statistical weights, $2l+1$. 
The latter assumption is not valid, as we show in detail later. The values obtained from the tables published by \cite{Storey1995} differ significantly from the \cloudy{} values in the outskirts ($\gtrsim$ 2~pc).
These tables assume that the population level in the ground state is sufficiently low that the collisional excitation rate from the ground state can be neglected (see \citealt{Hummer1987} \S~4 and \citealt{Storey1995} \S~2)
This is not a good approximation for the parameters in this test, as we show later. In short, where the other models differ from \cloudy{}, we have good reason to believe that they are inaccurate. 

The red lines in both panels are computed using \hylight{}, the model presented in this paper. The predictions from \hylight{} agree to within a few per cent with \cloudy{} throughout the cloud, both in terms of emissivity and level population.

Tables~\ref{tab:emis_comp_case_a} \& \ref{tab:emis_comp_case_b} compare some models for the shell setup for several hydrogen recombination lines, with luminosity scaled to that of the \cloudy{} reference model (assuming either Case~A or Case~B). The first column compares to the reference \cloudy{} model \lq Def\rq, for which we do not specify either Case~A or Case~B recombination - the differences are of order 10~per cent for the Balmer lines, but reach up to 25~per cent for Brackett lines. The {\sc LTE} values differ by up to an order of magnitude from those predicted by \cloudy{}: local thermodynamical equilibrium is clearly a poor approximation. The \cite{Storey1995} tabulated values agree with the \cloudy{} calculations at the per cent level in the worst case of Brackett~$\alpha$. The \cite{Raga2015} values differ at the 50~per cent level or more. Finally, the values computed using \hylight{} agree with \cloudy{} to better than one per cent for all lines shown.

We refer to the \cloudy{} model as the \lq baseline\rq model, as it captures most physical processes. We find that the results from \hylight{} compare favourably to those of the other models, and reproduce the \cloudy{} results very well. Using \cloudy{} models directly to post-process radiation-hydrodynamic simulations can be a daunting task: it requires tabulating population levels as a function of density, temperature and ionizing spectrum. In addition, the \cloudy{} calculation assumes that the gas is in photoionization equilibrium. Therefore, \cloudy{} cannot be used to account for any non-equilibrium effects, whereas the simulation may include these (e.g., the simulations performed using \sparc{} by \cite{Chan2026} which we analyse below).

\hylight{} uses level-resolved, temperature-dependent recombination rates and accounts for collisional excitations from the ground state. We show below that these two processes are dominant in setting the level population under typical nebular conditions. Given the ionization state of the gas - as obtained, for example, from a non-equilibrium radiation hydrodynamical simulation - \hylight{} computes the level population and the emissivity of the gas for common hydrogen lines. The model is not restricted to thermal or ionization equilibrium. It only assumes statistical equilibrium of the level population, i.e., the rate of populating a specific $nl$-state equals the rate at which the state is being depopulated - which is an extremely good approximation. We describe the model in more detail next. 
\begin{table*}
    \centering
    \caption{Line luminosities for selected recombination lines predicted by different models for the thin shell setup; Case~A recombination is imposed for all models, and luminosities are relative to the \cloudy{} reference model \lq Ref - ASh\rq. }
    \label{tab:emis_comp_case_a}
    \begin{tabular}{ccccccc}
    \hline
    Luminosity relative to Ref - ASh & Def - ASh & LTE & \citealt{Storey1995} Case A & \citealt{Raga2015} Case A & This work - Case A\\
    \hline
    Balmer $\alpha$ & 0.902 & 12.424 & 1.007 & 0.409 & 1.001\\
    Balmer $\beta$ & 0.962 & 7.441 & 1.005 & 0.552 & 1.001\\
    Balmer $\gamma$ & 0.980 & 5.788 & 1.004 & 0.536 & 0.999\\
    Paschen $\alpha$ & 0.825 & 4.538 & 1.009 & 0.337 & 1.000\\
    Paschen $\beta$ & 0.901 & 3.763 & 1.004 & 0.348 & 1.001\\
    Brackett $\alpha$ & 0.761 & 2.765 & 1.013 & 0.256 & 1.000\\
    \hline
    \end{tabular}
\end{table*}
\begin{table*}
    \centering
    \caption{Same as Table~\ref{tab:emis_comp_case_a}, but now imposing
    Case~B recombination, with luminosities relative to the \cloudy{} reference model \lq Ref - BSh\rq.}
    \label{tab:emis_comp_case_b}
    \begin{tabular}{cccccc} 
    \hline
    Luminosity relative to Ref - BSh & Def - BSh & LTE & \citealt{Storey1995} Case B& \citealt{Raga2015} Case B & This work - Case B\\
    \hline
    Balmer $\alpha$ & 0.977 & 8.157 & 1.005 & 0.608 & 0.997\\
    Balmer $\beta$ & 1.005 & 4.956 & 1.003 & 0.638 & 0.997\\
    Balmer $\gamma$ & 1.014 & 3.907 & 1.004 & 0.562 & 0.998\\
    Paschen $\alpha$ & 0.916 & 4.073 & 1.009 & 0.524 & 0.996\\
    Paschen $\beta$ & 0.967 & 3.308 & 1.006 & 0.476 & 1.000\\
    Brackett $\alpha$ & 0.868 & 2.631 & 1.013 & 0.379 & 0.993\\
    \hline
    \end{tabular}
\end{table*}

%% file: sections/model.tex
\section{Computing the level population in hydrogen: the \hylight{} model}
\label{sec:model}
\begin{figure*}
    \includegraphics[width=\textwidth]{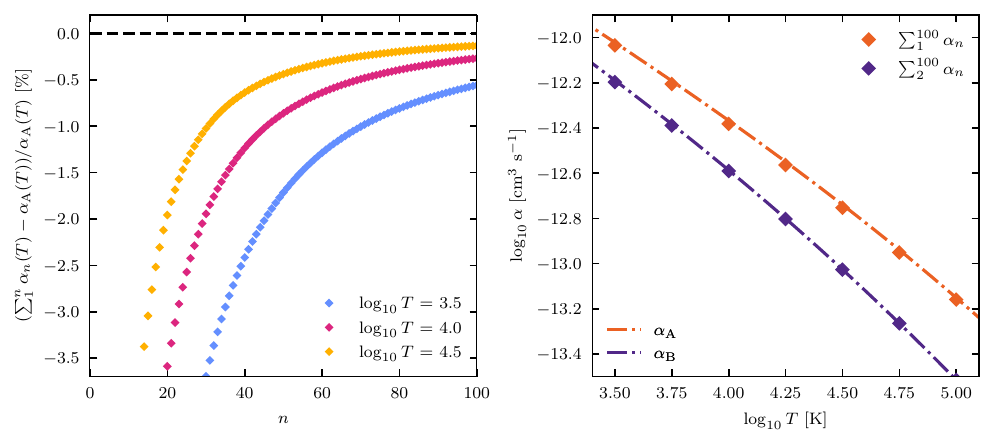}
    \caption{{\em Left panel:} Cumulative $nl$-resolved recombination rate, $\sum_{n'=1}^n\sum_{l'=0}^{n'}\alpha_{n'l'}(T)$, up to a given $n$, for various temperatures as per the legend. {\em Diamonds} are the difference in per cent of the cumulative rate compared to the total recombination coefficient, $\alpha_\mathrm{A}$. The black dashed line at 0 is drawn to guide the eye. For $T=10^{4}$~K, the cumulative rate for $n=100$ equals the total rate to better than 0.5~per cent, quantifying the level to which recombinations to levels with $n>100$ are unimportant in setting the recombination rate. The importance of recombinations to these higher energy levels increases with decreasing $T$.
    {\em Right panel:} Sum of the $nl$-resolved recombination rates up to and including level $n=100$ ({\em diamonds}) compared to the
    total recombination rate ({\em lines}) for Case~A ({\em purple}) and Case~B ({\em orange}) recombinations, as tabulated by \protect\citet{Ferland1992}. The summed level-resolved recombination rate up to $n=100$ agrees to be better than a per cent with the tabulated values for $T$ in the range $10^{3.5}-10^{5.5}$~K. }
    \label{fig:alphanl}
\end{figure*}
The ionization state of hydrogen gas in a nebula can be out of equilibrium - meaning the rate at which atoms are ionized differs from the rate at which they recombine. This can result from sudden changes in density or temperature of the gas (for example, due to shocks), or from a sudden increase or decrease in the ionizing flux (for example, because of variability of the source, or due to shadowing).  The timescale to reach equilibrium is usually either the photoionization timescale or the recombination timescale.

The level population of neutral hydrogen can also be out of equilibrium. However, the typical timescale to restore equilibrium is of order the inverse of the Einstein $A$ values, and hence very short ($\sim 10^{-6}-10^{-8}$~s). In astrophysical scenarios, it is therefore an extremely good approximation to assume that level populations {\em are} in equilibrium, even when the ionization levels are not.

The equilibrium abundance of a given level is found by equating the rate at which the level is populated to the rate it is depopulated, often referred to as statistical equilibrium in the literature. Levels are populated by (1) direct recombinations, described by the radiative recombination rate coefficients, (2) spontaneous transitions, described by the Einstein $A$ coefficients, (3) induced transitions, described by the Einstein $B$ coefficients, (4) absorption of photons, and (5) collisions. Levels are depopulated by the same processes, except for process (1). The relative importance of these processes depends on the physical properties of the system (density, temperature, ionization state, ambient radiation field). 

We now describe our implementation of a model to compute hydrogen emission lines, called \hylight{} (HYdrogen recombination LIne emission from ionized Gas in varying tHermal condiTions). The code is written in {\sc Python} and publicly available on GitHub\footnote{\href{https://github.com/YuankangLiu/HyLight}{https://github.com/YuankangLiu/HyLight}}. Of the processes that set the level population, \hylight{} neglects induced emission and absorption of photons from the source and the nebula (see \S~\ref{sec:limitations} for the effects of Lyman pumping from source), and further only includes the effects of collisions of hydrogen atoms in the ground state. We will test the impact of the omissions by comparing to \cloudy{}, which does not make these simplifications. In \hylight{}, the level population is obtained by solving the following set of coupled linear equations,
\begin{multline}
    \left[ n_p n_e \alpha_{nl}(T) + \sum_{n' > n}^{\infty} \sum_{l'=l\pm 1} n_{n'l'}A_{n'l',nl}\right]\\
    + \left[ n_\mathrm{H I} n_e 
    q_{1s,nl}(T)\right]
     = n_{nl} \sum_{n''=1}^{n-1} \sum_{l''=l \pm 1} A_{nl,n''l''}.
    \label{eq:equilibrium_eq_full}
\end{multline}
Here, $n_e$, $n_p$ and $n_\mathrm{H I}$ are the number densities of electrons, protons, and neutral hydrogen atoms, $\alpha_{nl}(T)$ is the temperature-dependent, level-resolved, radiative recombination rate coefficient, and $q_{1s, nl}$ is the temperature-dependent collisional excitation rate from the ground state to level $nl$; the $A$'s are the Einstein $A$ coefficients, as before. While dipole allowed transitions dominate radiative decay with non-zero Einstein $A$ coefficients, higher order multipole or forbidden transitions, though typically orders of magnitude weaker, exhibit non-zero $A$ values due to relativistic, spin-orbit, or electron correlation effects beyond the electric dipole approximation. In \hylight{}, we only take into account dipole-allowed transitions except for the 2$s$ to 1$s$ two-photon process (see \S~\ref{sec:two_photon_process}). The left hand side of Equation~(\ref{eq:equilibrium_eq_full}) is the rate at which level $nl$ is being populated by direct recombinations, and by spontaneous transitions from higher $n$ - the two terms in the first set of square brackets, and by collisions with hydrogen atoms in the ground state - the term in the second set of square brackets. The right-hand side is the rate at which level $nl$ is being depopulated by spontaneous transitions to states with lower $n$. We will first neglect the collisional term and determine the level population due to radiative transitions only. 
\subsection{The level population with radiative transitions only}
\label{sec:model_radiative}
Neglecting collisional processes yields the following equation for the level population,
\begin{align}
    n_p n_e \alpha_{nl}(T) + \sum_{n' > n}^{\infty} \sum_{l'=l\pm 1} n_{n'l'}A_{n'l',nl} &= n_{nl} \sum_{n''=1}^{n-1} \sum_{l''=l \pm 1} A_{nl,n''l''}.
    \label{eq:equilibrium_eq_rr}
\end{align}
Equation~(\ref{eq:equilibrium_eq_rr}) can be solved using the \lq Cascade Matrix Formalism\rq\ (henceforth \cmf{}, see e.g. \citealt{Seaton1959b} and \citealt{Osterbrock2006}). We start by computing the fraction of spontaneous transitions of a state $nl$ to another state $n'l'$ (with $n'<n$), denoted as $P_{nl, n'l'}$, as
\begin{align}
P_{nl, n'l'} &\equiv \frac{A_{nl, n'l'}}{\sum\limits_{n''=n_{\rm low}}^{n-1}\sum\limits_{l''=0}^{n''-1}A_{nl, n''l''}},
\label{eq:Pnl}
\end{align}
where $n_{\rm low}=1$ in the summation in the denominator (see later for why we introduce this variable). We compute the $P$'s using the $A$'s calculated with the code of \cite{Hoang-Binh2005}. The numerical values we find
are in excellent agreement with those listed in \citealt{Bethe1933} (table 19, page 446). 

The cascade matrix, $C_{nl, n'l'}$, is defined recursively in terms of the $P$'s, as 
\begin{align}
\label{eq:C_definition}
    C_{nl, n'l'} &= \sum_{n''=n'+1}^{n}\sum_{l''=l'\pm 1} C_{nl, n''l''} P_{n''l'', n'l'} \nonumber\\
    &= P_{nl, n'l'} + \sum_{l''=l'\pm 1}[C_{nl,n-1l''}P_{n-1l'',n'l'} + \nonumber\\
    & C_{nl,n-2l''}P_{n-2l'',n'l'} + \cdots],
\end{align}
with $C_{nl, n-1l'} = P_{nl, n-1l'}$ and $C_{nl,nl'}=\delta_{ll'}$ (the Kronecker delta function). The recurrence guarantees that $C_{nl, n'l'}$ is the fraction of spontaneous decays from the state $nl$ that end up in the state $n'l'$, either via a direct spontaneous decay or via a \lq cascade\rq\ of spontaneous decays through intermediate states. As noted before, we need to be satisfied with some maximum value $n_{\rm max}$ of $n$ that we will account for in the calculation of the $C$'s to start the recurrence.

We can now use the cascade matrix to compute the rate at which a level is populated, and Equation~(\ref{eq:equilibrium_eq_rr}) becomes
\begin{equation}
    n_p n_e \sum_{n'=n+1}^\infty\sum_{l'=0}^{n'-1} \alpha_{n'l'}(T) C_{n'l',nl} 
= n_{nl} \sum_{n''=n_{\rm low}}^{n-1} \sum_{l''=l \pm 1} A_{nl, n''l''}.
\label{eq:CMF}
\end{equation}
The left-hand side is the rate at which the state $nl$ is populated, the right-hand side the rate at which it is de-populated, and we can compute the $n_{nl}$ from the $C$'s, $A$'s and $\alpha$'s:
\begin{align}
    n_{nl}  &=  n_p n_e \frac{\sum\limits_{n'=n+1}^\infty\sum\limits_{l'=0}^{n'-1} \alpha_{n'l'}(T) C_{n'l',nl}}{\sum\limits_{n''=n_{\rm low}}^{n-1} \sum\limits_{l''=l \pm 1} A_{nl, n''l''}}\nonumber\\
    &\equiv n_p n_e R_{nl}(T).
\label{eq:n_nl}
\end{align}
Equation~(\ref{eq:n_nl}) shows that the level population $n_{nl}\propto n_{\rm p}n_e$, with a proportionality constant, $R_{nl}$, that depends on temperature. The numerator of $R_{nl}$ is the net rate at which level $nl$ is being populated by recombinations, either directly to state $nl$, or to a higher energy state which decays to level $nl$ by a series of spontaneous transitions, as described by the cascade matrix $C$. This term is sometimes called the effective recombination rate coefficient to level $nl$, $\alpha^{\rm eff}_{nl}(T)$,
\begin{align}
    R_{nl}(T) &= \frac{\alpha^{\rm eff}_{nl}(T)}{{\sum\limits_{n''=n_{\rm low}}^{n-1} \sum\limits_{l''=0}^{n''-1} A_{nl, n''l''}}}\\
    \alpha^{\rm eff}_{nl}(T) &= \sum\limits_{n'=n+1}^\infty\sum\limits_{l'=0}^{n'-1} \alpha_{n'l'}(T) C_{n'l',nl}.
    \label{eq:Rnl}
\end{align}
\noindent $\bullet$ {\em Evaluating the $\alpha_{nl}$'s.} 
When gas is in LTE with an ambient blackbody radiation field with temperature $T$, detailed balance ensures that the photoionization rate (from any level) is equal to the corresponding recombination rate (see \citealt{Rybicki1979} \S~10.5 and the appendix of \citealt{Ferland1992}). \cloudy{} uses the photoionization cross section from \citealt{Verner1996} to compute the level-resolved $\alpha_{nl}$'s~\citep[][]{Cloudy2017}. We extract these radiative recombination rate coefficients from \cloudy{}. Although these constants are derived using Milne relation assuming detailed balance, they are applicable to systems not in equilibrium. More details about such calculation can be found in appendix 2 of \citealt{Osterbrock2006}, chapter 7 of \citealt{Pradhan2011}, and \citealt{Nahar2021}. 
Fig.~\ref{fig:alphanl} (left panel) shows the difference in per cent of the sum of resolved recombinations up to some level, $n$, compared to the total rate, for various temperatures (different colours). At $T=10^{4}$~K, recombinations to levels $n>100$ contribute less than 0.5 per cent to the total rate. Fig.~\ref{fig:alphanl} (right panel) compares the result of summing the level-resolved recombination rates up to level $n=100$ starting from $n=1$ ($\alpha_{\rm A}$ or \lq Case A\rq\ recombination) or from $n=2$ ($\alpha_{\rm B}$ or \lq Case B\rq\ recombination) against the values tabulated in \citealt{Ferland1992}: the computed and tabulated rates agree to better than 1 per cent. 

\noindent $\bullet$ {\em Evaluating the Einstein $A$'s.} We evaluate the Einstein $A$ coefficients using a {\sc Fortran} 77 code written by \cite{Hoang-Binh2005}\footnote{The code is available at \href{https://data.mendeley.com/datasets/3drgznwty8/1}{https://data.mendeley.com/datasets/3drgznwty8/1}. }. We verified that the resulting values agree to typically better than 1 per cent with those tabulated in the NIST database\footnote{
\href{https://physics.nist.gov/cgi-bin/ASD/lines1.pl?spectra=H+I&output_type=0&low_w=&upp_w=&unit=1&submit=Retrieve+Data&de=0&plot_out=0&I_scale_type=1&format=0&line_out=0&en_unit=0&output=0&bibrefs=1&page_size=15&show_obs_wl=1&show_calc_wl=1&unc_out=1&order_out=0&max_low_enrg=&show_av=2&max_upp_enrg=&tsb_value=0&min_str=&A_out=0&intens_out=on&max_str=&allowed_out=1&forbid_out=1&min_accur=&min_intens=&conf_out=on&term_out=on&enrg_out=on&J_out=on}{NIST Atomic Spectra Database Lines Data for Atomic Hydrogen.}}. The {\sc Fortran} code also allows us to compute Einstein $A$ values not in the NIST database. 

Once the $\alpha$'s and Einstein $A$'s are known, the relation Equation~(\ref{eq:C_definition}) shows how to compute the $C_{nl, n'l'}$, provided the matrix elements are known for all larger $n$. Applying the recurrence relation therefore requires choosing a highest population level, $n=n_{\rm max}$, that we will account for, and setting $C=P$ for that highest level (i.e. truncating the cascade to transitions below $n_{\rm max}$). We can understand intuitively that the level population will converge with increasing $n_{\rm max}$, simply because the recombination rate $\alpha_{nl}$ decreases with increasing $n$. The right panel of Fig.~\ref{fig:alphanl} shows this convergence: the sum of the level-resolved $\alpha_{nl}$'s up to $n=100$ differs from the total $\alpha_{\rm A}$ by less than 1 per cent for all temperatures above $10^{3.5}$~K. By default, we start the recurrence from $n_{\rm max}=100$, meaning that we neglect recombinations to levels $n>100$. Below, we investigate in more detail the extent to which this approximation affects our results.

In summary, we compute the values of the Einstein coefficients following~\citealt{Hoang-Binh2005}, and use them to compute the cascade matrix, $C$. We then extract level-resolved recombination rates, $\alpha_{nl}$, from \cloudy{}. Both $C$ and $\alpha_{nl}$ are restricted to $n\le n_{\rm max}$, specified by the user. We can then compute the level population from Equation~(\ref{eq:n_nl}), for any $n\le n_{\rm max}$, for gas at a given temperature, $T$, with a given level of ionization specified by the electron and proton density, $n_e$ and $n_p$.

The left-hand side of Equation~(\ref{eq:CMF}) is reminiscent of the left-hand side in the equation for the neutral fraction, $x$, of gas with hydrogen density $n_{\rm H}$ in photoionization equilibrium,
\begin{align}
    n_{\rm HII} n_e \sum_{n=n_{\rm low}}^{\infty}\alpha_{nl}(T) &= x n_{\rm H}\Gamma,
    \label{eq:photo}
\end{align}
where $\Gamma$ is the photoionization rate. We exploit the similarity between these two equations to gain physical insight, as follows. Write the solution for the neutral fraction $x$ from Equation~(\ref{eq:photo}) in the form
\begin{align}
    x &\equiv \frac{n_{\rm HI}}{n_{\rm H}} = (1-x)^2 \frac{\tau_i}{\tau_r},
    \label{eq:x}
\end{align}
where $\tau_r$ is the recombination time and $\tau_i$ the ionization time,
\begin{align}
    \frac{1}{\tau_r} &= n_{\mathrm H} \sum_{n=n_{\rm low}}^\infty\sum_{l=0}^{n-1} \alpha_{nl}(T)\equiv \alpha_r(T);\quad \frac{1}{\tau_i} = \Gamma.
\end{align}
When elements other than hydrogen are neglected, $n_e=n_{\rm HII}=(1-x)n_{\rm H}$. Next we re-write Equation~(\ref{eq:CMF}) in the form
\begin{align}
    x_{nl} &\equiv \frac{n_{nl}}{(2l+1)\,n_{\rm H}} = (1-x)^2 \frac{\tau_{A, nl}}{\tau_{nl}},
    \label{eq:xnl}
\end{align}
where we defined 
 \begin{align}
    \frac{1}{\tau_{nl}} &= \frac{n_{\mathrm H}}{2l+1} \sum_{n'=n+1}^\infty\sum_{l'=0}^{n'-1} \alpha_{n'l'}(T)n C_{n'l',nl}\,\nonumber\\
    \frac{1}{\tau_{A, nl}} &= \sum_{n''=n_{\rm low}}^{n-1} \sum_{l''=0}^{n''-1} A_{nl, n''l''}.
\end{align}
Comparing Equations~(\ref{eq:x}) and (\ref{eq:xnl}) brings out the similarity between the two cases: when $x\ll 1$, the neutral fraction is set by the ratio $\tau_i/\tau_r$, where $\tau_i$ is the rate at which neutral gas is being depleted due to photo-ionization, and $\tau_r$ the rate at which it is formed due to recombinations. Similarly, for $x\ll 1$, the fraction of the gas in a given level, $x_{nl}$, is the ratio $\tau_{A, nl}/\tau_{nl}$, where
$\tau_{A, nl}$ is the rate at which the state $nl$ is depopulated due to spontaneous decay, and $\tau_{nl}$ is the rate at which it is populated due to the cascade from higher levels.

We can combine Equations~(\ref{eq:x}) and (\ref{eq:xnl}) to obtain a relation between the fraction of hydrogen atoms in a given state, the neutral fraction, and ratio's of characteristic times, valid when $x\ll 1$,
\begin{align}
    x_{nl} &= x \frac{\tau_r}{\tau_{nl}} \frac{\tau_{A, nl}}{\tau_i}.
\end{align}
We note that $\tau_r$ and $\tau_{nl}$ are both proportional to $1/n_{\rm H}$, so that their ratio is density independent; since $\tau_i$ and $\tau_{A, nl}$ also do not depend on
$n_{\rm H}$, $x_{nl}/x$ is density independent as well. Interestingly, the ratio $x_{nl}/x$ also depends only relatively weakly on temperature due to the weak dependence of recombination rate coefficient on temperature. 

\subsection{Convergence of the level population with $n_{\rm max}$}
\label{sec:cmf_validation}
\begin{table*}
    \centering
    \caption{A model variant to the Ref - ALSh model in Table~\ref{tab:cloudy_models}). In Ref - ALSh - RR, we disable all collisional processes, including collisional ionization, collisional excitation and $l$-mixing. }
    \label{tab:cloudy_rr_model}
    \begin{tabular}{ccccc} 
    \hline
    Models & Collisional ionization & Collisional excitation & $l$-mixing & Recombination top-off\\
    \hline
    Ref - ALSh & On & On & On & On\\
    Ref - ALSh - RR & Off & Off & Off & Off\\
    \hline
    \end{tabular}
\end{table*}

\begin{figure}
    \includegraphics[width=\columnwidth]{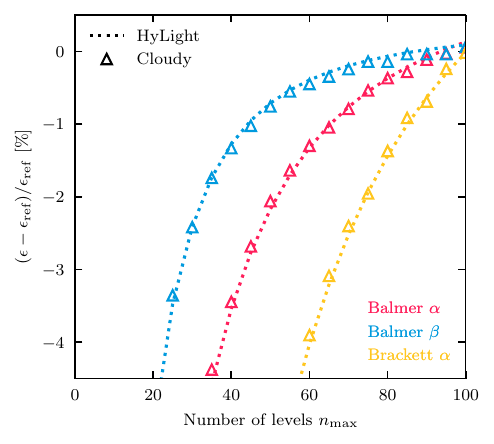}
    \caption{Convergence of the emissivity, $\epsilon$, for selected hydrogen transitions for different models in Case A. The reference setup is \lq Ref - ALSh\rq\, as shown in Table~\ref{tab:cloudy_models}. The horizontal axis is the number of resolved levels in the atomic model, while the vertical axis indicates the emissivity difference in per cent between a given model and the reference model. The dotted line shows the predictions from radiative-only \hylight{} predictions. The open triangles are predictions from the \lq LShA - Ref - RR\rq\ \cloudy{} model, which only includes radiative processes, as described in Table~\ref{tab:cloudy_rr_model}. There is excellent agreement between \cloudy{} and \hylight{}. As the number of levels increases, both models converge to the reference model. We conclude that 100 resolved levels is sufficient to produce accurate line emissivities. }
    \label{fig:ha_hb_bra_cvg_cmf_case_a}
\end{figure}
\begin{figure*}
    \includegraphics[width=\textwidth]{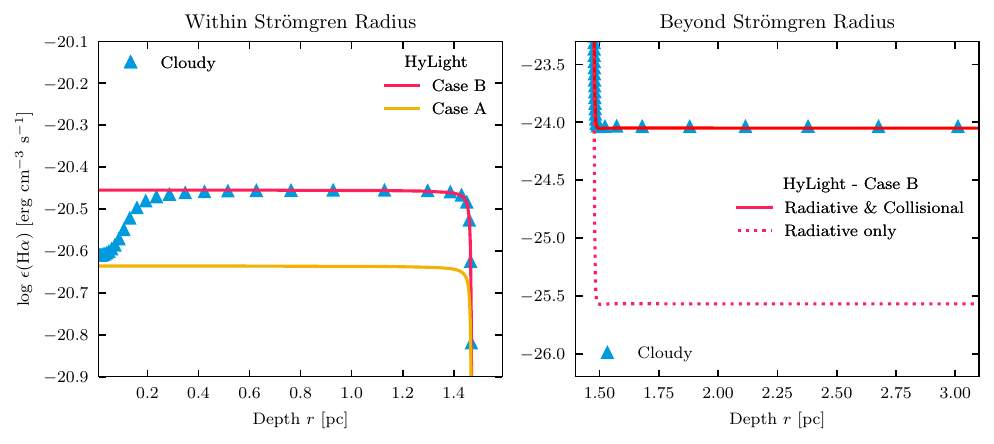}
    \caption{Profile of H$\alpha$ emissivity, $\epsilon({\rm H \alpha}) \equiv \epsilon_{3,2}$, of a gas cloud ionized by a laser beam. This \cloudy{} setup is labelled as \lq Ref - LSph\rq\ in Table~\ref{tab:cloudy_models}. The Str{\"o}mgren radius is approximately 1.5~pc. The solid curves are \hylight{} predictions when including radiative {\em and} collisional contributions. Within the Str{\"o}mgren radius, the H$\alpha$ emissivity gradually switches from Case~A to Case~B in the region $r \lesssim 0.4$ pc, a result of the finite optical depth of Lyman photons. Beyond the Str{\"o}mgren radius, the radiative-only approximation does not hold any more, since collisional excitation of the ground state atoms contributes dominantly: these are not included in the radiative-only calculation. The full \hylight{} model agrees well with \cloudy{}.}
    \label{fig:sph_comp_laser}
\end{figure*}

The recurrence calculation used to evaluate the cascade matrix $C$ using Equation~(\ref{eq:C_definition}) requires choosing a maximum $n$ level, $n_{\rm max}$, that is accounted for in the calculation. Here, we investigate the convergence of the results when increasing $n_{\rm max}$.

The setup is that of a thin shell with a source in the centre. Recombinations are forced to be in the Case~A limit. Rather than using a blackbody spectrum for the ionizing source as we did so far, we instead use a source which emits ionizing photons in a very narrow\footnote{The relative laser width is chosen to be 0.001. The detailed definition of relative laser width can be found in \cloudy{} documentations. The choice of laser width is discussed in Appendix~\ref{app:laser}. } energy range around $1.1~\unit{Ryd}$ (termed \lq laser\rq\ in \cloudy{}; the photon luminosity is $Q({\rm H})=10^{48}~\unit{s^{-1}}$). Such a sharply peaked spectrum avoids that hydrogen atoms in the ground state are excited by absorbing high-order Lyman photons from the source (a process not accounted for in our model). We investigate the importance of such \lq Lyman pumping\rq\ in \S~\ref{sec:limitations}. The model is labelled as \lq Ref - LShA\rq\ model in Table~\ref{tab:cloudy_models}.

We use \cloudy{} to compute the ionization state of the gas in the shell. We then compute the emissivity for a series of hydrogen lines (i) using \cloudy{} which includes collisional effects by default, (ii) using \cloudy{} but restricting its calculation to include radiative transitions only, and (iii) the \hylight{} model restricted to radiative transitions. For all three cases, we compute the emissivity while varying the number of levels (the number of $l$-resolved in \cloudy{}, $n_{\rm res}$, and the maximum number of levels included in our cascade matrix calculation, $n_{\rm max}$). The model (i) is labelled \lq Ref - ALSh\rq\ and the model (ii) is labelled \lq Ref - ALSh - RR\rq\ in Table~\ref{tab:cloudy_rr_model}. The table shows in detail which physical processes are enabled or disabled. We compare the results of models (ii) and (iii) to that of model (i) in Fig.~\ref{fig:ha_hb_bra_cvg_cmf_case_a}. Different colours correspond to Balmer~$\alpha$ (red), Balmer~$\beta$ (blue) and Brackett~$\alpha$ (yellow). 
The \cloudy{} model with radiative processes only is shown by open triangles; our \hylight{} model is shown by dotted lines. All emissivities are scaled to the prediction by the model (i), denoted by $\epsilon_{\rm ref}$ in Fig.~\ref{fig:ha_hb_bra_cvg_cmf_case_a}, which has all the physical processes enabled and an atomic model of 100 resolved levels and one collapsed level. Our radiative-only \hylight{} model and the radiative-only version of \cloudy{} agree very well for all $n$, both converging to better than a per cent to $\epsilon_{\rm ref}$ when $n_{\rm res}\ge 80$. The Bracket emissivity converges more slowly than the Balmer emissivities. The comparison between the model (i) (Ref - ALSh in Table~\ref{tab:cloudy_rr_model}) and the model(ii) (Ref - ALSh - RR in Table~\ref{tab:cloudy_rr_model}) is detailed in Appendix~\ref{app:cloudy_model_convergence}. 

We conclude from the above comparison that we need values of at least $n_{\rm max}\sim 100$ in our model (and of $n_{\rm res}$ in \cloudy{}) to obtain converged results in the computation of Balmer and Brackett lines. The value of this maximum level may be temperature dependent, and likely needs to be higher for the Paschen and even higher for the Brackett series. We note that the default value in \cloudy{} is $n_{\rm res}=10$, which may not be high enough to get accurate results for \ion{H}{ii}-like nebulae. More details on convergence are presented in Appendix~\ref{app:cloudy_model_convergence}. Interested readers can also refer to \citealt{Guzman2025} for more in-depth discussions. Increasing $n_{\rm res}$ increases the run-time and memory requirements of \cloudy{} significantly, since the computation of the cascade matrix scales as $\mathcal{O}(n^2)$ with the number of included levels. However, because in our implementation the cascade matrix is stored in memory and computed only once, the subsequent calculation time scales as $\mathcal{O}(n)$ with the number of physical condition entries. 

We further test our choice of $n_{\rm max}$ in the following more realistic setting. A spherical cloud (hydrogen only, density $n_{\rm H}=10^2~\unit{cm^{-3}}$, constant temperature $T=10^4~\unit{K}$) is ionized by the same laser-like central source as before. We compare our model for recombinations in the Case~A and Case~B limit (both with $n_{\rm max}=100$) to \cloudy{} (including collisions, not specifying either Case~A or Case~B, ignoring Lyman pumping from the source, and setting $n_{\rm res}=100$ with one unresolved level). This setup is labelled as \lq Ref - LSph\rq\ in Table~\ref{tab:cloudy_models}. 

The results are shown in Fig.~\ref{fig:sph_comp_laser}, with \cloudy{} results shown by filled blue triangles and \hylight{} shown with lines (red for Case~B, yellow for Case~A); the left panel shows the highly-ionized part of the nebula, up to the Str{\"o}mgren radius at $r\approx 1.5~\unit{pc}$. The left panel shows that our Case~B result agrees well with \cloudy{} for $r\ge 0.4~\unit{pc}$. However, close to the source, $r\le 0.1~\unit{pc}$, the optical depth in Lyman lines is small and \cloudy{}'s result is closer to our Case~A emissivity. Reproducing the resulting dip in emissivity close to the source requires performing radiative transfer of Lyman photons, which we don't do. The right panel shows the emissivity outside of the Str\"omgren radius where the gas is mostly neutral. Our calculation using radiative rates only (dotted red line labelled \lq radiative only\rq) yields an emissivity that is $\approx 1.5$~dex below that of \cloudy{}. Since the ionizing flux is very low, H$\alpha$ emission is a consequence of collisional excitations, which are not included in the radiative-only calculation. The solid red line is our model with collisional excitation included, and this reproduces the \cloudy{} results well. We describe these collisional excitation terms next (\S~\ref{sec:model_collisions}).
\subsection{The level population including collisional excitation}
\label{sec:model_collisions}
\begin{figure}
    \includegraphics[width=\columnwidth]{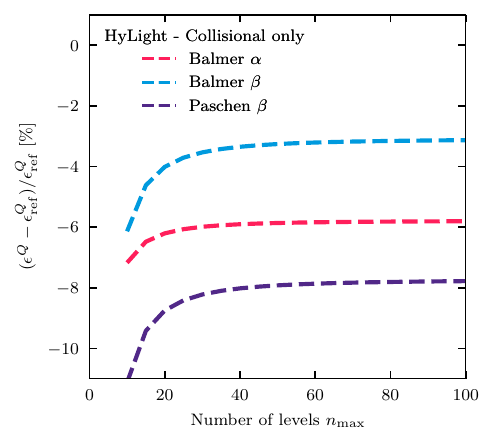}
    \caption{Convergence of the collisional contribution to H$\alpha$ emissivity as a function of the number of resolved levels, $n_{\rm max}$. The reference setup is \lq Ref - LSph\rq\, where the temperature is kept at 10$^4$~K throughout the cloud. The reference emissivity, $\epsilon_{\rm ref}$, is extracted at 2~pc from the source, where the gas is neutral and collisional excitation dominates over radiative recombination. As the number of levels increases in the model, the line emissivity converges. The difference between the converged answer and the \cloudy{} predictions is lower for low-order series (e.g., smaller than 6\% for H$\alpha$ and H$\beta$) but higher for higher series such as Paschen~$\beta$ (about 8\%). Given that the collisional contribution to recombination lines is only relevant in the partially neutral plasma and several orders of magnitude lower than that from radiative recombination within the Str{\"o}mgren radius, the difference of a few per cent is satisfactory. The differences come from disregard of collisional excitation and de-excitation between and from levels other than the ground state, as well as the exact calculation of collisional excitation rate for $nl$-resolved levels (see Appendix~\ref{app:collisions}). }
    \label{fig:ha_hb_brb_cvg_col_case_b}
\end{figure}

The level population when including collisions from the ground state, as well as radiative transitions, is
\begin{align}
    n_{nl} &= R_{nl}(T)n_e\,n_p + Q_{nl}(T)\,n_{\rm HI}n_e\nonumber\\
    Q_{nl}(T) &\equiv \frac{\sum\limits_{n'=n+1}^\infty\sum\limits_{l'=0}^{n'-1} q_{1s,n'l'}(T) C_{n'l',nl}}
    {\sum\limits_{n''=n_{\rm low}}^{n-1} \sum\limits_{l''=0}^{n''-1} A_{nl, n''l''}},
    \label{eq:nlQ}
\end{align}
where $R_{nl}(T)$ is defined in Equation~(\ref{eq:Rnl}). The level population is therefore a sum of a term due to radiative recombinations (first term) and a term due to collision excitations from the ground state (second term), including both excitations directly to the level $nl$ and excitations to higher levels that then cascade down to $nl$. This equation neglects excitations and de-excitation from states other than the ground state, and further makes the approximation that the vast majority of the neutral hydrogen atoms are in the ground state, so that $n_{1s}\approx n_{\rm HI}$\footnote{As an example, in the setup \lq Ref - Sph\rq, at 0.01 pc, the 3$s$ level population density is of the order of 10$^{-17}$~cm$^{-3}$, whereas the ground state 1$s$ level population density is $\approx 10^{-5}$~cm$^{-3}$, meaning the vast majority of the hydrogen atoms are indeed in the ground state. }. 

From Equation~(\ref{eq:nlQ}),  we show that the emissivity of a given hydrogen recombination line is also a sum of two terms, one due to recombinations, and one due to collisional excitations,
\begin{align}
    \epsilon_{n'n}(T) &= \epsilon_{n',n}^R(T) + \epsilon_{n',n}^Q(T)\nonumber\\
    \epsilon_{n'n}^R(T) &= n_e n_p h\nu_{n',n} \sum_{l'=\pm l} R_{n'l'}(T) A_{n'l', nl}\nonumber\\
    \epsilon_{n'n}^Q(T) &= n_e n_{\rm HI} h \nu_{n',n} \sum_{l'=\pm l} Q_{n'l'}(T) A_{n'l', nl} ,
\end{align}
where $h\nu_{n'n}$ is the energy of the photon emitted in a transition $n'\to n$.

We calculate the temperature-dependent $q_{1s, nl}$ rates based on \citealt{Anderson2000}, \citealt{Anderson2002} and \citealt{Lebedev1998} (see Appendix~\ref{app:collisions} for details). Summing these as per Equation~(\ref{eq:nlQ}) yields the collisional terms, $Q_{nl}(T)$.

Similar to the case of the radiative calculations, we need to choose a maximum value, $n_{\rm max}$, for the hydrogen principal quantum number that will be included in the rate calculations (see Equation~(\ref{eq:nlQ})). We examine the convergence with $n_{\rm max}$ for some hydrogen lines in Fig.~\ref{fig:ha_hb_brb_cvg_col_case_b}. The model simulated is
\lq Ref - LSph\rq\ from Table~\ref{tab:cloudy_models}, and we evaluate the emissivity at $2~\unit{pc}$ from the source, where the gas is mostly neutral and collisional excitation dominates the emissivity. 

The results of our calculations are compared to a \cloudy{} reference model which uses $n_{\rm res}=100$ resolved levels. We find that our rates converge for $n_{\rm max}\ge 40$, but the converged value differs from the \cloudy{} value by up to 8 per cent for the Paschen $\beta$ line. This difference is due to collisional excitation and de-excitation from levels other than the ground state, which are accounted for in \cloudy{} but not in \hylight{}. Furthermore, our treatment of converting $n$-resolved collisional excitation rate to $nl$-resolved rate is slightly different from \cloudy{} (see 
Appendix~\ref{app:collisions})

We conclude from this comparison that \hylight{} is not as accurate as \cloudy{} in predicting the emissivity of mostly neutral gas, where the emission is dominated by collisional excitation. Fortunately, the emission in nebular regions is very much dominated by the {\em ionized} component, for which the agreement with \cloudy{} is excellent.
\subsection{Modelling two-photon emission}
\label{sec:two_photon_process}
\begin{figure}
    \includegraphics[width=\columnwidth]{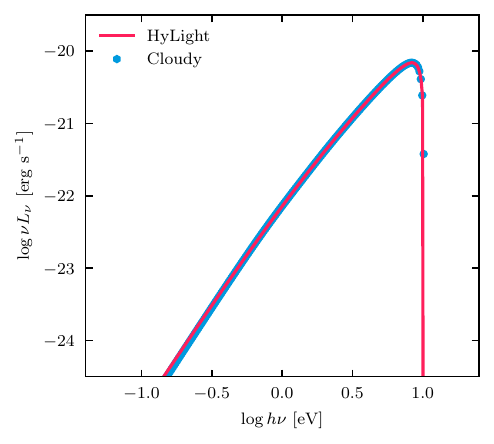}
    \caption{Spectrum of two-photon continuum for the setup of the model 
    \lq Ref - ASh\rq~of Table~\ref{tab:cloudy_models} (thin shell ionized by a blackbody). We find the agreement of total luminosity to be better than 0.05 per cent between the model of Equation~(\ref{eq:two_photon_emissivity}) and that of the reference \cloudy{} model.}
    \label{fig:two_photon_spectrum}
\end{figure}

The $2s \to 1s$ transition in \ion{H}{i} is dipole forbidden because it violates the $\Delta \, l=\pm 1$ selection rule. As a result, the 2$s$ state is metastable, with no allowed transition to the ground state. However, an \ion{H}{i} atom in the 2$s$ state can still decay to 1$s$ via a virtual intermediate state at a higher order in perturbation theory, a process that results in the emission of two photons. Energy conservation requires that the sum of the photon energies equals that of a Lyman $\alpha$ photon, $h\nu_1+h\nu_2=(3/4)~{\rm Ryd}$. In fact, two-photon transitions are currently the most accurate way to measure the value of the Rydberg constant in \unit{eV} \cite[e.g.][]{Grinin2020}.

\cite{nussbaumer1984} (see also \citealt{Bottorff2006} and references therein) approximate the rate and spectrum of the $2s\to1s$ two-photon transition as
\begin{align}
    A(y)&=C\left(
    y(1-y)\left[1-D\right]
    +\alpha\left[y(1-y)\right]^\beta D
    \right)\nonumber\\
    D &= \big(4y(1-y)\big)^\gamma,
\end{align}
with fitting parameters $C=202.0~\unit{s^{-1}}$, $\gamma=0.8$, $\alpha=0.88$, $\beta=1.53$. The variable $y=\nu/\nu_{2,1}$, where $\nu$ is the frequency of one of the two photons (and hence $0\le y\le 1$), and $\nu_{2,1}$ is the frequency of a Lyman~$\alpha$ photon. The total rate - {\em ie} the effective Einstein coefficient due to two-photon decay - is found by integrating over all $y$,
\begin{align}
    A^{\rm eff}_{2s, 1s} &=\frac{1}{2}\,
    \int_0^1 A(y) \mathrm{d}y = 8.224~\unit{s^{-1}},
    \label{eq:2photon}
\end{align}
(where the factor 1/2 corrects for double counting), which is close to the value $8.234~\unit{s^{-1}}$ quoted by~\cite{Labzowsky2006}. The calculation above does not account for the presence of an ambient radiation field~\citep[see the induced rate calculation in e.g.][]{Bottorff2006, Chluba2006}. 

The $2p\to 1s$ transition is dipole allowed. There is also a two-photon decay channel for this transition, but the corresponding rate is much lower than for the $2s\to 1s$ transition: the two-photon E1E2 decay and E1M1 decay transition probability are of the order of 10$^{-6}$~s$^{-1}$~\citep{Labzowsky2006}. For this reason, we suspect two-photon decay from 2$p$ to 1$s$ is not important in astrophysical situations. 

The specific emissivity per unit volume of the 2$s$ to 1$s$ two-photon process, $\epsilon_\nu$, is
given in terms of the function $A(y)$, by
\begin{equation}
    \label{eq:two_photon_emissivity}
    \epsilon_\nu \mathrm{d}\nu = h \nu A\bigg (\frac{\nu}{\nu_{2,1}} \bigg) n_{2s} \frac{\mathrm{d} \nu}{\nu_{2,1}},
\end{equation}
where $\nu$ is the frequency of the photon emitted in the two-photon process and $n_{2s}$ is the population density of the 2$s$ level.

We compare the specific emissivity computed from Equation~(\ref{eq:two_photon_emissivity}) for model \lq Ref - ASh\rq~of Table~\ref{tab:cloudy_models} against the reference \cloudy{} model in Fig.~\ref{fig:two_photon_spectrum}. The agreement is excellent. 

\subsection{Processes not included in \hylight{}}
\label{sec:limitations}
We briefly discuss the physical processes not included in \hylight{}.
\begin{figure}
    \includegraphics[width=\columnwidth]{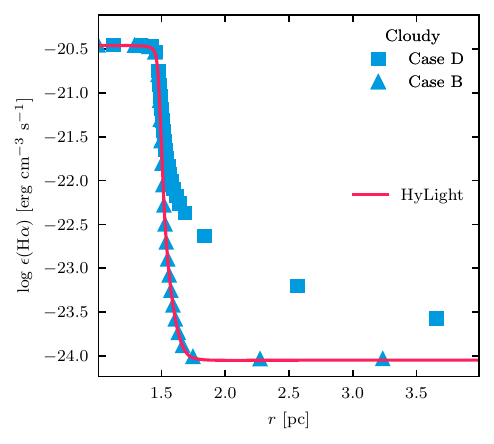}
    \caption{Test of the importance of Lyman photons from the source: H$\alpha$ emissivity profile of a spherical gas cloud ionized by a source with a blackbody spectrum. The model is referred to as \lq Ref - Sph\rq{}~in Table~\ref{tab:cloudy_models}. The \cloudy{} Case B model ({\em triangles}) does not account for Lyman photons from the source, whereas the Case D model ({\em squares}) does.
    Lyman photons from the source affect the H$\alpha$ emissivity by a few per cent outside the Str\"omgren radius, $r\approx 1.5$~pc. The \hylight{} emissivity ({\em red line}) does not account for Lyman photons from the source, and agrees well with the \cloudy{} Case B prediction.}
    \label{fig:sph_comp_case_b_d}
\end{figure}
\begin{enumerate}
\item {{\em Line pumping.}} Recombining hydrogen gas emits Lyman lines at a rate that can be computed as described in the previous sections. The optical depth to these lines is large in typical \ion{H}{ii} regions, in particular for the first few Lyman lines. The central optical depth for Lyman~$\alpha$ is
\begin{align}
    \label{eq:lyman_optical_depth}
    \tau_c &= N_{\rm HI}\,\left(\frac{3\pi\sigma_T}{8}\right)^{1/2} f_{1,2} \lambda_{1,2} c \left(\pi b^2\right)^{-1/2}\nonumber\\
    &\approx 1.0\,\left(\frac{N_{\rm HI}}{2.64\times 10^{13}~\unit{cm^{-2}}}\right) \left(\frac{20~\unit{km~s^{-1}}}{b}\right),
\end{align}
where $\sigma_T$ is the Thomson cross section, $f_{1,2}=0.4164$ is the oscillator strength, $\lambda_{1,2}=1215.67~\unit{\angstrom}$ is the laboratory wavelength, and $b$ is the line width. This equation yields values of $\tau_c$ of several thousand for a typical \ion{H}{ii} region (say $n_{\mathrm H}=10^2~\unit{cm^{-3}}$, extent $r=3~\unit{pc}$, and \ion{H}{i} neutral fraction $10^{-4}$). For this reason, Case~B recombination is usually the better choice. The reasonable agreement between \cloudy{} (which includes line pumping) and the Case~B \hylight{} calculation (which does not) of the emissivity - shows that the non-inclusion of line pumping is a reasonable approximation for \ion{H}{ii} regions.

Nebular gas may also absorb Lyman photons emitted by the {\em source}, sometimes referred to as Case~D. We investigate the impact of this process on the H$\alpha$ emissivity in Fig.~\ref{fig:sph_comp_case_b_d} for model \lq Ref - Sph\rq~from Table~\ref{tab:cloudy_models} (spherical cloud, uniform density and temperature, blackbody source). We compare \cloudy{} runs for Case~B and Case~D to the full \hylight{} model. Within the Str\"omgren radius, $r_S\approx 1.5~\unit{pc}$, the two \cloudy{} runs and the \hylight{} model agree very well. Outside $r_S$, \hylight{} agrees very well with the \cloudy{} Case~B run, whereas the \cloudy{} Case~D emissivity is a few per cent higher. We conclude that in this type of setup, line pumping by photons from the source is not very important. For more discussion on this topic, see e.g.~\citealt[][]{Luridiana2009}. In addition, we intend to use \hylight{} as a tool for post-processing hydrodynamic simulations; hence, it would be challenging to capture Lyman pumping, which would require us to follow the radiative transfer of Lyman photons. 

\item {\em Stimulated emission.} \hylight{} does not include stimulated emission, whereas \cloudy{} does by default. Again, the fact that the Case~B \hylight{} emissivity agrees well with \cloudy{} in Fig.~\ref{fig:sph_comp_case_b_d} implies that stimulated emission is not important in \ion{H}{ii} regions. Similarly, in terms of post-processing radiation hydrodynamic simulation, it would be challenging to include stimulated emission as it would require coupling the computation of the level populations to the radiative transfer of photons from all sources in the simulation. 

\item {\em Collisional processes.} \hylight{} includes the impact of collisions with electrons on the level population when they excite \ion{H}{i} atoms that are in the ground state (1$s$). However, \hylight{} neglects collisional excitation from states with $n>1$ and, in particular, of collisions that change $l$ without changing $n$. In this case, the $l$-levels are solutions of Equation~(\ref{eq:equilibrium_eq_full}). We have demonstrated that this limitation only leads to a few per cent difference compared to the full collisional modelling in \S~\ref{sec:model_collisions}. 

\begin{figure}
    \includegraphics[width=\columnwidth]{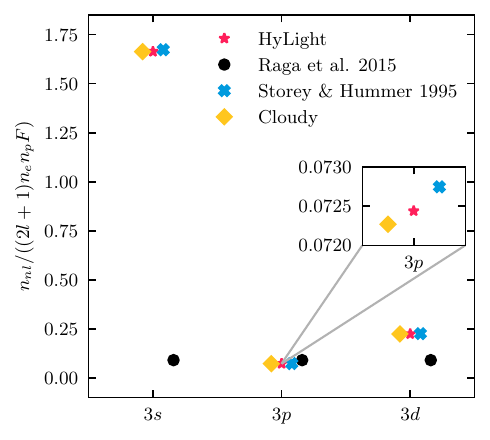}
    \caption{Scaled level population for level $n_{3l}$ as a function of $l$ in setup Ref-ASh, the factor $F$ is the mean level population calculated using Equation~(\ref{eq:factor_F}). Results are shown for \hylight{} ({\em red stars}), \protect\citet{Raga2015} ({\em black circles}), \protect\citet{Storey1988} ({\em blue crosses}) and \cloudy{} ({\em yellow diamonds}). All models agree well, except for that of \protect\citet{Raga2015} which assumes that the population level is set by statistical equilibrium, $n_{3l}\propto (2l+1)$. The inset visualizes the good level of agreement. Symbols are slightly offset in the horizontal direction to avoid overlap.}
    \label{fig:lvlpop_n3}
\end{figure}

The model of \citet{Raga2015} assumes\footnote{This is not the only difference; for example, that model also uses $n_{\rm max}=5$ energy levels.} that $l$-levels are populated according to their statistical weight, i.e. $n_{nl}\propto (2l+1)$ at given $n$. Fig.~\ref{fig:lvlpop_n3} shows the level populations scaled by $2l+1$ for the $3s$, $3p$ and $3d$ states for different models, using the setup \lq Ref - ASh\rq. The level population is also scaled by $n_e$, $n_p$, and $F$, where $F$ is the average population density for a given $n$-level:
\begin{equation}
    F = \frac{1}{n} \frac{\sum\limits_{n'=n+1}^\infty\sum\limits_{l'=0}^{n'-1} \alpha_{n'l'}(T) C_{n'l',nl}}{\sum\limits_{n''=n_{\rm low}}^{n-1} \sum\limits_{l''=0}^{n''-1} A_{nl, n'l'}}.
    \label{eq:factor_F}
\end{equation}
The values in the \cite{Raga2015} model (filled black circles) are then independent of $l$ by construction. The values obtained with \cloudy{} (yellow diamonds), \cite{Storey1995} (blue crosses) and \hylight{} (red stars) agree very well with each other, and show that the population of the $3s$ state is much higher than it would be under the assumption that $l$-levels are populated according to $2l+1$ statistical weights. In conclusion, the model of \cite{Raga2015} are not intended to be used for low-density nebular regions but dense planetary nebulae where the gas density is high enough to populate the $l$-levels within an $n$-level according to $2l+1$ statistical weights. 

Collisions with protons can change $l$ for a given $n$ \cite[e.g.][]{Pengelly1964b, Vrinceanu2001}, often referred to as $l$-mixing. Such collisions do indeed tend to populate $l$-levels according to their statistical weights. The rate of $l$-mixing depends on the gas metallicity, temperature, and element abundances, and is strongly dependent on density~\citep{Pengelly1964b}. By comparing the characteristic $l$-mixing timescale to the radiative timescale using the method by \cite{Vrinceanu2019}, we can estimate the effect of $l$-mixing on each $n$-level. At the typical condition for \ion{H}{ii} regions, for highly excited states, the $l$-mixing timescale is much shorter than the radiative timescale, hence the $l$-states tend to be in statistical equilibrium. For lower levels, the time it takes to redistribute the $l$-states is much larger than that for the radiative process. Therefore, within a given $n$-level, the $l$-states will not follow the statistical weight and are simply set by Equation~(\ref{eq:equilibrium_eq_full}). Given the strong density dependence, the impact of collisions is typically formulated in terms of a critical density below which collisions can be safely ignored, with the value of this critical density depending on $n$. 

\begin{figure*}
    \includegraphics[width=\textwidth]{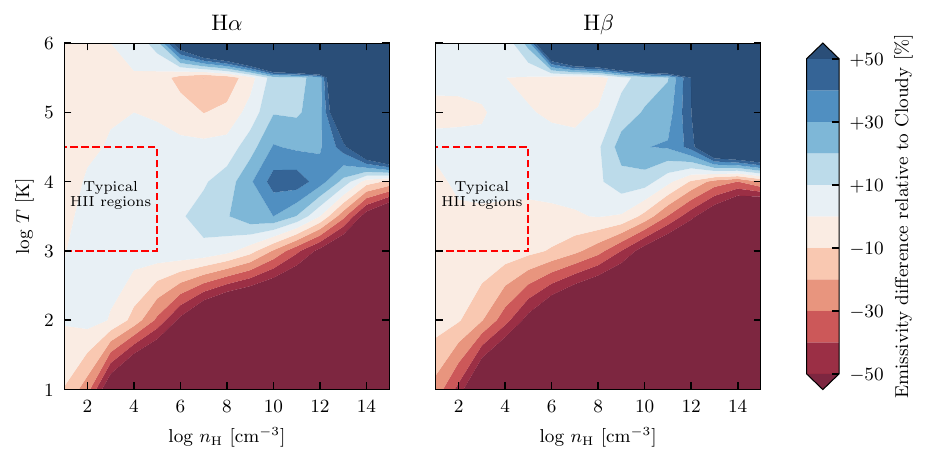}
    \caption{Relative difference between line emissivity predicted by \hylight{} and by \cloudy{} for an extended range of temperatures and gas densities. Under typical \ion{H}{ii} region conditions (enclosed in the red dashed line), the relative difference between H$\alpha$ (left panel) and H$\beta$ (right panel) emissivity predicted by \hylight{} and those by \cloudy{} is well below 10\%. At high densities, differences of up to 50 per cent arise because \hylight{} does not include $l$-mixing collisions. Comparison for other common recombination lines is shown in Appendix~\ref{app:hylight_limitation}. }
    \label{fig:hylight_limit_Ha_Hb}
\end{figure*}

Ignoring $l$-mixing collisions limits the accuracy of
\hylight{} under certain physical conditions. We illustrate these limitations 
by plotting the relative difference between the line emissivity of H$\alpha$ and H$\beta$ predicted by \hylight{} and predicted by \cloudy{} in Fig.~\ref{fig:hylight_limit_Ha_Hb}. The \hylight{} model is not accurate in the high-density and high-temperature regime due to its neglect of collisional processes, in particular, of $l$-mixing collisions. Fortunately, \hylight{} and \cloudy{} agree to a few per cent at typical \ion{H}{ii} region conditions (delineated by a red dashed line in Fig.~\ref{fig:hylight_limit_Ha_Hb}). The comparison of other typical recombination lines can be found in Appendix~\ref{app:hylight_limitation}, where similar conclusions can be drawn.

\begin{figure*}
    \includegraphics[width=\textwidth]{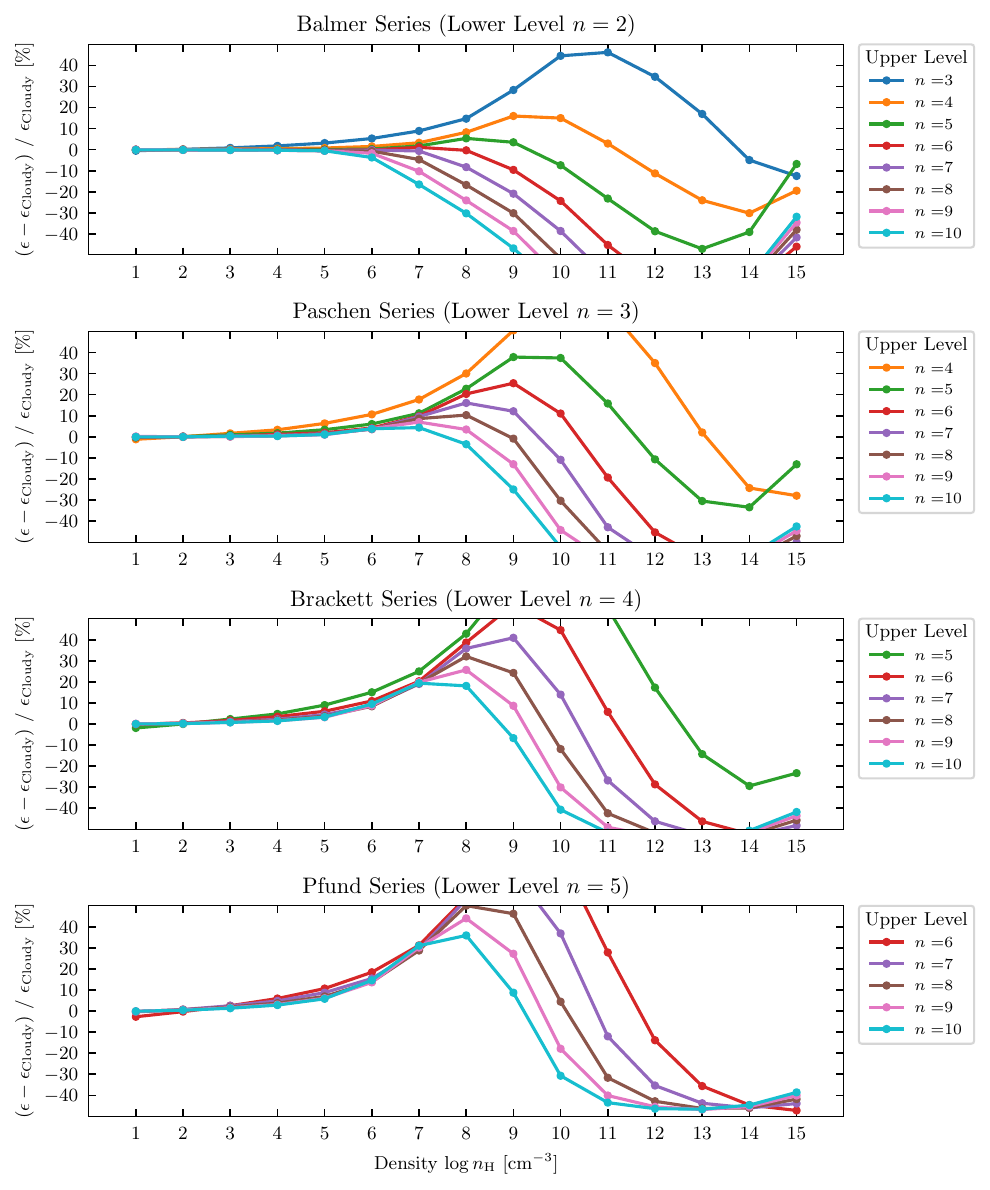}
    \caption{Relative emissivity difference in per cent between \hylight{} and \cloudy{} for various transitions at a fixed gas temperature $T=10^4$~K across a range of gas density ($1 < \log n_{\rm H} < 15$). The inputs for \hylight{} in this comparison (hydrogen density and ionization fractions, electron density, and gas temperature) are all taken from \cloudy{} outputs. At $T \approx 10^4$~K, the difference in emissivity prediction between \hylight{} and \cloudy{} is at the per cent level for typical \ion{H}{ii} region density (1 < $\log n_{\rm H}$ < 4). Due to incomplete modelling of collisional processes, the difference between \hylight{} and \cloudy{} becomes more significant in high-density regions. We conclude that \hylight{} is able to predict typical recombination lines accurately for typical \ion{H}{ii} region conditions. }
    \label{fig:hylight_limit_10}
\end{figure*}

For a typical \ion{H}{ii} region temperature of $T = 10^4$ K, we show the relative difference in line emissivity of common recombination lines between \hylight{} and \cloudy{} in Fig.~\ref{fig:hylight_limit_10}. At typical \ion{H}{ii} region densities ($10 < n_{\rm H} < 10^4$~cm$^{-3}$), the difference between \hylight{} and \cloudy{} is only a few per cent for typical recombination lines. Therefore, we conclude that \hylight{} model is accurate for typical \ion{H}{ii} region conditions. 

\begin{figure}
    \includegraphics[width=\columnwidth]{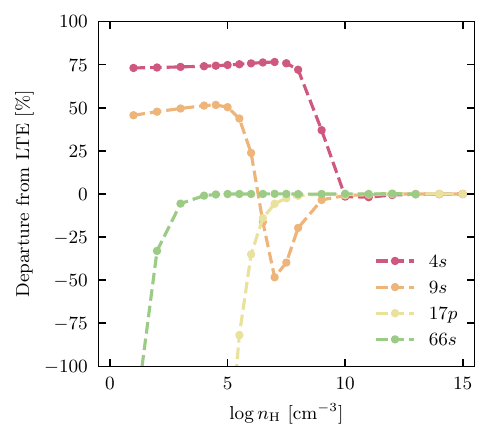}
    \caption{Departure from LTE (relative difference in level population density from LTE predictions in per cent) for different states (4$s$ in red, 9$s$ in orange, 17$p$ in yellow and 66$s$ in green) at a fixed temperature of 10$^4$~K as a function of gas density. The level populations are extracted from \cloudy{} \lq Ref - ASh\rq\ setup. The vertical axis shows the deviation from LTE (see Equation~(\ref{eq:levelpop_LTE})) in per cent. Lower levels are far from the LTE predictions at low densities. As the gas density increases, all the levels approach LTE. }
    \label{fig:departure_from_LTE}
\end{figure}

Without the $l$-mixing collisions, the $l$-states within an $n$ level are generally {\em not} distributed according to their statistical weights, even at high density when the system should be in LTE. We illustrate the density dependence of the departure from LTE for four example states (red4$s$, 9$s$, 17$p$, and 66$s$) in Fig.~\ref{fig:departure_from_LTE}. Using the setup of \lq Ref - ASh\rq{}~in Table~\ref{tab:cloudy_models}, we compute the level population with the reference \cloudy{} code and compare to the level population values assuming LTE (see Equation~(\ref{eq:levelpop_LTE}) in Appendix~\ref{sec:appendix_depar_coeff}). At the lowest density of $n_{\rm H} = 10~\unit{cm^{-3}}$, all three states are far from LTE, whereas at high density, $n_{\rm H} = 10^{15}~\unit{cm^{-3}}$, they are all in LTE. The critical density - the value where the level population transitions to LTE - is $n_{\rm H} \approx 10^{3}~\unit{cm^{-3}}$ for level $66s$, but $n_{\rm H} \approx 10^{10}~\unit{cm^{-3}}$ for $4s$. 

\end{enumerate}

%% file: sections/application.tex
\section{Example applications of the \hylight{} model}
\label{sec:application}
We illustrate the usefulness of the \hylight{} model with some examples.
The first example shows how \hylight{} branching ratios can be used to relate the emissivity in hydrogen lines to each other or to the ionization rate. The second example uses \hylight{} to compute \ifu\ data from a 3-dimensional radiation hydrodynamic simulation. 
\begin{table*}
\caption{Radiative-only branching ratio, $B_{2,1}^R$, computed using Equation~(\ref{eq:BRs}),  
for various temperatures, assuming Case~A and Case~B recombinations.}
\begin{tabular}{@{}ccccccccccc@{}}
\toprule
\multicolumn{1}{c}{\multirow{2}{*}{Line}} & \multicolumn{5}{c}{Case A} & \multicolumn{5}{c}{Case B} \\ \cmidrule(l){2-6} \cmidrule(l){7-11}
\multicolumn{1}{c}{} & $\log T$ = 3.0 & $\log T$ = 3.5 & $\log T$ = 4.0 & $\log T$ = 4.5 & $\log T$ = 5.0 & $\log T$ = 3.0 & $\log T$ = 3.5 & $\log T$ = 4.0 & $\log T$ = 4.5 & $\log T$ = 5.0 \\ \midrule
Balmer~$\alpha$   & 0.452 & 0.374 & 0.298 & 0.236 & 0.197 & 0.575 & 0.514 & 0.451 & 0.397 & 0.356 \\ 
Balmer~$\beta$    & 0.086 & 0.085 & 0.078 & 0.068 & 0.058 & 0.115 & 0.119 & 0.117 & 0.109 & 0.100 \\ 
Balmer~$\gamma$   & 0.032 & 0.034 & 0.033 & 0.030 & 0.026 & 0.044 & 0.048 & 0.049 & 0.047 & 0.043 \\ 
Paschen~$\alpha$  & 0.280 & 0.202 & 0.137 & 0.091 & 0.064 & 0.291 & 0.215 & 0.152 & 0.107 & 0.080 \\ 
Paschen~$\beta$   & 0.067 & 0.057 & 0.044 & 0.032 & 0.024 & 0.071 & 0.063 & 0.050 & 0.039 & 0.030 \\
Brackett~$\alpha$ & 0.185 & 0.121 & 0.074 & 0.045 & 0.029 & 0.187 & 0.124 & 0.077 & 0.048 & 0.032 \\
\bottomrule
\end{tabular}
\label{tab:branching_ratio}
\end{table*}

\subsection{Branching ratios and hydrogen line luminosities}
\begin{figure}
    \includegraphics[width=\columnwidth]{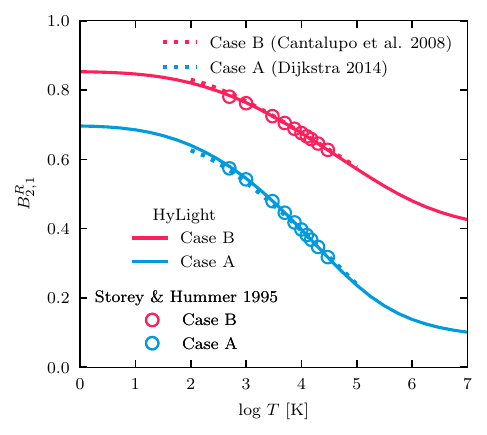}
    \caption{Fraction $B^R_{2,1}$ of the total number of recombinations that result in the emission of a Lyman~$\alpha$ photon as a function of temperature, assuming Case~A recombinations ({\em blue}), or Case~B recombinations ({\em red}). The \hylight{} model is shown by {\em solid lines}; the interpolation functions from \protect\cite{Cantalupo2008} and \protect\cite{Dijkstra2014} are shown with {\em dotted lines}; discrete \protect\cite{Storey1995} values are shown with {\em open circles}. All models, including \hylight{}, ignore collisional excitation, in which case $B^R_{2,1}$ is independent of density. All models agree well with each other. A comparison against \cloudy{} can be found in Appendix~\ref{app:cloudy_consistency}. }
    \label{fig:lymna_alpha_conversion_coeff}
\end{figure}

\begin{figure}
    \includegraphics[width=\columnwidth]{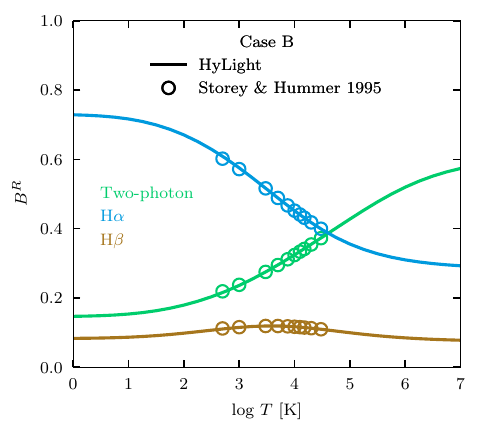}
    \caption{Fraction $B^R$ of the total number of recombinations that result in the emission of selected lines as a function of temperature in Case B. The \hylight{} model is shown as {\em solid lines}; the results from \protect\cite{Storey1995} are shown with {\em open circles}. Ignoring collisional excitation from the ground state, both models are in excellent agreement for all three chosen transitions, namely two-photon emission from the metastable 2$s$ state ({\em green}), H$\alpha$ ({\em blue}), and H$\beta$ (\em brown). }
    \label{fig:br_2photon_halpha_hbeta}
\end{figure}

We mean by \lq branching ratio\rq, $B_{nl, n'l'}$ of a particular transition\footnote{Note that the definition here is different from \citet[e.g.][\S~4.3]{Spitzer1978}. }, $nl\to n'l'$, the ratio of the rate at which this transition occurs over the total recombination rate. We similarly define the total branching ratio, $B_{n, n'}$, as the branching ratio for transitions from $n\to n'$ - summed over all $l$. As an example, the total branching ratio for the H$\alpha$ transition is
\begin{align}
    B_{3,2}(T) &= \frac{\mathrm{d}n_{3\to 2}(T)}{\mathrm{d}t}\frac{1}{n_e n_p \alpha_{\rm tot}(T)}\nonumber\\
    &=\frac{A_{3s, 2p} n_{3s}(T) + A_{3p, 2s} n_{3p}(T) + A_{3d, 2p} n_{3d}(T)}
    {n_e n_p \alpha_{\rm tot}(T)},
    \label{eq:B32}
\end{align}
where $\alpha_{\rm tot}(T)$ is the total recombination coefficient (which we can choose to be in the Case~A or Case~B approximation). 

We can compute the branching ratio starting from Equation~(\ref{eq:nlQ}) for the level population, so that in general
\begin{align}
    B_{nl, n'l'} (T) &= A_{nl, n'l'}\frac{R_{nl}(T)}{\alpha_{\rm tot}(T)} +
    A_{nl, n'l'}\frac{Q_{nl}(T)n_{\rm HI}}{n_p\alpha_{\rm tot}(T)}\nonumber\\
    &\equiv B^R_{nl, n'l'}(T) + B^Q_{nl, n'l'}(T)\frac{n_{\rm HI}}{n_p},
\end{align}
which shows that the branching ratio is the sum of a term 
due to recombinations (first term) and a term due to collisions (second term). Summing over $l$ yields the total branching ratio, $B_{n,n'}$, as
\begin{align}
    B_{n,n'}(T) &= \sum_{l=0}^{n-1} \sum_{l'=l\pm 1} B_{nl, n'l'} (T)\nonumber\\
    &\equiv B^R_{n, n'}(T) + B^Q_{n, n'} (T) \frac{n_{\rm HI}}{n_p}\,,
    \label{eq:BRs}
\end{align}
with the second line again separating recombinations from collisional excitations. 

When collisional excitations are ignored, the branching ratio relates the emissivity, $\epsilon_{n,n'}$, to the effective recombination rate, $\alpha^{\rm eff}_{n, n'}$:
\begin{align}
    \epsilon_{n,n'} &= h\nu_{n,n'} B^R_{n,n'}(T) \alpha_{\rm tot} n_e n_p\nonumber\\
       &\equiv h\nu_{n,n'}\alpha^{\rm eff}_{n,n'} n_e n_p\nonumber\\
    \alpha^{\rm eff}_{n,n'} &\equiv B^R_{n,n'}(T) \alpha_{\rm tot}.
    \label{eq:alpha_nn_eff}
\end{align}
The relation between the line emissivity, $\epsilon$, and the effective recombination rate, $\alpha^{\rm eff}$, was discussed by \citet{Osterbrock1974} (\S~4.3) and then formally derived by~\citet{Spitzer1978} (\S~9.1, table 9.1). Both have used two-photon emission as an example, but the spirit remains the same. With more accurate atomic data available, \citet{Cantalupo2008} derived a fitting formula for Lyman~$\alpha$, i.e. $B^R_{2,1}(T)$, in Case~B (accurate in the temperature range $10^2~{\unit K}\le T \le 10^5~\unit{K}$ based on \citealt{Pengelly1964} and \citealt{Martin1988}). The Case~A result is also quoted by \citealt{Dijkstra2014}.

It is straightforward to calculate $B^R_{n,n'}(T)$ and hence the effective recombination coefficient using \hylight{}. We plot the result in Fig.~\ref{fig:lymna_alpha_conversion_coeff} and Fig.~\ref{fig:br_2photon_halpha_hbeta} along with the results from \citealt{Cantalupo2008}, \citealt{Dijkstra2014} and \citealt{Storey1995}. 
The \citealt{Storey1995} values are evaluated from line emissivities (referred to as {\rm R\_NU} in the original table) based on Equation~(\ref{eq:alpha_nn_eff}). We also report some numerical values in Table~\ref{tab:branching_ratio} for reference. We find excellent agreement between \hylight{} and those calculated from \cite{Storey1995}. 

The branching ratio can be used to relate the line luminosity to the total ionization rate for gas in ionization equilibrium. Taking the example of H$\alpha$, the line luminosity is found by integrating the emissivity over the volume of the ionized region,
\begin{align}
L_{3,2} 
&= h\nu_{3,2} \int B_{3,2} n_e n_p \alpha_{\rm tot} \mathrm{d}V\nonumber\\
&= h\nu_{3,2} B^R_{3,2} \int \alpha_{\rm tot} n_e n_p \mathrm{d}V
+ h\nu_{3,2} B^Q_{3,2} \int \alpha_{\rm tot} n_e n_{\mathrm{HI}} \mathrm{d}V\nonumber\\
&\approx h\nu_{3,2} B^R_{3,2} \dot N_\gamma.
\end{align}
The last line assumes that the gas is highly ionized, the nebula is radiation bounded, and $B^R_{3,2}n_p\gg B^Q_{3,2}n_{\rm HI}$ so that the collisional term can be neglected. In addition, it is assumed that $B^R_{3,2}$ is a constant throughout the ionized region, which would be a good approximation if the temperature is approximately uniform. The remaining volume integral is then the total recombination rate in the cloud, which equals the rate at which the source emits ionizing photons, $\dot N_\gamma$, {\it ie} that the gas is in photoionization equilibrium. We can substitute numerical values to find that
\begin{align}
\dot N_\gamma &= 2.22\times 10^{45} {\rm s}^{-1}\frac{L_{3,2}}{L_\odot}=3.47\times 10^{44}{\rm s}^{-1} \frac{L_{2,1}}{L_\odot},
    \label{eq:Ha}
\end{align}
thereby relating the ionization rate to the H$\alpha$ luminosity ($L_{3,2}$) or Lyman~$\alpha$ luminosity ($L_{2,1}$). The numerical values used Case~B recombinations and a gas temperature of $10^4~\unit{K}$. Arguably, calculating the Lyman~$\alpha$ emissivity in Case~B is inconsistent, since Case~B assumes that the nebula is optically thick to all Lyman lines.
However, the Lyman~$\alpha$ emissivity can also be written as
\begin{align}
L_{2,1} &= h\nu_{2,1} \alpha^{\rm eff}_{{\rm B}, 1s} n_e n_p,
\label{eq:Lya}
\end{align}
where $\alpha^{\rm eff}_{{\rm B}, 1s}$ is the effective recombination rate for Case~B recombinations from Equation~(\ref{eq:Rnl}). The value of $\alpha^{\rm eff}_{{\rm B}, 1s}$ indeed assumes that the nebula is optically thick to all Lyman lines, but the expression for $L_{2,1}$ assumes that, nevertheless, the 2$p$ state decays to 1$s$ by emitting a Lyman~$\alpha$ photon. Equation~(\ref{eq:Lya}) is useful for comparing different models with each other, but is not applicable to a physical system (see also \S 10.3 of the \cloudy{} manual).

We can make the model that relates $\dot N_\gamma$ to $L_{3,2}$ more applicable to real \ion{H}{ii} regions by accounting for dust \cite[see also][]{Hirashita2003}. Consider spherical dust grains with radius $r$, uniform density by number $n$, and cross section for absorption of ionizing and H$\alpha$ photons $\sigma_{\rm UV}$ and $\sigma_{\rm IR}$ respectively; the source emits ionizing photons at a rate $\dot N_\gamma$ with a spectrum with mean energy of an ionizing photon $h\nu_{\rm UV}$. We further assume that gas is fully ionized inside the Str\"omgren radius $R_S$, with no ionizing photons escaping beyond $R_S$. The fraction of UV photons absorbed by dust is
\begin{align}
    d_{\rm UV} &= 1 - \exp(-\tau_{\rm UV}(R_S)),
\end{align}
and the fraction of UV-photons that ionize \ion{H}{i}, is then
$1-d_{\rm UV}$.
The H$\alpha$ luminosity is therefore
\begin{align}
    L_{3,2} &= (1-d_{\rm UV})\,(1-d_{\rm IR})\,B^R_{3,2}\dot N_\gamma h\nu_{3,2},
\end{align}
where $d_{\rm IR}$ is the fraction of H$\alpha$ photons absorbed by dust.
The dust luminosity is
\begin{align}
    L_{\rm dust} &= d_{\rm UV} \dot N_\gamma h\nu_{\rm UV},
\end{align}
where we neglect the dust heating by H$\alpha$ photons compared to that by UV photons. The ratio $L_{3,2}/L_{\rm dust}$, is
\begin{align}
\frac{L_{3,2}}{L_{\rm dust}} &= \frac{(1-d_{\rm UV})(1-d_{\rm IR})}{d_{\rm UV}}\,\frac{B_{3,2}\,\nu_{3,2}}{\nu_{\rm UV}}\equiv {\cal R}\frac{B_{3,2}\,\nu_{3,2}}{\nu_{\rm UV}}.
\end{align}
The dimensionless variable ${\cal R}$ depends on how efficiently the dust absorbs UV-photons compared to H$\alpha$ photons, but also on the geometry of the nebula. Finally, we can write $\dot N_\gamma$ in terms of the H$\alpha$ or dust luminosity, as
\begin{align}
    \dot N_{\gamma} &= \frac{L_{3,2}}{{\cal R} d_{\rm UV} B_{3,2} h\nu_{3,2}}\nonumber\\
    &= \frac{L_{\rm dust}}{d_{\rm UV} h\nu_{\rm UV}}.
\end{align}
And hence the total IR luminosity emitted by this dust is
\begin{align}
    L_{\rm dust} &= d_{\rm UV} \dot N_\gamma h \nu_{3,2} + (1-d_{\rm UV}) d_{\rm IR} \dot N_\gamma h\nu_{3,2}.
\end{align}

\subsection{Mock integral field data calculated from a simulation}
As a second example of \hylight{}, we post-process the snapshots of a non-equilibrium radiation-hydrodynamical simulation to make mock \ifu\ data cubes in hydrogen emission lines. We choose two setups: (1) a spherical, uniform density \ion{H}{ii} region with a single central source, with the calculation imposing ionization equilibrium (\lq idealised \ion{H}{ii} region\rq), and (2) a cubic simulation volume with an initially turbulent density field, irradiated from one side with ionizing radiation and accounting for non-equilibrium processes (\lq turbulent box\rq). We compare our results for the idealised \ion{H}{ii} region to \cloudy{}, and use the turbulent box simply as an illustration of the power of the atomic model in generating synthetic observations. We begin by briefly reviewing the \sparc{} code used to perform the simulations.

\subsubsection{\sparc{}: a multi-frequency radiation hydrodynamical code}
\sparc{} is an implementation of a two-moment method for radiative transfer in the \swift{} smoothed-particle hydrodynamics (SPH) code~\citep{Schaller2024} as described by \cite{Chan2021} and \cite{Chan2026}. This two-moment method, also known as the \lq M1\rq\ scheme, integrates the first two moments of the radiative transfer equation together with an \lq M1\rq\ closure relation~\citep{Minerbo1978, Levermore1984}. Radiation is transported between gas particles and each particle stores the local photon flux and density in a set of spectral bins. The bins are typically chosen to include important ionization thresholds (typically for hydrogen and helium, together with a spectral bin for non-ionizing photons).

\sparc{} uses the \chimes{}\footnote{\href{https://richings.bitbucket.io/chimes/home.html}{https://richings.bitbucket.io/chimes/home.html}} chemistry module~\citep{Richings2014a, Richings2014b} to update the photon densities, ionization fractions for all species, and gas temperature. This chemistry module tracks the ionization states of 157 species, including common molecules. When the ionization/recombination rates are high, \chimes{} sub-cycles time-steps, integrating the equations with the {\sc Cvode} solver \citep{Cohen1996}\footnote{\href{https://computing.llnl.gov/projects/sundials/cvode}{https://computing.llnl.gov/projects/sundials/cvode}} for coupled, numerically \lq stiff\rq\ differential equations. Full details of the coupling between radiation and thermochemistry, including tests of the method, are presented by \citet{Chan2026}. 

\subsubsection{Post-processing a simulation with \radmc{}}
\label{sec:pipeline}
Given the ionization state, density and temperature of each gas particle, as computed using \sparc{}, we use \hylight{} to compute the level population density of the hydrogen atoms for each particle. We can then calculate the number of particles in a specific excited state using Equation~(\ref{eq:level_population_number}). We further feed the information to \radmc{}\footnote{\href{https://www.ita.uni-heidelberg.de/~dullemond/software/radmc-3d/}{https://www.ita.uni-heidelberg.de/~dullemond/software/radmc-3d/}}~\citep{Dullemond2012}.

\radmc{} computes the specific intensities emerging from a nebula using Monte Carlo radiative transfer on an adaptively refined mesh, accounting for self-absorption and dust attenuation. We interpolate the physical properties of the SPH particles (density, temperature, abundance, level population density) to \radmc{}'s adaptively refined mesh. Note that we map the number of hydrogen atoms in the excited states onto an adaptively oct-tree refined grid instead of directly calculating the level population density on the grid (with the interpolated electron density, \ion{H}{ii} and \ion{H}{i} density) to ensure the convergence of total luminosity. The convergence originates from the conservation of particle mass. Refinement of the mesh occurs whenever the minimum smoothing length\footnote{The smoothing length is a measure of the resolution of the SPH~\cite[see e.g.][]{Monaghan1988, Schaller2024}. } of the SPH particles within a cell is larger than a fraction of the cell size. This fraction is specified by the user. We verified the convergence of the results in Appendix~\ref{app:convergence_refinement}. \radmc{} then performs dust radiative transfer and non-LTE line transfer for specified lines, outputting the result in the form of a mock IFU datacube. The spatial pixel size and wavelength resolution are both specified by the user. 

\subsubsection{Mocks of an idealized \ion{H}{ii} region}
\label{sect:HII}
\begin{figure}
    \includegraphics[width=\columnwidth]{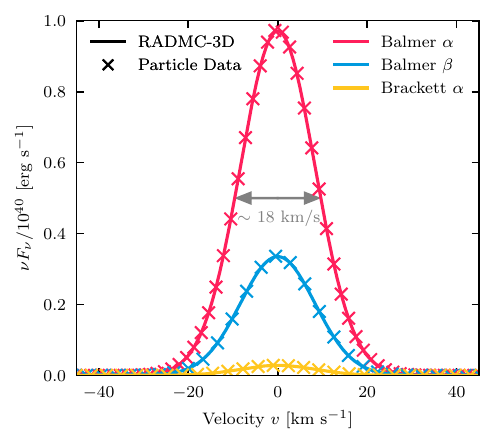}
    \caption{Spectrum of the H$\alpha$, H$\beta$, and Brackett~$\alpha$ line (shown in red, blue and yellow) for the idealised \ion{H}{ii} region described in \S~\ref{sect:HII}. The result from combining the output from the \sparc{} simulation with \hylight{} and \radmc{} is shown with {\em solid lines}, the line profile computed directly from the particle data is shown by {\em crosses}.
    The total luminosity estimated by \radmc{} agrees to better than 0.5~per cent with that from the particle data. This demonstrates the consistency of our pipeline for post-processing the simulations with \radmc{}. A line width ($\sim$ 18 km s$^{-1}$) corresponding to gas that emits at a temperature $10^4~\unit{K}$ is indicated in the figure to guide the eye.}
    \label{fig:radmc3d_spectrum_HHeZ}
\end{figure}

As a first test case of the combination of \sparc{}, \hylight{} and \radmc{}, we initialise a cubic computational volume with uniform density and uniform temperature gas (linear extent $22~\unit{pc}$, hydrogen density $n_{\rm H}=10~\unit{cm^{-3}}$, gas temperature 10$^4$~K, initially neutral). The gas is represented with $128^3$ SPH particles. A central source emits ionizing photons with a blackbody spectrum (ionization rate $\dot N_\gamma=5\times 10^{47}~\unit{s^{-1}}$, blackbody temperature $10^5~\unit{K}$). With these parameters, the \ion{H}{ii} region is radiation bounded with Str\"omgren radius $R_S\approx 5.4~{\unit{pc}}$. This is the same setup as described in \S~3.3.2 \lq Multi-frequency \ion{H}{ii} region with hydrogen, helium, and metals\rq\ of \citealt{Chan2026}, except for the gas temperature\footnote{The gas temperature is not fixed in the example in \S~3.3.2 of \citealt{Chan2026}. }. The simulations track the following elements and their ions: H, He, C, N, O, Ne, Mg, Si, S, Ca and Fe, with solar abundances as given in table~1 of \cite{Wiersma2009}.

Radiative transport is performed using \sparc{} with eight frequency bins (with edges [13.60, 18.00, 24.59, 35.50, 54.42, 68.02, 81.62, 95.22, $\infty$]~\unit{eV}\footnote{These values correspond or approximately correspond to ionization energies of different ionization stages of H, He, C, N, O.}). We turn off gas dynamics, dust, cosmic rays, and molecular physics for simplicity. The simulation is run for $1.2~\unit{Myr}$, by which time the ionization state is nearly in equilibrium and the ionization front is close to the Str{\"o}mgren radius. During the simulation, we keep the density of the gas constant. The test also neglects hydrodynamical effects and examines the cloud at a constant temperature of 10$^4$~K, to enable comparison to \cloudy{}. 

We follow the workflow described in \S~\ref{sec:pipeline} to prepare the input files for \radmc{}. Then we execute line RT in \radmc{} to obtain a mock spectrum and image of the simulated \ion{H}{ii} region\footnote{We made use of the \texttt{spectrum} and \texttt{image} commands in \radmc{}.}. 

Given our choice of parameters, self-shielding of H$\alpha$ is negligible and since we don't include dust, the luminosity of the cloud is equal to the integral of the emissivity over its volume. We use this to predict the line profile from the particle data by accounting for the temperature of each gas particle, and summing the emission lines of all particles (the line width is approximately $18~\unit{km~s^{-1}}$). The gas in the simulation is static, and we do not include an unresolved \lq micro-turbulence\rq\ component and hence the line width is purely due to thermal broadening. This way, we predict the total luminosity and the line profile directly from the particle data. Comparing this to \radmc{} provides a test of the interface between \sparc{}, \hylight{} and \radmc{}. 

We compute the luminosity of the H$\alpha$ line for a single SPH particle, $i$, as
\begin{align}
    L_{3,2, i} &= \int \sum_{l=0}^{2} \sum_{l' = l \pm 1} n_{3l, i} A_{3l, 2l'} h \nu_{3,2} \mathrm{d}V \nonumber\\
    &\approx \frac{h \nu_{3,2}}{n_{\rm H, i}} \sum_{l=0}^{2} \sum_{l' = l \pm 1} n_{3l, i} A_{3l, 2l'} \int n_{\rm H, i} \mathrm{d}V \nonumber\\
    &\approx \frac{h \nu_{3,2}}{n_{\rm H, i}} \sum_{l=0}^{2} \sum_{l' = l \pm 1} n_{3l, i} A_{3l, 2l'} m_{{\rm H},i}. 
\end{align}
Here, $n_{\rm H, i}$ is the hydrogen density of the particle, $n_{3l, i}$ is the population density of the particle in the $n=3$ excited state (including 3$s$, 3$p$, and 3$d$) - as computed by \hylight{} - and the integral $\int n_{\rm H, i} \mathrm{d}V\approx m_{{\rm H},i}$ is the total hydrogen mass of the particle, $m_{{\rm H},i}$. Summing over all dipole allowed radiative transitions yields the total H$\alpha$ luminosity of the particle, and summing that over particles yields the \lq particle\rq\ estimate of the H$\alpha$ luminosity of the cloud. We assume that the shape of the emission line for each particle is Gaussian with a variance set by the temperature of the particle. Summing over particles yields the spectrum of the cloud. More details on the methodology can be found in Appendix~\ref{app:lum_from_part}. We compare the spectrum computed from the particles to that obtained using \radmc{} in Fig.~\ref{fig:radmc3d_spectrum_HHeZ}, finding excellent agreement. 

We compare the luminosity obtained by (i) the total H$\alpha$ luminosity calculated from particles; (ii) relating the ionization rate to the line luminosity using the branching ratio's described in the previous section; (iii) integrating the luminosity over the line profile computed by \radmc{}; and (iv) the total luminosity predicted by \cloudy{}\footnote{We set up a \cloudy{} model with the same ionization source and same element abundance used in \sparc{} but with only 30 resolved levels and 10 collapsed due to limitation of computing resources. }. The total luminosity in the first three cases converges to $6.8 \times 10^{35}$~erg~s$^{-1}$. The relative difference between (i) and (ii) is smaller than 0.5 per cent; The relative difference between (i) and (iii) is smaller than 1 per cent, whereas the difference between (i) and (iv) is about 5 per cent. The difference comes from the two facts: first, in the setup of \cloudy{} used for this comparison, we were not using 100 resolved levels for hydrogen due to the limitation of the computing resources; and second, the simulation is not evolved for long enough for the gas to be in ionization equilibrium everywhere. This causes small differences in the level prediction between \hylight{} (which does not impose ionization equilibrium) and \cloudy{} (which does).

\begin{figure}
    \includegraphics[width=\columnwidth]{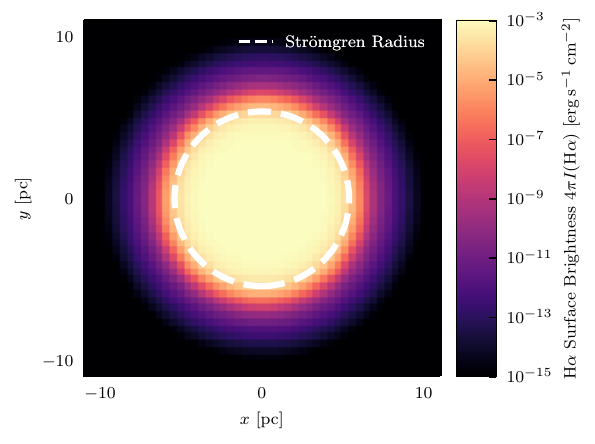}
    \caption{H$\alpha$ surface brightness, $S^{\mathrm H \alpha}$, of the idealised \ion{H}{ii} region simulation described in \S~\ref{sect:HII}. The ionisation state of the gas is computed using \sparc{}, the level population using \hylight{} and the specific intensity using \radmc{}. The surface brightness is obtained by integrating the \radmc{} IFU cube over frequency (Equation~(\ref{eq:SR})). As expected, H$\alpha$ originates predominantly from the highly ionized gas inside the Str\"omgren radius ($R_S\approx 5.4~\unit{pc}$), which is indicated by the {\em dashed white circle}. }
    \label{fig:radmc3d_image_HHeZ}
\end{figure}

\begin{figure}
    \includegraphics[width=\columnwidth]{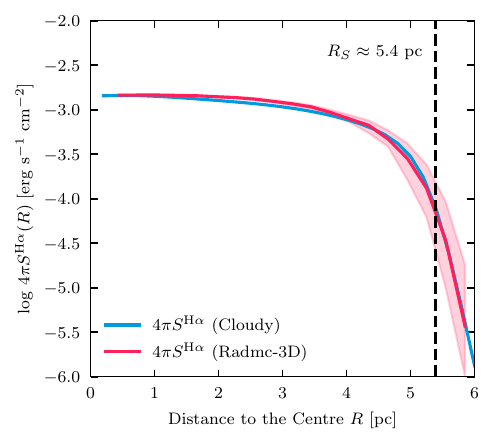}
    \caption{The spherically-averaged surface brightness profile of the idealised \ion{H}{ii} region shown in Fig.~\ref{fig:radmc3d_image_HHeZ}. The result from combining \sparc{} with \hylight{} and \radmc{} is shown in {\em red} (where the solid line is the median profile, the shaded region encompasses the 10$^{\rm th}$ $\sim$ 90$^{\rm th}$ percentiles), the \cloudy{} result, computed using Equation~(\ref{eq:SC}), is shown in {\em blue}. The two models agree well with each other inside the Str{\"o}mgren radius $R_S$, which is shown as the vertical dashed black line. Small differences outside $R_S$ arise because the gas in the simulation is not yet fully in photoionization equilibrium.}
    \label{fig:sb_Halpha_comp}
\end{figure}

Next we compare surface brightness profiles inferred by feeding the \hylight{} population levels to \radmc{}, with the profile computed using \cloudy{}. \radmc{} computes the specific intensity, $I_\nu(x,y; {\bf n})$, for each position $(x,y)$ in the image plane in the direction ${\bf n}$ of a specified \lq observer\rq. It does so by ray-tracing the emissivity, accounting for self-absorption and absorption by dust. We compute the corresponding \lq observed\rq\ surface brightness, $S$, for a particular line, e.g. H$\alpha$, by integrating over frequency,
\begin{align}
    S^{\mathrm H \alpha} (x,y) &= \int_{\rm H\alpha} \mathrm{d}\nu I_\nu (x, y; {\bf n}).
    \label{eq:SR}
\end{align}

\cloudy{} computes the emissivity in a line as a function of the distance $r$ 
to the centre of the cloud. We compute the corresponding surface brightness 
at location $(x,y)$ by integrating the emissivity along a sight line
with impact parameter $R$,
\begin{align}
S^{\rm H \alpha} (R) &= 2\,\int_{R}^{R_\mathrm{max}} \mathrm{d}r \frac{r}{\sqrt{r^2 - R^2}} \frac{\epsilon_{3,2} (r)}{4\pi},
\label{eq:SC}
\end{align}
where $R=((x-x_0)^2+(y-y_0)^2)^{1/2}$ is the distance to the centre $(x_0,y_0)$ of the image, and we set $R_{\rm max}=8$~pc (outside the Str\"omgren radius, $R_S$). Unlike the \radmc{} calculation, this calculation does not account for self-shielding or dust absorption. However, the population of the $n=2$ level is low, so self-absorption is negligible, and we do not include dust in the calculation. The \cloudy{} model is perfectly spherically symmetric.

We compare the circularly-averaged profile of $S$ computed by \radmc{} using \hylight{} from Equation~(\ref{eq:SR}), and $S$ computed with \cloudy{}, Equation~(\ref{eq:SC}), in Fig.~\ref{fig:sb_Halpha_comp} (solid red and solid blue), finding excellent agreement. Deviations from spherical symmetry appear in the simulation due to sampling. In particular, small number statistics means that the radiation field close to the source, where it is sampled by few particles, is not spherically symmetric. Such deviations propagate to larger distances, and we illustrate this scatter by plotting the 10 - 90$^{\rm th}$ percentile of the surface brightness in bins of $R$ as the red shaded area. Close to and outside $R_S$, the mean curves differ because the simulated nebula is not yet fully in ionization equilibrium.

In summary: we have computed the emissivity of a simple \ion{H}{ii} region with \sparc{}, post-processing the simulation with \hylight{} and \radmc{}. We find excellent agreement with \cloudy{} in terms of the line luminosities of hydrogen recombination lines, and of the corresponding surface brightness profile. 

\subsubsection{Mocks of an illuminated turbulent density field}
\label{sec:turbulent_HII_region}
We illustrate the flexibility of our modelling by the following more realistic version of a small patch of a dynamical \ion{H}{ii} region with an initially inhomogeneous density structure, and gravity. The setup is described in more detail in section 4 of \citealt{Chan2026}. Briefly, we set up a cubic simulation volume with a linear extent of $1~\unit{pc}$ with 128$^3$ gas particles, initially at uniform density and uniform temperature, (hydrogen number density ⟨$n_{\rm H}\rangle=50~\unit{cm^{-3}}$, $T=15~\unit{K}$) and the same solar abundance pattern as in the previous section. The gas is initially neutral. 

Inhomogeneities are introduced by driving isothermal solenoidal turbulence with a Mach number of five, using the {\sc Phantom} SPH code described by \citet{Price2018}. The turbulence is driven over five crossing times, and we use the particle distribution at the end of the {\sc Phantom} run as the initial condition for \sparc{}; the initial gas velocity is set to zero everywhere.

We inject ionizing radiation with a blackbody spectrum (temperature $3\times 10^4~\unit{K}$) in the form of plane-parallel radiation that is propagating perpendicular to the $x=0$ plane of the computational volume. The flux of ionizing photons at the edge is taken to be $2\times 10^{10}~\unit{cm^{-2}~s^{-1}}$. We also include low-energy cosmic ray heating and ionization, both of which are also implemented in \chimes{}. 

\begin{figure*}
    \includegraphics[width=\textwidth]{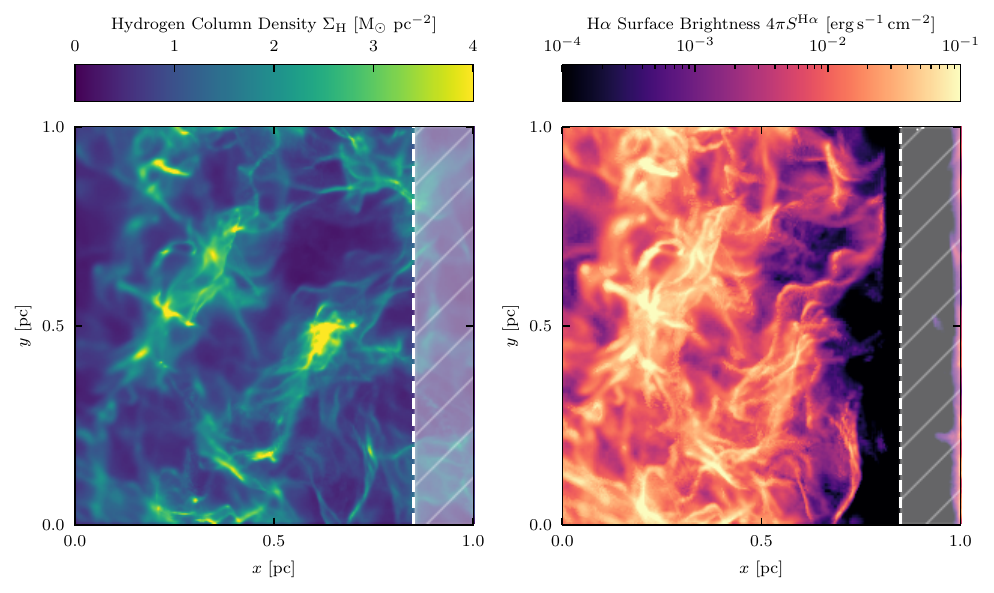}
    \caption{Column density of hydrogen gas (\textit{left panel}) and H$\alpha$ surface brightness (\textit{right panel}) for the case of an initially turbulent density field. Ionizing radiation is injected perpendicular to the $x=0$ plane, with \sparc{} propagating the radiation to the right. Photons are removed from the radiation field for $x>0.8~\unit{pc}$ ({\em grey hashed regions in both panels}) to avoid effects of periodic boundary conditions. The hydrogen column density is obtained by integrating the total hydrogen density (the sum of \ion{H}{i} and \ion{H}{ii}) along the $z$-axis. The H$\alpha$ surface brightness is calculated by combining \hylight{} with \radmc{}. The filamentary pattern resulting from the turbulent driving seen in the column density map ({\em left panel}) is reflected in the H$\alpha$ surface brightness map ({\em right panel}). }
    \label{fig:radmc3d_turbbox}
\end{figure*}

\begin{figure}
    \includegraphics[width=\columnwidth]{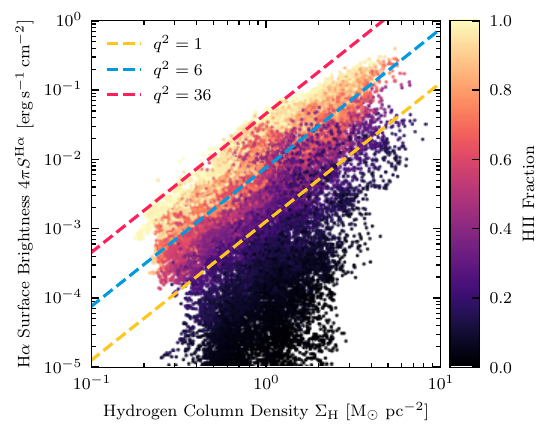}
    \caption{Relation between hydrogen column density, $\Sigma_{\rm H}$, and H$\alpha$ surface brightness, $4 \pi S^{\rm H\alpha}$, for the simulation of Fig.~\ref{fig:radmc3d_turbbox}. Each point corresponds to a pixel, with points coloured by the mean hydrogen ionized fraction along the sight line. The dashed lines indicate the relation of Equation~(\ref{eq:sb_sd}), which are only applicable in highly ionized regions, with $q$ being a dimensionless parameter measuring the mean ionized fraction. In highly ionized regions, these trend lines capture the colour shading of the pixels. }
    \label{fig:radmc3d_turbbox_sb_sd}
\end{figure}
Our results are illustrated by Fig.~\ref{fig:radmc3d_turbbox}, where the grey buffer region at $x>0.8~\unit{pc}$ is where we remove photons to avoid the periodic boundary conditions artificially enhancing the radiation inside the volume. The left panel illustrates the total hydrogen column density at the start of the simulation, calculated as
\begin{equation}
    \Sigma_{\rm H} = \int n_{\rm H} m_{\rm H} \mathrm{d}z, 
\end{equation}
where $n_{\rm H}$ is the total hydrogen number density (the sum of the \ion{H}{i} and \ion{H}{ii} densities), $m_{\rm H}$ is the mass of the hydrogen atom, and the integral is along the direction of projection (here, the $z$ axis). Given the ionization states computed by \sparc{}, we use \hylight{} to compute the level population of \ion{H}{i}, and finally \radmc{} to compute the H$\alpha$ surface brightness calculated from Equation~(\ref{eq:SR}). The result is plotted in the right panel of the Figure.

The turbulent driving used in the initial conditions creates filamentary regions of high $\Sigma_{\rm H}$, apparent in the left panel. Comparing to the right panel shows that these create a corresponding filamentary appearance of regions of high H$\alpha$ surface brightness, in the right panel - and vice versa. Therefore, turbulent driving results in the emergence of filaments of high H$\alpha$ emissivity.

We investigate the relation between column density and H$\alpha$ surface brightness in more detail by plotting $\Sigma_{\rm H}$ versus $S_{\rm H\alpha}$ in Fig.~\ref{fig:radmc3d_turbbox_sb_sd}. The colour of each pixel is a measure of the mean ionized fraction along the sight line. We note that pixels of equal colour tend to fall along approximately diagonal lines in this plot, when the gas is sufficiently ionised. 

To explain the emergence of this correlation, we overplot the power law, 
\begin{align}
    4 \pi S^{\rm H\alpha} = \frac{q^2\,h\nu_{3,2} \langle\alpha_{3,2}^{\rm eff}\rangle}
    {m_\mathrm{H}^2 L} \Sigma_\mathrm{H}^2,
    \label{eq:sb_sd}
\end{align}
as dashed lines, where $\langle\alpha_{3,2}^{\rm eff}\rangle$ is the average effective rate of 
emission of H$\alpha$ photons defined by Equation~(\ref{eq:alpha_nn_eff}), and $L=1~\unit{pc}$ is the box size of the simulation volume; different lines correspond to different values of the dimensionless parameter $q$. These lines capture the trend seen in the coloured points of H$\alpha$ surface brightness increasing linearly with the square of the hydrogen column density, at constant neutral fraction.

The origin of the correlation expressed by Equation~(\ref{eq:sb_sd}) can be understood as follows: firstly, the H$\alpha$ surface brightness is an integral of the H$\alpha$ emission coefficient, $j_{3,2}$, over the line of sight, i.e. $z$-axis,
\begin{align}
    4 \pi S^{\rm H\alpha} &= 4 \pi \int j_{3,2} \mathrm{d} z\nonumber\\
    &= \int \epsilon_{3,2} \mathrm{d} z \nonumber\\
    &= h\nu_{3,2} \int n_p n_e \alpha_{3,2}^{\rm eff} \mathrm{d} z\nonumber\\
    &\approx h\nu_{3,2} \langle x_{\rm HII}\rangle^2 \langle\alpha_{3,2}^{\rm eff}\rangle \int n^2_\mathrm{H} \mathrm{d}z\nonumber\\
    &\equiv h\nu_{3,2} \langle x_{\rm HII}\rangle^2 \langle\alpha_{3,2}^{\rm eff}\rangle L\langle n^2_\mathrm{H}\rangle.
    \label{eq:derivation_I}
\end{align}
The first step uses Equation~(\ref{eq:alpha_nn_eff}), the second step
defines $\langle x_{\ion{H}{ii}}\rangle$ to be the mean ionized fraction and
$\langle\alpha_{3,2}^{\rm eff}\rangle$ the mean effective rate of emission of H$\alpha$ photons, and the final step defines
\begin{align}
    \langle n^2_\mathrm{H} \rangle &= \frac{1}{L} \int n^2_\mathrm{H} \mathrm{d}z.
\end{align}
Defining
\begin{align}
    \langle n_\mathrm{H} \rangle &= \frac{1}{L} \int n_\mathrm{H} \mathrm{d}z = \frac{\Sigma_\mathrm{H}}{L},
\end{align}
and setting $q^2=\langle x_{\rm HII}\rangle^2\,\langle n^2_\mathrm{H} \rangle/\langle n_\mathrm{H} \rangle^2$ yields Equation~(\ref{eq:sb_sd}).

Equation~(\ref{eq:sb_sd}) shows that, at a constant $q$ (given level of ionization, given level of density inhomogeneity, $\langle n^2_\mathrm{H} \rangle/\langle n_\mathrm{H} \rangle^2$), $4 \pi S^{\rm H\alpha}\propto \Sigma^2_\mathrm{H}$. Decreasing the level of inhomogeneity, or the level of ionization, decreases $4 \pi S^{H\alpha}$ for a given value of  $\Sigma^2_\mathrm{H}$ - this is what we see in Fig.~\ref{fig:radmc3d_turbbox_sb_sd}. Note that in our derivation of Equation~(\ref{eq:derivation_I}), we assume the gas is highly ionized so that $n_p \approx n_e \approx n_{\rm H}$. Therefore, Equation~(\ref{eq:sb_sd}) only applies to highly ionized regions. This is confirmed by Fig.~\ref{fig:radmc3d_turbbox_sb_sd}, the scaling relation (dashed lines) only applies to highly ionized regions (red and blue, not yellow).)
\end{document}

%% file: sections/conclusion.tex
\section{Conclusions}
Hydrogen recombination lines play a crucial role in the emission line diagnostics used to infer physical conditions of nebular regions and the ISM of galaxies. Forward models of the ISM, based on hydrodynamical simulations that include radiative transfer, need to be post-processed to compute the strength of the hydrogen emission lines to enable a direct comparison to observations. This post-processing step requires a model that predicts the level population in \ion{H}{i}, since hydrogen emission lines arise from transitions from excited states to more bound states.

We have compared predictions for the level population as a function of density and temperature between several models (such as that of \citealt{Raga2015} and \cloudy{} \citep{Cloudy2017, Cloudy2023, Cloudy2025}), as well as values from published tables \citep{Storey1995}, and find values that differ by factors of several, even for common lines such as H$\alpha$ and H$\beta$ (see \S~\ref{sec:motivation}). We also find that the \cloudy{} prediction using the default choice of parameters for the number of energy levels included in the calculation, can be off from the numerically converged answer by 10s of per cent.
Tabulated values cannot be used to investigate the level population in case the gas is not in ionisation equilibrium, and the same holds for the \cloudy{} predictions, which assume ionisation balance. However, numerical simulations are able to integrate the rate equations and can hence account for non-equilibrium effects. The disagreement of energy levels between different models, and the fact that most of these models cannot account for non-equilibrium effects, motivated us to develop \hylight{}. This {\sc Python} code determines the level population of \ion{H}{i}, given the temperature, density, and ionisation state of the gas. The latter can be computed by radiation-hydrodynamic simulation that includes non-equilibrium effects, for example, using the \sparc{} code described by \cite{Chan2026}.

The \hylight{} model accounts for the dominant mechanisms that determine the level population under nebular conditions, namely recombinations and collisional excitation from the ground state: the model is described in \S~\ref{sec:model}. \hylight{} is implemented in {\sc Python} and is publicly available. The level population and emissivity predicted by \hylight{} agree with those from \cloudy{} (version C23.02) at the per cent level or better, when the gas is assumed to be in ionisation equilibrium. \S~\ref{sec:model} also investigates the minimum number of levels the model should include in order to converge. \hylight{} serves as a stand-alone package for the accurate calculation of hydrogen line emission, and is easy to integrate into other codes or post-processing pipelines for analysing simulation snapshots. 

We present two applications of \hylight{} in \S~\ref{sec:application}. We first use the model to compute branching ratios - the ratio of the rates at which \ion{H}{i} atomic transition between two levels $n\to n'$ (where $n$ is the principal quantum number) over the total recombination rate. Such a branching ratio can be used to relate the photo-ionisation rate in a radiation-bound \ion{H}{ii} region to line luminosity such as that of the H$\alpha$ line (see Equation~(\ref{eq:Ha})). 

In the second application, we post-process simulations performed with the \sparc{} code \citep{Chan2026}. One simulation is that of a spherically symmetric, uniform, radiation-bound \ion{H}{ii} region, which is ionised by a time-independent central source. We feed the simulation output to \hylight{} to determine the level population, and feed the output of that calculation into \radmc{}~\citep{Dullemond2012} to compute an H$\alpha$ surface intensity map. Because the setup is simple, we can compute the H$\alpha$ surface brightness profile with \cloudy{} as well, and we find excellent agreement between the two methods.

In a second simulation, we use \sparc{} to compute the evolution of a patch of ISM, with initial density perturbations generated by turbulent driving, which is overrun by an initially plane-parallel ionisation front. The calculation is performed in full non-equilibrium, by taking advantage of the \chimes{} \citep{Richings2014a, Richings2014b} photo-chemistry solver coupled to radiative transfer in \sparc{}. We show how regions of high H$\alpha$ surface brightness correspond to regions with high hydrogen column density, as might have been expected. Such calculations can be used to investigate the extent to which density structure in a cloud affects the interpretation of line ratio diagrams, where different lines may originate from spatially distinct parts of the cloud.

\hylight{} paves the way for examining emission line diagnostics using simulations. It enables examining the impact of feedback and the presence of metals on the emission line diagnostics in the ISM. The method also enables us to compare galaxy-scale simulations against observations (see \citealt{Howatson2025} for example) and future simulations against spatially resolved observations of nearby star-forming regions~\citep[][]{McLeod2019}. The mock IFU data cubes from our novel simulations contain rich information about the gas properties and will be compared to observations, allowing us to investigate the spatially resolved observations from a new perspective. In addition, our calculation is not restricted to equilibrium solutions of the system. The method applies to studies of the non-equilibrium evolution of the ISM and even on-the-fly radiative cooling in the future.

%% file: sections/appendices.tex
\section{Convergence in \cloudy{} models}
\label{app:cloudy_model_convergence}
\begin{figure}
    \includegraphics[width=\columnwidth]{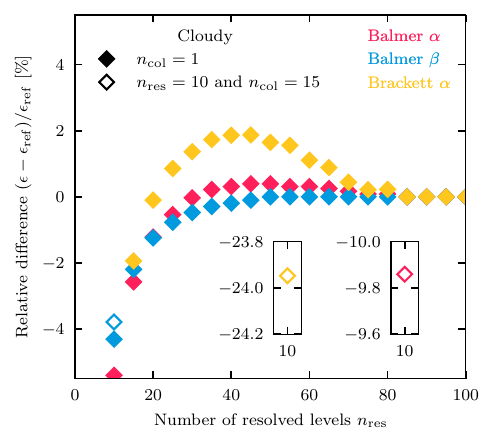}
    \caption{Convergence of emissivity with the number of resolved levels for selected transitions for different \cloudy{} models in Case A. The reference setup is \lq Ref - ALSh\rq\ with 100 resolved levels. The solid symbols indicate models with different numbers of resolved levels while keeping only one collapsed level. The open symbols indicate the default setup, i.e. 15 resolved levels and 10 collapsed ones. The emissivities of the selected lines gradually converge to the reference model. We conclude that 100 resolves levels in \cloudy{} are sufficient for convergence to better than 1 per cent. Note that the default model can differ significantly: for H$\alpha$, the difference is about 10\%; for H$\beta$, the difference is around 4\%; for Brackett~$\alpha$, the difference can be as high as 24\%. }
    \label{fig:emis_conv_cloudy}
\end{figure}

We have used \cloudy{} models to investigate the hydrogen level population and recombination line emissivity in photoionized regions. We investigate in this Appendix how \cloudy{} models converge as the number of resolved levels in the built-in atomic model increases. In Fig.~\ref{fig:emis_conv_cloudy}, we show the convergence for selected transitions when the number of resolved levels increases. We choose the reference \lq Ref - ALSh\rq\ setup in Table~\ref{tab:cloudy_models} but vary the number of resolved levels, $n_{\rm res}$. Note that we have used the same reference model as in Fig.~\ref{fig:ha_hb_bra_cvg_cmf_case_a}. As the number of resolved levels increases, the emissivities (for all three lines) converge to the reference model. We also indicate the emissivity predictions from the default atomic model (\lq Def - ALSh\rq) which includes 10 resolved levels and 15 collapsed levels. A significant difference between the default model and the reference model can be seen: for Brackett~$\alpha$, the difference is approximately 24\%. This finding is similar to that in~\citealt{Guzman2025}. 

\begin{figure*}
    \includegraphics[width=\textwidth]{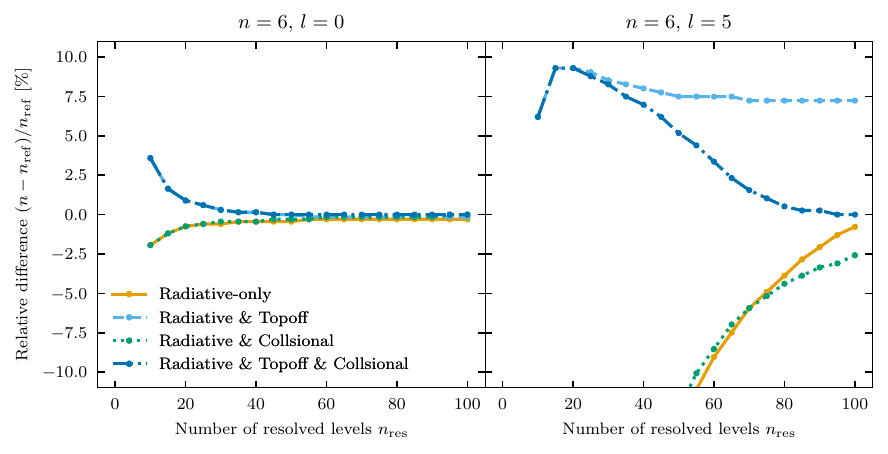}
    \caption{Convergence of level populations predictions in the \lq Ref - ALSh\rq~model (Table~\ref{tab:cloudy_models}) for various settings. The reference model involves 100 resolved and 1 collapsed level, as in all other figures. The vertical axis shows the relative difference between the level populations predicted by different setups and that of the reference model. The \lq Radiative-only\rq\ model takes into account only radiation recombination and radiative cascade. The \lq Radiative \& Topoff\rq\ setup includes the conservation of recombination coefficients on top of the \lq Radiative-only\rq\ model, while the \lq Radiative \& Collisional\rq\ setup contains collisional processes modelling (collisional excitation, collisional de-excitation, $l$-changing collision) on top of the \lq Radiative-only\rq\ model. The \lq Radiative \& Topoff \& Collisional\rq\ setup contains all four processes. Different $l$-states require different numbers of resolved levels to achieve convergence, and various physical processes have an impact on the convergence. }
    \label{fig:lvlpop_n6_model_variants}
\end{figure*}

The convergence of \lq Ref - ALSh\rq\ model is different from that of \lq Ref - ALSh - RR\rq~(see Fig.~\ref{fig:ha_hb_bra_cvg_cmf_case_a}). The differences stem from the different processes included in the model. In \cloudy{}, users have the freedom to turn on or off different physical processes depending on the aim of the research. Yet these processes have an impact on the level populations of atomic states, leading to different emissivity predictions. Here, we show such an impact by studying two major processes - collisional processes (collisional excitation, collisional de-excitation and $l$-changing collision) and the \lq topoff\rq\ process. In the case of running \cloudy{} models with a limited number of resolved levels, the residual recombination coefficients of the higher levels (i.e. collapsed levels) may not be negligible. The \lq topoff\rq\ command in \cloudy{} adds the missing recombinations to the model atom as described in the \cloudy{} Hazy1 manual. Interested readers can find more details in \citealt{Bauman2005}. We show the convergence of level population predictions as a function of the number of resolved levels for various setups in Fig.~\ref{fig:lvlpop_n6_model_variants}. Four different setups are shown: the radiative-only calculations include only radiative processes; the \lq radiative \& topoff\rq\ setup conserves the total recombination coefficient on top of radiative-only calculations; the \lq radiative \& collisional\rq\  setup adds collisional processes on top of radiative processes; finally, \lq radiative \& topoff \& collisional\rq\ contains all four processes. In Fig.~\ref{fig:lvlpop_n6_model_variants}, the 6$s$ level population converges for all four setups while the 6$h$ level population does not, despite there being 100 resolved levels already. We conclude that various physical processes (radiative, collisional and top-off) have an impact on the convergence of the level population, and different $l$-states require different numbers of resolved levels to achieve convergence. 

\begin{figure*}
    \includegraphics[width=\textwidth]{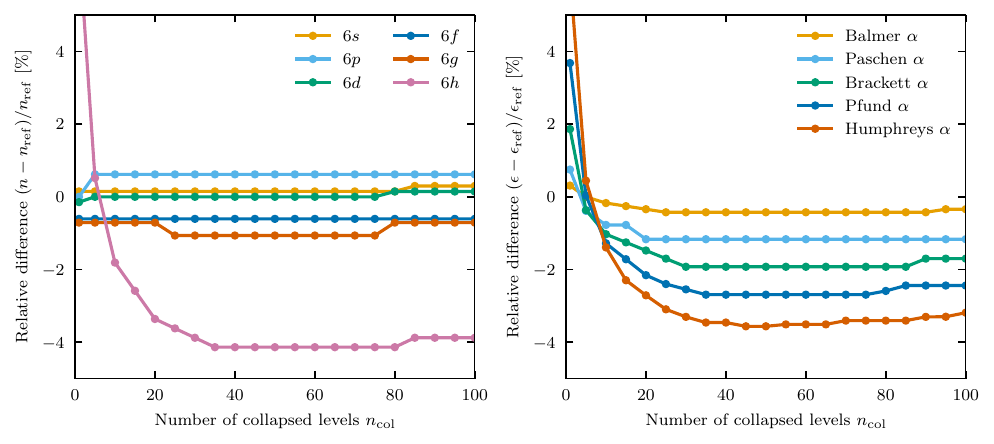}
    \caption{Level population and emissivity predictions from \cloudy{} in Case A as a function of the number of collapsed levels for a fixed number of resolved levels of 40. The vertical axis shows the relative difference, either level population ({\em left}) or line emissivity ({\em right}), between a given setup and the reference setup (\lq Ref - ALSh\rq). Not all $l$-states in $n=6$ achieve convergence, even though there are 100 collapsed levels. All the line emissivities are not converged as well, since they are dependent on the level populations. }
    \label{fig:conv_ncol_cloudy_40}
\end{figure*}

\begin{figure*}
    \includegraphics[width=\textwidth]{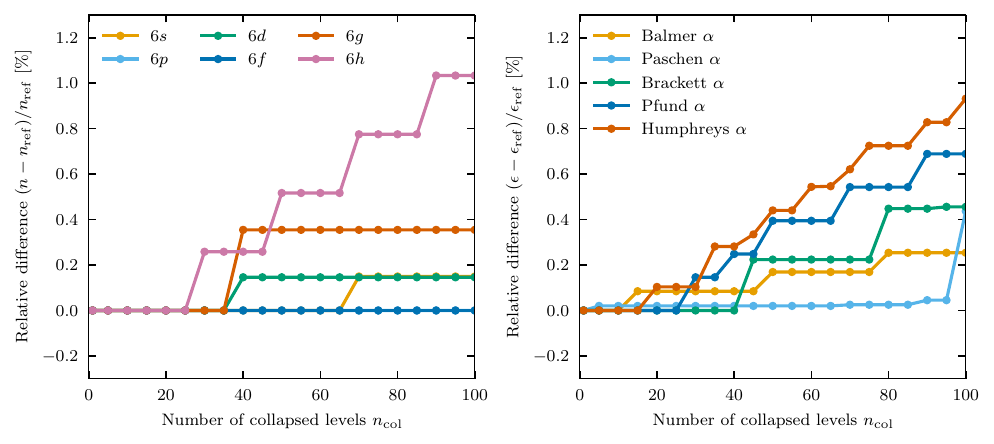}
    \caption{Same as Fig.~\ref{fig:conv_ncol_cloudy_40}, but with 100 resolved levels. }
    \label{fig:conv_ncol_cloudy_100}
\end{figure*}

Furthermore, we study the impact of the number of collapsed levels on the line emissivity and level population. In Fig.~\ref{fig:conv_ncol_cloudy_40} and Fig.~\ref{fig:conv_ncol_cloudy_100}, we show the convergence of the level population of all the $l$-states in $n=6$ and the emissivity of several transitions as a function of the number of collapsed levels. In both figures, we use \lq Ref - ALSh\rq~(see Table~\ref{tab:cloudy_models}) with 100 resolved levels and 1 collapsed level as the reference model. In Fig.~\ref{fig:conv_ncol_cloudy_40}, we fix the number of resolved levels to 40, while in Fig.~\ref{fig:conv_ncol_cloudy_100} we use 100 resolved levels. Not all the level populations have reached convergence with more collapsed levels. Since the emissivity is dependent on various $l$-states, none of the line emissivities has reached convergence. Increasing the number of resolved levels does not help ease the issue, as is shown in Fig.~\ref{fig:conv_ncol_cloudy_100}, where we have 100 resolved levels. 

\section{Departure coefficients}
\label{sec:appendix_depar_coeff}
Level population in LTE can be calculated analytically by combining the Saha equation and the Boltzmann equation. At a local temperature $T$, electron density $n_e$, and proton density $n_p$, the Saha equation in LTE for ionization fraction reads 
\begin{equation}
    \frac{n_e n_p}{n_{1s}} = \Big ( \frac{2 \pi m k T}{h^2} \Big )^{3/2} \exp (- h \nu_0 / k_B T), 
    \label{eq:saha_equation}
\end{equation}
where $h \nu_0$ is the ionization energy of a neutral hydrogen atom into ionized hydrogen and $k_B$ is the Boltzmann constant. 
For the level population in an excited state, the Boltzmann equation yields
\begin{equation}
    \frac{n_{nl}}{n_{1s}} = (2l + 1) \exp (- E_{1,n} / k_B T),
    \label{eq:boltzmann_equation}
\end{equation}
where $E_{1,n}$ is the energy difference between the ground state and the excited state $n$. Combining Equation~(\ref{eq:saha_equation}) with Equation~(\ref{eq:boltzmann_equation}), yields the level population in LTE,
\begin{equation}
    n_{nl} = (2 l + 1) \Big( \frac{2 \pi m k_B T}{h^2} \Big)^{3/2} \exp(E_n / k_B T),
    \label{eq:levelpop_LTE}
\end{equation}
where $E_n$ is the ionization potential of the level $n$,
\begin{equation}
    E_n = h \nu_0 - E_{1,n}. 
\end{equation}
Historically, \citet{Menzel1937a} and \citet{Menzel1937b} proposed to express the level population in terms of the dimensionless factor $b_{nl}$\footnote{At that time, the model did not resolve $l$-states. }, often referred to as the departure coefficient, to describe deviations from LTE. Therefore, the level population can then be written in terms of the $b_{nl}$ and the level population in LTE, 
\begin{equation}
    n_{nl} = b_{nl} (2 l + 1) \Big( \frac{2 \pi m k T}{h^2} \Big)^{3/2} \exp(E_n / k_B T). 
    \label{eq:levelpop_nonLTE}
\end{equation}
In the case of LTE, the $b_{nl}$'s are equal to one.

Assuming that recombinations and radiative decays are the only processes that are responsible for setting the level population, i.e., Equation~(\ref{eq:equilibrium_eq_rr}) is valid, substituting Equation~(\ref{eq:levelpop_nonLTE}) into Equation~(\ref{eq:CMF}) gives
\begin{multline}
    \alpha_{nl} \frac{1}{2 l + 1} \Big( \frac{h^2}{2 \pi m k T} \Big)^{3/2} \exp(- E_n / k_BT) \\+ \sum^{\infty}_{n'>n} \sum_{l"} b_{n'l'} A_{n'l',nl} \Big( \frac{2l' + 1}{2l + 1} \Big) \exp[(E_{n'} - E_n) / k_BT] \\ = b_{nl} \sum^{n-1}_{n"=1} \sum_{l"} A_{nl,n"l"}.
    \label{eq:b_nl}.
\end{multline}
It is clear that the departure coefficients $b_{nl}$ are independent of density in this case, as used in \S~\ref{eq:CMF}. However, if we take into account other processes when calculating the level population, e.g., collisional excitation or $l$-changing collisions, the $b_{nl}$'s will depend on density\citep[see e.g.][equation~2.6]{Brocklehurst1971}. 

\citet{Storey1995} tabulated emissivities of recombination lines when neglecting collisional excitations from the ground state: in their calculation, \lq the ground state population was not calculated, but instead was assumed to be sufficiently low that collisional excitation from the ground state was negligible\rq\ (see \citealt{Hummer1987} \S~4 and \citealt{Storey1995} \S~2). Therefore, their tabulated values are only applicable to photonionized regions, but not to partially ionized plasma, e.g., the partially ionized region beyond the Str{\"o}mgren radius in the right panel of Fig.~\ref{fig:sph_comp_laser}. As long as the line emissivity comes from radiative recombination predominantly, it is safe to quote their departure coefficients, as are most cases in the literature. 

\section{Choice of laser width}
\label{app:laser}
\begin{figure}
    \includegraphics[width=\columnwidth]{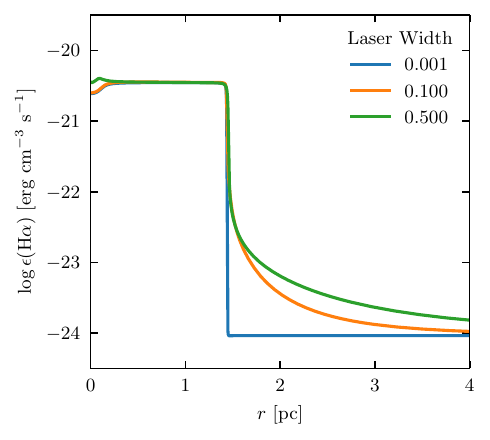}
    \caption{Profile of H$\alpha$ emissivity, $\epsilon({\rm H \alpha}) \equiv \epsilon_{3,2}$, of a gas cloud ionized by a laser source with different laser widths. We use the setup \lq Ref - LSph\rq\ in Table~\ref{tab:cloudy_models} with varying laser widths. Larger widths lead to the inclusion of high-order Lyman photons which have a finite optical depth. Such photons will hence be converted to low-order series and give rise to the H$\alpha$ emission. }
    \label{fig:lyman_pumping_from_source}
\end{figure}
In this section, we illustrate why we have chosen a laser width of 0.001 throughout this paper. In Fig.~\ref{fig:sph_comp_laser}, we have used a laser beam of 1.1 Ryd with a width of 0.001 to shine on the gas cloud, leading to a very sharp transition between ionized and neutral gas, which is reflected in the sudden drop in H$\alpha$ emissivity. Increasing the width of the laser will inevitably include Lyman photons that are just below 1 Ryd. These photons have a finite optical depth which is close to the continuum optical depth. Such photons from the source will hence be converted to lower-order series, such as H$\alpha$. More on the finite optical depth of Lyman photons can be found in \citealt{Hummer1992}. We illustrate this process by varying the laser widths in Fig.~\ref{fig:lyman_pumping_from_source}. We show the H$\alpha$ emissivity profile ionized by the same laser with three choices of laser width: 0.001, 0.100, and 0.500. A laser with a larger width will contain more high-order Lyman series and hence be more converted to lower-order series (e.g. H$\alpha$), resulting in a larger excess in H$\alpha$ emission beyond the Str{\"o}mgren radius ($\sim$ 1.5 pc).

\section{Electron collisional excitation rate coefficient}
\label{app:collisions}
In a partially ionized plasma, the collisional excitation from the ground state can alter the population density of the excited states. The ground state is populated much more than any other excited level. Hence, for simplicity, we only model the collisional contribution to the recombination line emissivity from the ground state.
In this section, we show how the electron collisional excitation rate coefficients are calculated and how they differ from \citet{Raga2015}. In general, our approach follows closely the implementation in \cloudy{}. 

\begin{figure*}
    \includegraphics[width=\textwidth]{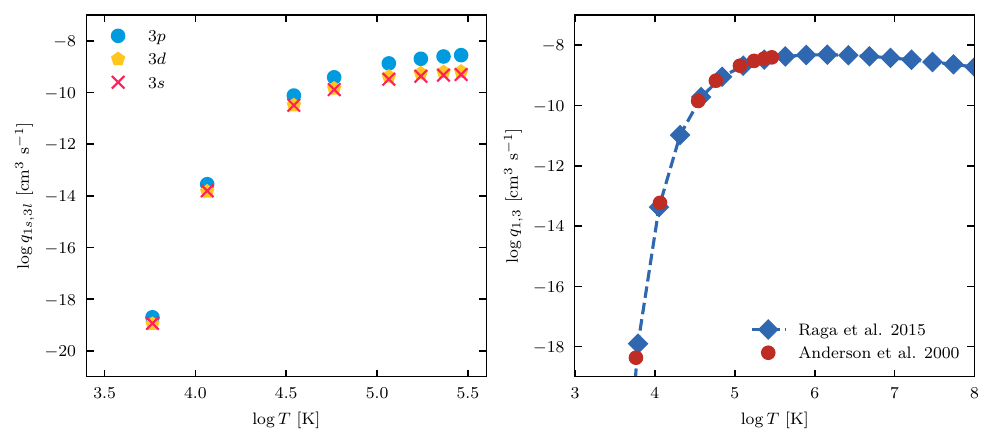}
    \caption{Collisional rate coefficient for excitations by electron collisions as a function of temperature. {\em Left}: Collisional rate coefficient obtained by Equation~(\ref{eq:q_nl1s}) with effective collisional strengths listed in \citealt{Anderson2000}. These rates are $l$-resolved. {\em Right}: Comparison of collisional rate coefficient, $q_{1,3}$, between the values used in \citealt{Raga2015} and those obtained using \citealt{Anderson2000}. The $q_{1,3}$ for \citealt{Anderson2000} is simply the sum of the three components: $q_{1s,3s}$, $q_{1s,3p}$, and $q_{1s,3d}$. We see excellent agreement between \citealt{Anderson2000} and \citealt{Raga2015} within the temperature regime that is valid for \citealt{Anderson2000}. }
    \label{fig:q_lu_comp}
\end{figure*}

For all the levels $n \le 5$, the Maxwell-averaged effective collisional strengths between the levels, $\Upsilon_{nl, n'l'}$, were computed by \cite{Anderson2000} and \cite{Anderson2002} using the $R$-matrix method. Their results are tabulated as a function of electron energies in the range from 0.5 eV to 25.0 eV. We use the standard relation between the effective collision strength and the rate coefficients to calculate the collisional de-excitation and excitation rate coefficients (see chapter 3 of~\citealt{Osterbrock2006}), $q_{nl, 1}$ and $q_{1, nl}$ (with dimension cm$^{-3}$~s$^{-1}$), 
\begin{equation}
\label{eq:q_nl1s}
    q_{nl, 1s} = \frac{8.628 \times 10^{-6}}{g_{1s} T^{1/2}}\Upsilon_{nl, 1s}(T) \exp \bigg(-\frac{\Delta E_{nl, 1s}}{k_B T} \bigg), 
\end{equation}
and
\begin{equation}
\label{eq:q_1snl}
    q_{1s, nl} = \frac{g_{1s}}{g_{nl}} \exp \bigg(\frac{\Delta E_{nl, 1s}}{k_B T}\bigg) q_{nl, 1s}, 
\end{equation}
where the statistical weight of the ground state is $g_{1s} = 2$. For all the values between the tabulated data points, we use a fifth-order polynomial fit to the effective collisional strength as a function of temperature, then calculate $q_{nl, 1s}$ and $q_{1s, nl}$ using Equation~(\ref{eq:q_nl1s}) and Equation~(\ref{eq:q_1snl}). This polynomial fit also allows us to extrapolate to the lower temperature regime, below 0.5 eV. Note that this approach is slightly different from that in \cloudy{}~\citep[see \S~2.4 of][]{Lykins2015}. The values for $q_{1s,3s}$, $q_{1s,3p}$, and $q_{1s,3d}$ are shown in the left panel of Fig.~\ref{fig:q_lu_comp}. 

For all the levels with $n > 5$, we use the semi-classical straight-trajectory Born approximation proposed by~\citet{Lebedev1998} to calculate the \textit{ab initio} excitation rate coefficients. This approach is similar to \citealt{guzman2019a}. The de-excitation coefficient for $n' \to n$ is
\begin{align}
    q_{n', n} &= \frac{g_n}{g_{n'}} \exp \left( \frac{\Delta E_{nn'}}{k_B T} \right) q_{n \to n'}\nonumber\\
    &= \frac{g_n}{g_{n'}} 2 \sqrt{\pi} a_0^2 \alpha c n \left[ \frac{n'}{Z (n' - n)} \right]^3 \frac{f(\theta) \psi}{\sqrt{\theta}},
\end{align}
where
\begin{align}
    \psi = \, & \frac{2n'^2 n^2}{(n' + n)^4 (n' - n)^2} [4 (n' - n) - 1] \exp \left(  \frac{E_n}{k_BT} \right) E_1 \left( \frac{E_n}{k_B T}\right) \nonumber\\
    &+ \frac{8 n^3}{(n' + n)^2 (n' - n) n^2 n'^2} (n' - n - 0.6) \left(\frac{4}{3} + n^2 (n' - n) \right) \nonumber\\
    &\times \left[ 1 - \frac{E_n}{k_B T} \exp \left( \frac{E_n}{k_B T} \right) E_1 \left( \frac{E_n}{k_B T} \right) \right],
\end{align}
\begin{equation}
    f(\theta) = \frac{\ln \left(1 + \frac{n \theta}{Z (n' - n) \sqrt{\theta} + 2.5} \right)}{\ln \left(1 + \frac{n \sqrt{\theta}}{Z (n' - n)} \right)}, \\
\end{equation}
\begin{equation}
    \theta = \frac{k_B T}{Z^2 I_\mathrm{H}}.
\end{equation}
In the equations above, $\alpha$ is the fine structure constant, $I_\mathrm{H}$ the Rydberg energy, $Z$ the atomic charge of the atom ($Z = 1$ for hydrogen atom), and $E_1(x)$ the first exponential integral~\citep[see][]{arfken2013}. 

Note that there are other methods for calculating collisional excitation coefficients~\citep[see][]{VanRegemorter1962, Vriens1980}, the impact of which on the recombination line emissivity has been studied in~\citealt{guzman2019a}. 

For reference, let us now examine the method proposed by \citet{Raga2015}. The temperature dependence of the collisional strengths of the $1 \rightarrow k$ transitions is described by a least-squares polynomial fit of the form
\begin{equation}
    \Omega_{1, k}(T) = \sum_{p=0}^{5}  a_p \log_{10} \bigg (\frac{T}{10^4~\mathrm{K}} \bigg )^k, 
\end{equation}
where the $a_p$'s are the fitting coefficients reported in appendix~A of \citealt{Raga2015}. Here, we use a different symbol for the collisional strength only to distinguish different methods. Then the collisional excitation coefficient, $q_{1, k}$, can be calculated using Equation~(\ref{eq:q_nl1s}) and Equation~(\ref{eq:q_1snl}). The total collisional excitation rate coefficient for a given $n$ level is the sum of contributions from all the $l$-states. We compare the $q_{1,3}$ from \citealt{Raga2015} to that calculated from \citealt{Anderson2000} (by summing three components: $q_{1s,3s}$, $q_{1s,3p}$, and $q_{1s,3d}$) in the right panel of Fig.~\ref{fig:q_lu_comp} - they are in excellent agreement. In \citealt{Raga2015}, they consider the $l$-levels within each $n$-level to be populated according to their statistical weights, which is inaccurate in typical \ion{H}{ii} region conditions. 

We follow \S~3.2.4 in \citealt{Hummer1987} to convert $n$-resolved collisional excitation rate to $nl$-resolved, which is slightly different from \cloudy{}'s implementation~\citep[see \S~2.4 in][]{Guzman2025}. 

\section{Limitations of \hylight{}}
\label{app:hylight_limitation}
\begin{figure*}
\includegraphics[width=\textwidth]{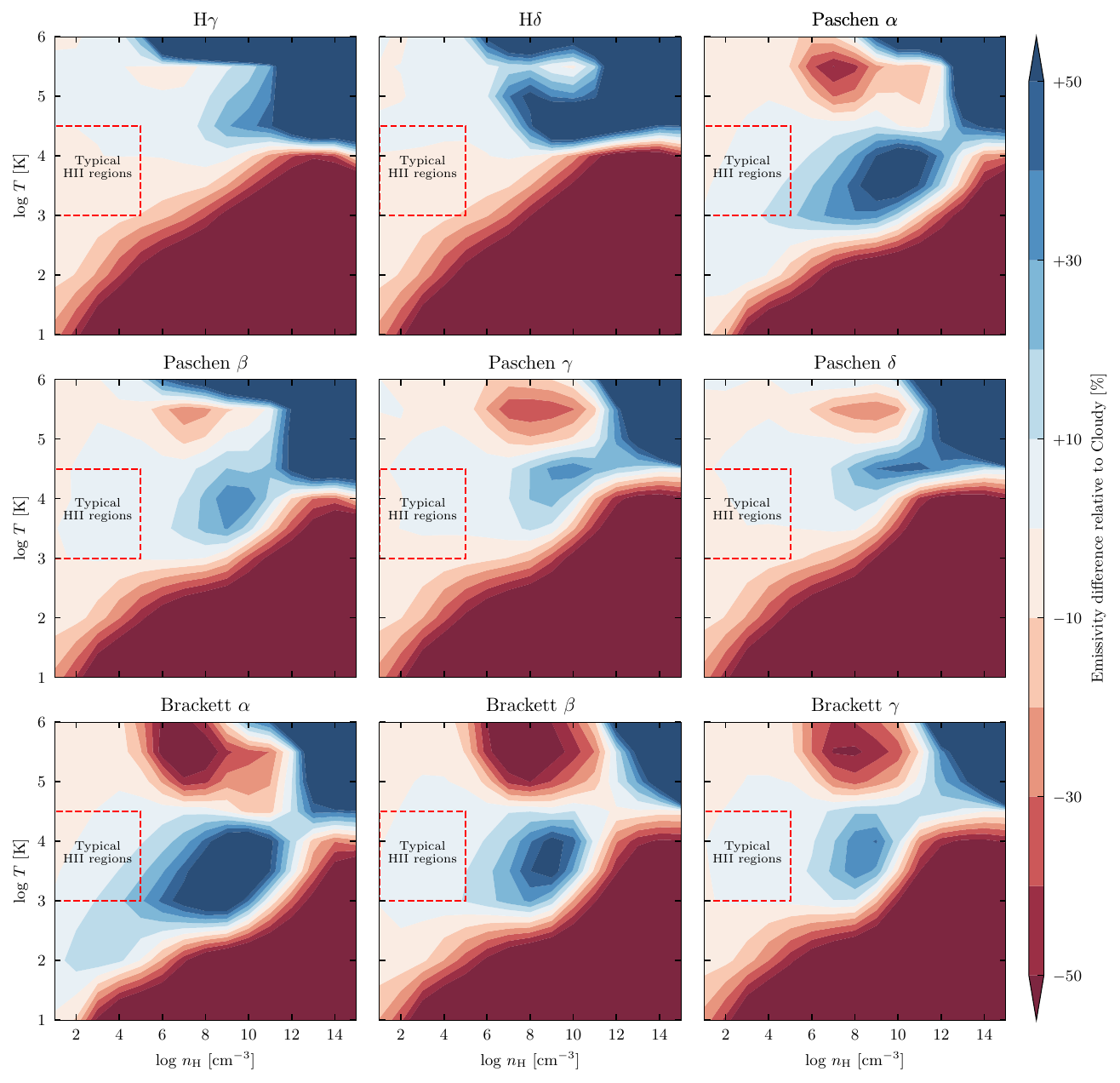}
\caption{Same as Fig.~\ref{fig:hylight_limit_Ha_Hb}, but for other common recombination lines. }
\label{fig:hylight_limit_appendix}
\end{figure*}
Since \hylight{} does not fully account for collisional processes, especially the $l$-mixing collisions, the \hylight{} predictions for line emissivity at high-density and/or high-temperature regions become inaccurate, as discussed in \S~\ref{sec:limitations}). In this section, we show the contour plot of relative difference between \hylight{} and \cloudy{}'s for several commonly observed recombination lines in Fig.~\ref{app:hylight_limitation}. Despite our neglect of $l$-mixing collisions, the \hylight{} predictions at typical \ion{H}{ii} conditions are robust. 

\section{Spectrum and luminosity from SPH particle data}
\label{app:lum_from_part}
This section outlines our method to estimate the total luminosity of a smoothed-particle hydrodynamic (SPH) snapshot. We illustrate the methodology using H$\alpha$ as an example. 
The snapshot contains the density, temperature, and ionization fraction of each SPH particle $i$. Given the mass of the particle, we calculate its total mass in neutral hydrogen, $m_{\mathrm{HI}, i}$. \hylight{} then calculates the level population of individual excited states of \ion{H}{i}, e.g. $n_{3s,i}$, $n_{3p,i}$ and $n_{3d,i}$. 
Then H$\alpha$ luminosity of the SPH particle $i$, is then 
\begin{equation}
    L_i = \left(N_{3s, i} A_{3s, 2p} + N_{3p, i} A_{3p, 2s} + N_{3d, i} A_{3d, 2p}\right) h\nu_{3,2},
\end{equation}
where $\nu_{3\rightarrow2}$ is the photon frequency corresponding to the H$\alpha$ emission. The $N_{3s,i}$ term can be calculated from the integral, 
\begin{align}
\label{eq:level_population_number}
    N_{3s,i} = \int n_{3s,i} \mathrm{d}V = \frac{n_{3s,i}}{n_{{\rm H},i}} \int n_{{\rm H},i} \mathrm{d}V = \frac{n_{3s,i}}{n_{{\rm H},i}} N_{{\rm H},i},
\end{align}
where $n_{{\rm H}, i}$ is the number density of neutral hydrogen of the particle and $\mathrm{d}V$ its volume, $N_{{\rm H},i}=m_{\mathrm{HI}, i}/m_{\rm H}$ is the ratio of the neutral hydrogen mass of the particle over the mass of the hydrogen atom.

We can also calculate the spectrum for each SPH particle. Assuming thermal motion and turbulence are the only physical processes that contribute to line broadening, we can write the line profile $\phi(\nu)$ as
\begin{equation}
    \phi(\nu) = \frac{1}{\sigma \sqrt{2\pi}} \exp \Big( -\frac{(\nu -\nu_0)^2}{2\sigma^2} \Big),
\end{equation}
where $\nu_0$ is the central frequency and $\sigma$ is the dispersion given by summing up the thermal broadening, $\sigma_{\rm thermal}$, and turbulence broadening, $\sigma_{\rm turbulence}$:
\begin{equation}
    \sigma^2 = \sigma^2_{\rm thermal} + \sigma^2_{\rm turbulence}.
\end{equation}
The thermal component can be calculated by 
\begin{equation}
    \sigma_{\rm thermal} = \frac{\nu_0}{c} \sqrt{\frac{k_B T}{m_{\rm H}}},
\end{equation}
where $c$ is the speed of light, $k_B$ is the Boltzmann constant, $T$ is the gas temperature, and $m_{\rm H}$ is the mass of hydrogen atom. In our simulation setup, we do not account for unresolved turbulence; hence, we set $\sigma_{\rm turbulence} = 0$. The line luminosity of a SPH particle is then
\begin{equation}
    L_i(\nu) = L_i \phi(\nu).
\end{equation}
Summing over all the SPH particles gives the total line luminosity, $L_{\rm total}$:
\begin{equation}
    L_{\rm total}(\nu) = \sum_i L_i(\nu).
\end{equation}

\section{Comparing \hylight{} branching ratios with \cloudy{}}
\label{app:cloudy_consistency}
\begin{figure}
\includegraphics[width=\columnwidth]{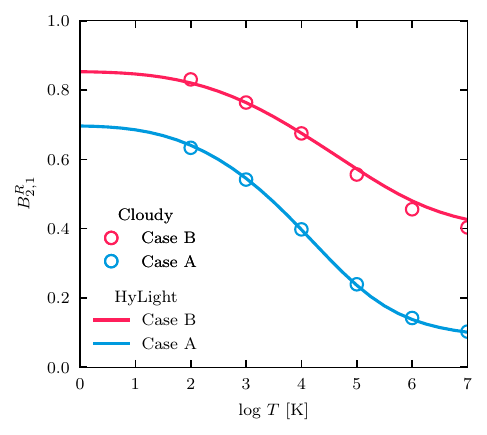}
\caption{Fraction $B^R_{2,1}$ of the total number of recombinations that result in the emission of a Lyman~$\alpha$ photon as a function of temperature, assuming Case~A recombinations ({\em blue}), or Case~B recombinations ({\em red}). The \hylight{} model is shown as {\em drawn lines}; discrete \cloudy{} values are shown with {\em open circles}. In \cloudy{}, full collisional modelling (including $l$-mixing) is accounted for, whilst in \hylight{}, only collisional excitation from the ground state is included, resulting in some tiny difference at high temperature regime. }
\label{fig:cloudy_consistency}
\end{figure}

Similar to Fig.~\ref{fig:lymna_alpha_conversion_coeff}, we report the $B_{2,1}^R$ values calculated from \cloudy{} models in Fig.~\ref{fig:cloudy_consistency}. We see excellent agreement at around $T = 10^4$ K. The small differences in the high-temperature region are due to the incomplete collisional process modelling in \hylight{}. 

\section{\radmc{} grid construction: refinement criterion}
\label{app:convergence_refinement}
\begin{figure*}
\includegraphics[width=\textwidth]{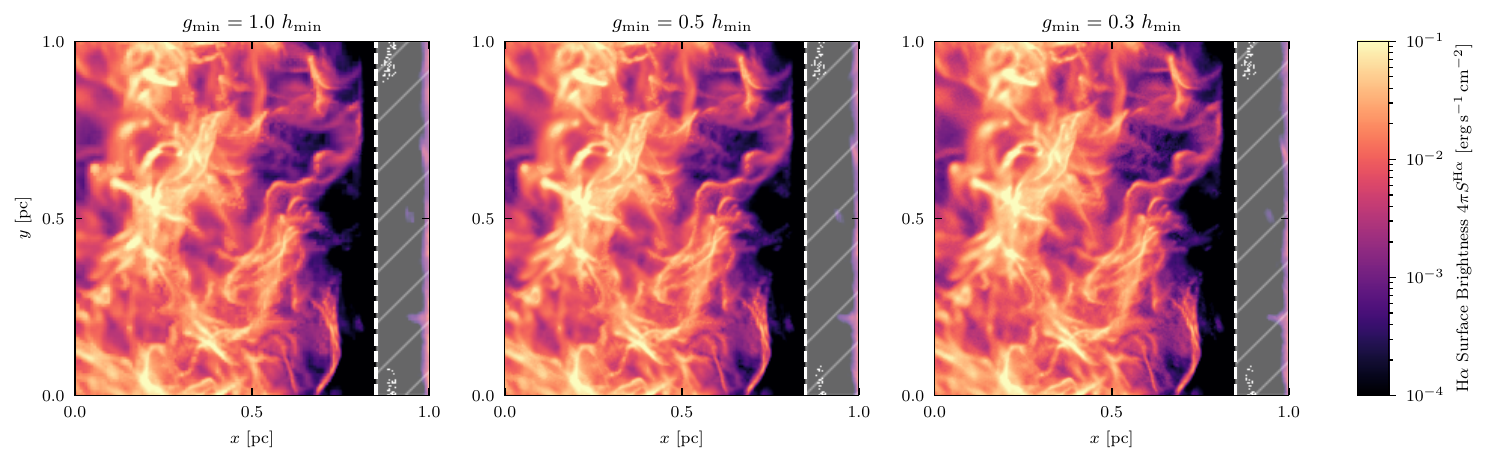}
\caption{H$\alpha$ surface brightness maps of a turbulent \ion{H}{ii} region, computed using \hylight{} and \radmc{}. Panels from left to right show the effect of improving the \radmc{} resolution cell size, from cell size 1.0~$h_{\rm min}$ to 0.3~$h_{\rm min}$. The total luminosity calculated from \radmc{} converges to the same value, 1.17 $\times$ 10$^{35}$~erg~s$^{-1}$, which is consistent with the value obtained by summing over individual particles. The large-scale structures are similar across the different resolutions. }
\label{fig:refinement_comparison}
\end{figure*}

\begin{figure}
\includegraphics[width=\columnwidth]{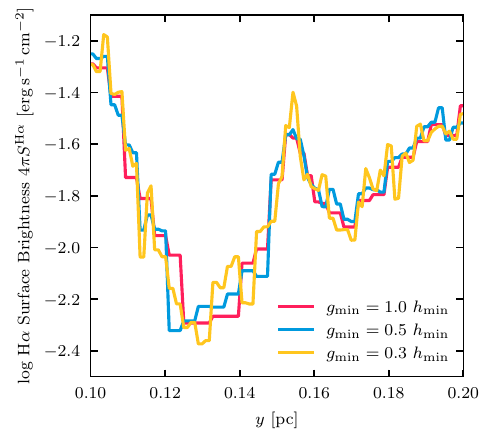}
\caption{H$\alpha$ surface brightness variations along the sightline at $x = 0.5$ pc for different refinement criteria. The higher-resolution map with the smallest cell size shows fluctuations on smaller scales, but the overall structure is similar.}
\label{fig:refinement_comparison_sightline}
\end{figure}

In our post-processing pipeline, SPH quantities are mapped onto an adaptively oct-tree refined grid that can then be fed into \radmc{}. The adaptive refinement criterion is set in a way that the smallest grid size, $g_{\rm min}$, is a fraction of the minimum smoothing length, denoted by $h_{\rm min}$, of the particles in that grid. The convergence of the refinement choice is shown in Fig.~\ref{fig:refinement_comparison}, where we post-processed a realistic \ion{H}{ii} region with a turbulent density field (see \S~\ref{sec:turbulent_HII_region}) with three refinement criteria: $g_{\rm min} = 1.0~h_{\rm min}$, $g_{\rm min} = 0.5~h_{\rm min}$, and $g_{\rm min} = 0.3~h_{\rm min}$. The mock H$\alpha$ intensity maps in Fig.~\ref{fig:refinement_comparison} shows convergence in terms of the total H$\alpha$ luminosity: the \radmc{} total luminosity (summed over all grid cells) converges to the value calculated directly from the particles, 1.17 $\times$ 10$^{35}$~erg~s$^{-1}$. A \radmc{} run with higher refinement level (e.g. $g_{\rm min}$ = 0.3~$h_{\rm min}$) reveals more small-scale structures, as expected. We show in Fig.~\ref{fig:refinement_comparison_sightline} the H$\alpha$ intensity along a sightline through the centre of the box for various level of refinement as a more quantitative comparison. The intensity fluctuations are more visible in the case of $g_{\rm min} = 0.3~h_{\rm min}$, where the dense regions are better refined than the other two cases. 

%% file: references.bib
@article{Cloudy2025,
  title = {The 2025 {{Release}} of {{Cloudy}}},
  author = {Gunasekera, Chamani M. and Van Hoof, Peter A. M. and Dehghanian, Maryam and Chakraborty, Priyanka and Shaw, Gargi and Bianchi, Stefano and Chatzikos, Marios and Tsujimoto, Masahiro and Ferland, Gary J.},
  year = 2025,
  month = nov,
  journal = {Revista Mexicana de Astronom\'ia y Astrof\'isica},
  volume = {61},
  number = {03},
  pages = {120--133},
  issn = {3061-8649},
  doi = {10.22201/ia.01851101p.2025.61.03.01},
  urldate = {2026-02-23},
  copyright = {https://creativecommons.org/licenses/by-nc/4.0/}
}

@book{Lebedev1998,
  title = {Physics of {{Highly Excited Atoms}} and {{Ions}}},
  author = {Lebedev, Vladimir S. and Beigman, Israel L.},
  year = {1998},
  publisher = {Springer Berlin Heidelberg},
  address = {Berlin, Heidelberg},
  doi = {10.1007/978-3-642-72175-5},
  urldate = {2024-11-21},
  copyright = {http://www.springer.com/tdm},
  isbn = {978-3-642-72177-9 978-3-642-72175-5}
}

@article{nussbaumer1984,
	title = {The hydrogenic 2s-1s two-photon emission},
	volume = {138},
	issn = {0004-6361},
	url = {https://ui.adsabs.harvard.edu/abs/1984A&A...138..495N},
	urldate = {2024-07-29},
	journal = {Astronomy and Astrophysics},
	author = {Nussbaumer, H. and Schmutz, W.},
	month = sep,
	year = {1984},
	note = {ADS Bibcode: 1984A\&A...138..495N},
	pages = {495},
}

@book{arfken2013,
  title = {Mathematical Methods for Physicists: A Comprehensive Guide},
  shorttitle = {Mathematical Methods for Physicists},
  author = {Arfken, George and Weber, Hans-Jurgen and Harris, Frank E.},
  year = {2013},
  edition = {7th ed},
  publisher = {Elsevier},
  address = {Amsterdam Boston},
  isbn = {978-0-12-384654-9},
  langid = {english},
  lccn = {530.15}
}

@article{guzman2019a,
  title = {H-, {{He-like}} Recombination Spectra -- {{III}}. n-Changing Collisions in Highly Excited {{Rydberg}} States and Their Impact on the Radio, {{IR}}, and Optical Recombination Lines},
  author = {Guzm{\'a}n, F and Chatzikos, M and {van~Hoof}, P A M and Balser, Dana S and Dehghanian, M and Badnell, N R and Ferland, G J},
  year = {2019},
  month = jun,
  journal = {Monthly Notices of the Royal Astronomical Society},
  volume = {486},
  number = {1},
  pages = {1003--1018},
  issn = {0035-8711, 1365-2966},
  doi = {10.1093/mnras/stz857},
  urldate = {2024-09-03},
  copyright = {https://academic.oup.com/journals/pages/open\_access/funder\_policies/chorus/standard\_publication\_model},
  langid = {english}
}

@article{Vriens1980,
  title = {Cross-Section and Rate Formulas for Electron-Impact Ionization, Excitation, Deexcitation, and Total Depopulation of Excited Atoms},
  author = {Vriens, L. and Smeets, A. H. M.},
  year = {1980},
  month = sep,
  journal = {Physical Review A},
  volume = {22},
  number = {3},
  pages = {940--951},
  issn = {0556-2791},
  doi = {10.1103/PhysRevA.22.940},
  urldate = {2023-11-04},
  langid = {english}
}

@article{Storey1995,
  title = {Recombination Line Intensities for Hydrogenic Ions-{{IV}}. {{Total}} Recombination Coefficients and Machine-Readable Tables for {{Z}}=1 to 8},
  author = {Storey, P. J. and Hummer, D. G.},
  year = {1995},
  month = jan,
  journal = {Monthly Notices of the Royal Astronomical Society},
  volume = {272},
  number = {1},
  pages = {41--48},
  issn = {0035-8711, 1365-2966},
  doi = {10.1093/mnras/272.1.41},
  urldate = {2023-09-18},
  langid = {english}
}

@article{Brocklehurst1971,
  title = {Calculations of {{Level Populations}} for the {{Low Levels}} of {{Hydrogenic Ions}} in {{Gaseous Nebulae}}},
  author = {Brocklehurst, M.},
  year = {1971},
  month = sep,
  journal = {Monthly Notices of the Royal Astronomical Society},
  volume = {153},
  number = {4},
  pages = {471--490},
  issn = {0035-8711, 1365-2966},
  doi = {10.1093/mnras/153.4.471},
  urldate = {2024-03-13},
  langid = {english}
}

@article{luridiana2009,
  title = {{{FLUORESCENT EXCITATION OF BALMER LINES IN GASEOUS NEBULAE}}: Case {{D}}},
  shorttitle = {{{FLUORESCENT EXCITATION OF BALMER LINES IN GASEOUS NEBULAE}}},
  author = {Luridiana, V. and {Sim{\'o}n-D{\'i}az}, S. and Cervi{\~n}o, M. and Delgado, R. M. Gonz{\'a}lez and Porter, R. L. and Ferland, G. J.},
  year = {2009},
  month = feb,
  journal = {The Astrophysical Journal},
  volume = {691},
  number = {2},
  pages = {1712--1728},
  issn = {0004-637X, 1538-4357},
  doi = {10.1088/0004-637X/691/2/1712},
  urldate = {2023-11-15}
}

@article{Baldwin1981,
  title = {Classification Parameters for the Emission-Line Spectra of Extragalactic Objects},
  author = {Baldwin, J. A. and Phillips, M. M. and Terlevich, R.},
  year = {1981},
  month = feb,
  journal = {Publications of the Astronomical Society of the Pacific},
  volume = {93},
  pages = {5},
  issn = {0004-6280, 1538-3873},
  doi = {10.1086/130766},
  urldate = {2023-07-03},
  langid = {english}
}

@article{Kewley2019,
  title = {Understanding {{Galaxy Evolution Through Emission Lines}}},
  author = {Kewley, Lisa J. and Nicholls, David C. and Sutherland, Ralph S.},
  year = {2019},
  month = aug,
  journal = {Annual Review of Astronomy and Astrophysics},
  volume = {57},
  number = {1},
  pages = {511--570},
  issn = {0066-4146, 1545-4282},
  doi = {10.1146/annurev-astro-081817-051832},
  urldate = {2023-04-14},
  langid = {english}
}

@article{Kennicutt1998,
  title = {{{STAR FORMATION IN GALAXIES ALONG THE HUBBLE SEQUENCE}}},
  author = {Kennicutt, Robert C.},
  year = {1998},
  month = sep,
  journal = {Annual Review of Astronomy and Astrophysics},
  volume = {36},
  number = {1},
  pages = {189--231},
  issn = {0066-4146, 1545-4282},
  doi = {10.1146/annurev.astro.36.1.189},
  urldate = {2025-02-18},
  langid = {english}
}

@article{Menzel1937a,
  title = {Physical {{Processes}} in {{Gaseous Nebulae}}. {{I}}.},
  author = {Menzel, Donald H.},
  year = {1937},
  month = may,
  journal = {The Astrophysical Journal},
  volume = {85},
  pages = {330},
  issn = {0004-637X},
  doi = {10.1086/143827},
  urldate = {2024-03-04}
}

@article{Menzel1937b,
  title = {Physical {{Processes}} in {{Gaseous Nebulae}}. {{II}}. {{Theory}} of the {{Balmer Decrement}}},
  author = {Menzel, Donald H. and Baker, James G.},
  year = {1937},
  month = jul,
  journal = {The Astrophysical Journal},
  volume = {86},
  pages = {70},
  issn = {0004-637X},
  doi = {10.1086/143844},
  urldate = {2024-03-04}
}

@article{Baker1938,
  title = {Physical {{Processes}} in {{Gaseous Nebulae}}. {{III}}. {{The Balmer Decrement}}.},
  author = {Baker, James G. and Menzel, Donald H.},
  year = {1938},
  month = jul,
  journal = {The Astrophysical Journal},
  volume = {88},
  pages = {52},
  issn = {0004-637X, 1538-4357},
  doi = {10.1086/143959},
  urldate = {2024-02-27},
  langid = {english}
}

@article{Martin1988,
  title = {Hydrogenic Radiative Recombination at Low Temperature and Density},
  author = {Martin, P. G.},
  year = {1988},
  month = feb,
  journal = {The Astrophysical Journal Supplement Series},
  volume = {66},
  pages = {125},
  issn = {0067-0049, 1538-4365},
  doi = {10.1086/191249},
  urldate = {2023-11-10},
  langid = {english}
}

@article{Storey1988,
  title = {Recombination Line Intensities for Hydrogenic Ions - {{II}}. {{Case B}} Calculations for {{C VI}}, {{N VII}} and {{O VIII}}},
  author = {Storey, P. J. and Hummer, D. G.},
  year = {1988},
  month = apr,
  journal = {Monthly Notices of the Royal Astronomical Society},
  volume = {231},
  number = {4},
  pages = {1139--1144},
  issn = {0035-8711, 1365-2966},
  doi = {10.1093/mnras/231.4.1139},
  urldate = {2023-11-08},
  langid = {english}
}

@article{Hummer1987,
  title = {Recombination-Line Intensities for Hydrogenic Ions - {{I}}. {{Case B}} Calculations for {{H I}} and {{He II}}.},
  author = {Hummer, D. G. and Storey, P. J.},
  year = {1987},
  month = feb,
  journal = {Monthly Notices of the Royal Astronomical Society},
  volume = {224},
  pages = {801--820},
  issn = {0035-8711},
  doi = {10.1093/mnras/224.3.801},
  urldate = {2023-09-20}
}

@article{Raga2015,
  title = {Recombination and Collisionally Excited {{Balmer}} Lines},
  author = {Raga, A. C. and {Castellanos-Ram{\'i}rez}, A. and Esquivel, A. and {Rodr{\'i}guez-Gonz{\'a}lez}, A. and Vel{\'a}zquez, P. F.},
  year = {2015},
  month = oct,
  journal = {Revista Mexicana de Astronomia y Astrofisica},
  volume = {51},
  pages = {231},
  issn = {0185-1101},
  urldate = {2023-09-12}
}

@article{Hirschmann2017,
  title = {Synthetic Nebular Emission from Massive Galaxies -- {{I}}: Origin of the Cosmic Evolution of Optical Emission-Line Ratios},
  shorttitle = {Synthetic Nebular Emission from Massive Galaxies -- {{I}}},
  author = {Hirschmann, Michaela and Charlot, Stephane and Feltre, Anna and Naab, Thorsten and Choi, Ena and Ostriker, Jeremiah P. and Somerville, Rachel S.},
  year = {2017},
  month = dec,
  journal = {Monthly Notices of the Royal Astronomical Society},
  volume = {472},
  number = {2},
  pages = {2468--2495},
  issn = {0035-8711, 1365-2966},
  doi = {10.1093/mnras/stx2180},
  urldate = {2023-07-18},
  langid = {english}
}

@article{Hirschmann2019,
  title = {Synthetic Nebular Emission from Massive Galaxies -- {{II}}. {{Ultraviolet-line}} Diagnostics of Dominant Ionizing Sources},
  author = {Hirschmann, Michaela and Charlot, Stephane and Feltre, Anna and Naab, Thorsten and Somerville, Rachel S and Choi, Ena},
  year = {2019},
  month = jul,
  journal = {Monthly Notices of the Royal Astronomical Society},
  volume = {487},
  number = {1},
  pages = {333--353},
  issn = {0035-8711, 1365-2966},
  doi = {10.1093/mnras/stz1256},
  urldate = {2023-07-18},
  langid = {english}
}

@article{Hirschmann2023,
  title = {Emission-Line Properties of {{IllustrisTNG}} Galaxies: From Local Diagnostic Diagrams to High-Redshift Predictions for {{{\emph{JWST}}}}},
  shorttitle = {Emission-Line Properties of {{IllustrisTNG}} Galaxies},
  author = {Hirschmann, Michaela and Charlot, Stephane and Feltre, Anna and {Curtis-Lake}, Emma and Somerville, Rachel S and Chevallard, Jacopo and Choi, Ena and Nelson, Dylan and Morisset, Christophe and Plat, Adele and {Vidal-Garcia}, Alba},
  year = {2023},
  month = oct,
  journal = {Monthly Notices of the Royal Astronomical Society},
  volume = {526},
  number = {3},
  pages = {3610--3636},
  issn = {0035-8711, 1365-2966},
  doi = {10.1093/mnras/stad2955},
  urldate = {2025-02-20},
  copyright = {https://creativecommons.org/licenses/by/4.0/},
  langid = {english}
}

@article{Kewley2001,
  title = {Theoretical {{Modeling}} of {{Starburst Galaxies}}},
  author = {Kewley, L. J. and Dopita, M. A. and Sutherland, R. S. and Heisler, C. A. and Trevena, J.},
  year = {2001},
  month = jul,
  journal = {The Astrophysical Journal},
  volume = {556},
  number = {1},
  pages = {121--140},
  issn = {0004-637X, 1538-4357},
  doi = {10.1086/321545},
  urldate = {2023-07-12},
  langid = {english}
}

@article{Kauffmann2003,
  title = {The Host Galaxies of Active Galactic Nuclei},
  author = {Kauffmann, Guinevere and Heckman, Timothy M. and Tremonti, Christy and Brinchmann, Jarle and Charlot, St{\'e}phane and White, Simon D. M. and Ridgway, Susan E. and Brinkmann, Jon and Fukugita, Masataka and Hall, Patrick B. and Ivezi{\'c}, {\v Z}eljko and Richards, Gordon T. and Schneider, Donald P.},
  year = {2003},
  month = dec,
  journal = {Monthly Notices of the Royal Astronomical Society},
  volume = {346},
  number = {4},
  pages = {1055--1077},
  issn = {00358711, 13652966},
  doi = {10.1111/j.1365-2966.2003.07154.x},
  urldate = {2023-07-12},
  langid = {english}
}

@article{Cloudy2023,
  title = {The 23.01 {{Release}} of {{Cloudy}}},
  author = {Gunasekera, Chamani M. and Van Hoof, Peter A. M. and Chatzikos, Marios and Ferland, Gary J.},
  year = {2023},
  month = nov,
  journal = {Research Notes of the AAS},
  volume = {7},
  number = {11},
  pages = {246},
  issn = {2515-5172},
  doi = {10.3847/2515-5172/ad0e75},
  urldate = {2025-02-21}
}

@article{Dopita1996,
  title = {Spectral {{Signatures}} of {{Fast Shocks}}. {{I}}. {{Low-Density Model Grid}}},
  author = {Dopita, Michael A. and Sutherland, Ralph S.},
  year = {1996},
  month = jan,
  journal = {The Astrophysical Journal Supplement Series},
  volume = {102},
  pages = {161},
  issn = {0067-0049, 1538-4365},
  doi = {10.1086/192255},
  urldate = {2025-02-21},
  langid = {english}
}

@article{Morisset2013,
  title = {{{pyCloudy}}: {{Tools}} to Manage Astronomical {{Cloudy}} Photoionization Code},
  shorttitle = {{{pyCloudy}}},
  author = {Morisset, Christophe},
  year = {2013},
  month = apr,
  journal = {Astrophysics Source Code Library},
  pages = {ascl:1304.020},
  urldate = {2025-02-21}
}

@article{McLeod2016,
  title = {Connecting the Dots: A Correlation between Ionizing Radiation and Cloud Mass-Loss Rate Traced by Optical Integral Field Spectroscopy},
  shorttitle = {Connecting the Dots},
  author = {McLeod, A. F. and Gritschneder, M. and Dale, J. E. and Ginsburg, A. and Klaassen, P. D. and Mottram, J. C. and Preibisch, T. and Ramsay, S. and Reiter, M. and Testi, L.},
  year = {2016},
  month = nov,
  journal = {Monthly Notices of the Royal Astronomical Society},
  volume = {462},
  number = {4},
  pages = {3537--3569},
  issn = {0035-8711, 1365-2966},
  doi = {10.1093/mnras/stw1864},
  urldate = {2023-07-13},
  langid = {english}
}

@article{Simpson2004,
  title = {On the {{Measurement}} of {{Elemental Abundance Ratios}} in {{Inner Galaxy H}} {\textsc{Ii}} {{Regions}}},
  shorttitle = {On the {{Measurement}} of {{Elemental Abundance Ratios}} in {{Inner Galaxy H}}},
  author = {Simpson, Janet P. and Rubin, Robert H. and Colgan, Sean W. J. and Erickson, Edwin F. and Haas, Michael R.},
  year = {2004},
  month = aug,
  journal = {The Astrophysical Journal},
  volume = {611},
  number = {1},
  pages = {338--352},
  issn = {0004-637X, 1538-4357},
  doi = {10.1086/422028},
  urldate = {2025-02-25},
  langid = {english}
}

@article{Perez2001,
  title = {Density Structure of the Giant {{HII}} Region {{NGC}} 2363},
  author = {P{\'e}rez, Enrique and Gonz{\'a}lez Delgado, Rosa and V{\'i}lchez, Jos{\'e} M.},
  year = {2001},
  journal = {Astrophysics and Space Science},
  volume = {277},
  number = {1suppl},
  pages = {83--86},
  issn = {0004640X},
  doi = {10.1023/A:1012735812409},
  urldate = {2025-02-25}
}

@article{Jin2022,
  title = {Messenger {{Monte Carlo MAPPINGS V}} ({{M}} {\textsuperscript{3}} )---{{A Self-consistent}}, {{Three-dimensional Photoionization Code}}},
  author = {Jin, Yifei and Kewley, Lisa J. and Sutherland, Ralph},
  year = {2022},
  month = mar,
  journal = {The Astrophysical Journal},
  volume = {927},
  number = {1},
  pages = {37},
  issn = {0004-637X, 1538-4357},
  doi = {10.3847/1538-4357/ac48f3},
  urldate = {2023-09-26}
}

@article{Chan2021,
  title = {Smoothed Particle Radiation Hydrodynamics: Two-Moment Method with Local {{Eddington}} Tensor Closure},
  shorttitle = {Smoothed Particle Radiation Hydrodynamics},
  author = {Chan, T K and Theuns, Tom and Bower, Richard and Frenk, Carlos},
  year = {2021},
  month = jul,
  journal = {Monthly Notices of the Royal Astronomical Society},
  volume = {505},
  number = {4},
  pages = {5784--5814},
  issn = {0035-8711, 1365-2966},
  doi = {10.1093/mnras/stab1686},
  urldate = {2023-04-26},
  langid = {english}
}

@article{Richings2021,
  title = {Unravelling the Physics of Multiphase {{AGN}} Winds through Emission Line Tracers},
  author = {Richings, Alexander J and {Faucher-Gigu{\`e}re}, Claude-Andr{\'e} and Stern, Jonathan},
  year = {2021},
  month = mar,
  journal = {Monthly Notices of the Royal Astronomical Society},
  volume = {503},
  number = {2},
  pages = {1568--1585},
  issn = {0035-8711, 1365-2966},
  doi = {10.1093/mnras/stab556},
  urldate = {2025-02-25},
  copyright = {http://creativecommons.org/licenses/by/4.0/},
  langid = {english}
}

@article{McLeod2015,
  title = {The {{Pillars}} of {{Creation}} Revisited with {{MUSE}}: Gas Kinematics and High-Mass Stellar Feedback Traced by Optical Spectroscopy},
  shorttitle = {The {{Pillars}} of {{Creation}} Revisited with {{MUSE}}},
  author = {McLeod, A. F. and Dale, J. E. and Ginsburg, A. and Ercolano, B. and Gritschneder, M. and Ramsay, S. and Testi, L.},
  year = {2015},
  month = jun,
  journal = {Monthly Notices of the Royal Astronomical Society},
  volume = {450},
  number = {1},
  pages = {1057--1076},
  issn = {0035-8711, 1365-2966},
  doi = {10.1093/mnras/stv680},
  urldate = {2023-06-30},
  langid = {english}
}

@article{McLeod2019,
  title = {Feedback from Massive Stars at Low Metallicities: {{MUSE}} Observations of {{N44}} and {{N180}} in the {{Large Magellanic Cloud}}},
  shorttitle = {Feedback from Massive Stars at Low Metallicities},
  author = {McLeod, A F and Dale, J E and Evans, C J and Ginsburg, A and Kruijssen, J M D and Pellegrini, E W and Ramsay, S K and Testi, L},
  year = {2019},
  month = jul,
  journal = {Monthly Notices of the Royal Astronomical Society},
  volume = {486},
  number = {4},
  pages = {5263--5288},
  issn = {0035-8711, 1365-2966},
  doi = {10.1093/mnras/sty2696},
  urldate = {2023-06-23},
  langid = {english}
}

@article{Veilleux1987,
  title = {Spectral Classification of Emission-Line Galaxies},
  author = {Veilleux, Sylvain and Osterbrock, Donald E.},
  year = {1987},
  month = feb,
  journal = {The Astrophysical Journal Supplement Series},
  volume = {63},
  pages = {295},
  issn = {0067-0049, 1538-4365},
  doi = {10.1086/191166},
  urldate = {2023-08-06},
  langid = {english}
}

@article{Marino2013,
  title = {The {{O3N2}} and {{N2}} Abundance Indicators Revisited: Improved Calibrations Based on {{CALIFA}} and {{{\emph{T}}}}{\textsubscript{e}} -Based Literature Data},
  shorttitle = {The {{O3N2}} and {{N2}} Abundance Indicators Revisited},
  author = {Marino, R. A. and {Rosales-Ortega}, F. F. and S{\'a}nchez, S. F. and Gil De Paz, A. and V{\'i}lchez, J. and {Miralles-Caballero}, D. and Kehrig, C. and {P{\'e}rez-Montero}, E. and Stanishev, V. and {Iglesias-P{\'a}ramo}, J. and D{\'i}az, A. I. and {Castillo-Morales}, A. and Kennicutt, R. and {L{\'o}pez-S{\'a}nchez}, A. R. and Galbany, L. and {Garc{\'i}a-Benito}, R. and Mast, D. and {Mendez-Abreu}, J. and {Monreal-Ibero}, A. and Husemann, B. and Walcher, C. J. and {Garc{\'i}a-Lorenzo}, B. and Masegosa, J. and Del Olmo Orozco, A. and Mour{\~a}o, A. M. and Ziegler, B. and Moll{\'a}, M. and Papaderos, P. and {S{\'a}nchez-Bl{\'a}zquez}, P. and Gonz{\'a}lez Delgado, R. M. and {Falc{\'o}n-Barroso}, J. and Roth, M. M. and Van De Ven, G. and Team, Califa},
  year = {2013},
  month = nov,
  journal = {Astronomy \& Astrophysics},
  volume = {559},
  pages = {A114},
  issn = {0004-6361, 1432-0746},
  doi = {10.1051/0004-6361/201321956},
  urldate = {2025-02-25}
}

@article{Rerez-Montero2014,
  title = {Deriving Model-Based {{Te-consistent}} Chemical Abundances in Ionized Gaseous Nebulae},
  author = {{Perez-Montero}, E.},
  year = {2014},
  month = may,
  journal = {Monthly Notices of the Royal Astronomical Society},
  volume = {441},
  number = {3},
  pages = {2663--2675},
  issn = {0035-8711, 1365-2966},
  doi = {10.1093/mnras/stu753},
  urldate = {2024-01-30},
  langid = {english}
}

@article{Perez-Montero2017,
  title = {Ionized {{Gaseous Nebulae Abundance Determination}} from the {{Direct Method}}},
  author = {{P{\'e}rez-Montero}, Enrique},
  year = {2017},
  month = apr,
  journal = {Publications of the Astronomical Society of the Pacific},
  volume = {129},
  number = {974},
  pages = {043001},
  issn = {0004-6280, 1538-3873},
  doi = {10.1088/1538-3873/aa5abb},
  urldate = {2025-02-25}
}

@article{Kewley2002,
  title = {Using {{Strong Lines}} to {{Estimate Abundances}} in {{Extragalactic H}} {\textsc{Ii}} {{Regions}} and {{Starburst Galaxies}}},
  shorttitle = {Using {{Strong Lines}} to {{Estimate Abundances}} in {{Extragalactic H}}},
  author = {Kewley, L. J. and Dopita, M. A.},
  year = {2002},
  month = sep,
  journal = {The Astrophysical Journal Supplement Series},
  volume = {142},
  number = {1},
  pages = {35--52},
  issn = {0067-0049, 1538-4365},
  doi = {10.1086/341326},
  urldate = {2023-04-13},
  langid = {english}
}

@article{Katz2022,
  title = {{{RAMSES-RTZ}}: Non-Equilibrium Metal Chemistry and Cooling Coupled to on-the-Fly Radiation Hydrodynamics},
  shorttitle = {{{RAMSES-RTZ}}},
  author = {Katz, Harley},
  year = {2022},
  month = mar,
  journal = {Monthly Notices of the Royal Astronomical Society},
  volume = {512},
  number = {1},
  pages = {348--365},
  issn = {0035-8711, 1365-2966},
  doi = {10.1093/mnras/stac423},
  urldate = {2025-02-25},
  copyright = {https://creativecommons.org/licenses/by/4.0/},
  langid = {english}
}

@article{Richings2022,
  title = {The Effects of Local Stellar Radiation and Dust Depletion on Non-Equilibrium Interstellar Chemistry},
  author = {Richings, Alexander J and {Faucher-Gigu{\`e}re}, Claude-Andr{\'e} and Gurvich, Alexander B and Schaye, Joop and Hayward, Christopher C},
  year = {2022},
  month = oct,
  journal = {Monthly Notices of the Royal Astronomical Society},
  volume = {517},
  number = {2},
  pages = {1557--1583},
  issn = {0035-8711, 1365-2966},
  doi = {10.1093/mnras/stac2338},
  urldate = {2025-02-25},
  copyright = {https://creativecommons.org/licenses/by/4.0/},
  langid = {english}
}

@article{Dagostino2019,
  title = {Comparison of {{Theoretical Starburst Photoionization Models}} for {{Optical Diagnostics}}},
  author = {D'Agostino, Joshua J. and Kewley, Lisa J. and Groves, Brent and Byler, Nell and Sutherland, Ralph S. and Nicholls, David and Leitherer, Claus and Stanway, Elizabeth R.},
  year = {2019},
  month = jun,
  journal = {The Astrophysical Journal},
  volume = {878},
  number = {1},
  pages = {2},
  issn = {1538-4357},
  doi = {10.3847/1538-4357/ab1d5e},
  urldate = {2023-07-16}
}

@article{Richings2014a,
  title = {Non-Equilibrium Chemistry and Cooling in the Diffuse Interstellar Medium - {{I}}. {{Optically}} Thin Regime},
  author = {Richings, A. J. and Schaye, J. and Oppenheimer, B. D.},
  year = {2014},
  month = apr,
  journal = {Monthly Notices of the Royal Astronomical Society},
  volume = {440},
  number = {4},
  pages = {3349--3369},
  issn = {0035-8711, 1365-2966},
  doi = {10.1093/mnras/stu525},
  urldate = {2023-07-17},
  langid = {english}
}

@article{Richings2014b,
  title = {Non-Equilibrium Chemistry and Cooling in the Diffuse Interstellar Medium -- {{II}}. {{Shielded}} Gas},
  author = {Richings, A. J. and Schaye, J. and Oppenheimer, B. D.},
  year = {2014},
  month = aug,
  journal = {Monthly Notices of the Royal Astronomical Society},
  volume = {442},
  number = {3},
  pages = {2780--2796},
  issn = {1365-2966, 0035-8711},
  doi = {10.1093/mnras/stu1046},
  urldate = {2023-07-17},
  langid = {english}
}

@article{Pengelly1964,
  title = {Recombination {{Spectra}}: {{I}}. {{Calculations}} for {{Hydrogenic Ions}} in the {{Limit}} of {{Low Densities}}},
  shorttitle = {Recombination {{Spectra}}},
  author = {Pengelly, R. M. and Seaton, M. J.},
  year = {1964},
  month = jan,
  journal = {Monthly Notices of the Royal Astronomical Society},
  volume = {127},
  number = {2},
  pages = {145--163},
  issn = {0035-8711, 1365-2966},
  doi = {10.1093/mnras/127.2.145},
  urldate = {2025-03-20},
  langid = {english}
}

@book{Osterbrock1974,
  title = {Astrophysics of Gaseous Nebulae},
  author = {Osterbrock, Donald E.},
  year = {1974},
  month = jan,
  journal = {A Series of Books in Astronomy and Astrophysics},
  urldate = {2025-03-20}
}

@BOOK{Spitzer1978,
       author = {{Spitzer}, Lyman},
        title = "{Physical processes in the interstellar medium}",
         year = 1978,
          doi = {10.1002/9783527617722},
       adsurl = {https://ui.adsabs.harvard.edu/abs/1978ppim.book.....S}
}

@article{Cantalupo2008,
  title = {Mapping {{Neutral Hydrogen}} during {{Reionization}} with the {{Ly$\alpha$ Emission}} from {{Quasar Ionization Fronts}}},
  author = {Cantalupo, Sebastiano and Porciani, Cristiano and Lilly, Simon J.},
  year = {2008},
  month = jan,
  journal = {The Astrophysical Journal},
  volume = {672},
  number = {1},
  pages = {48--58},
  issn = {0004-637X, 1538-4357},
  doi = {10.1086/523298},
  urldate = {2025-03-17},
  langid = {english}
}

@article{Dijkstra2014,
  title = {Ly{$\alpha$} {{Emitting Galaxies}} as a {{Probe}} of {{Reionisation}}},
  author = {Dijkstra, Mark},
  year = {2014},
  journal = {Publications of the Astronomical Society of Australia},
  volume = {31},
  pages = {e040},
  issn = {1323-3580, 1448-6083},
  doi = {10.1017/pasa.2014.33},
  urldate = {2025-03-17},
  copyright = {https://www.cambridge.org/core/terms},
  langid = {english}
}

@article{Seaton1959b,
  title = {The {{Solution}} of {{Capture-Cascade Equations}} for {{Hydrogen}}},
  author = {Seaton, M. J.},
  year = {1959},
  month = apr,
  journal = {Monthly Notices of the Royal Astronomical Society},
  volume = {119},
  number = {2},
  pages = {90--97},
  issn = {0035-8711, 1365-2966},
  doi = {10.1093/mnras/119.2.90},
  urldate = {2025-04-02},
  langid = {english}
}

@book{Bethe1933,
  title = {{Quantentheorie}},
  editor = {Bethe, H. and Hund, F. and Mott, N. F. and Pauli, W. and Rubinowicz, A. and Wentzel, G. and Smekal, A.},
  year = {1933},
  publisher = {Springer Berlin Heidelberg},
  address = {Berlin, Heidelberg},
  doi = {10.1007/978-3-642-52619-0},
  urldate = {2025-04-03},
  copyright = {http://www.springer.com/tdm},
  isbn = {978-3-642-52565-0 978-3-642-52619-0},
  langid = {ngerman}
}

@BOOK{Osterbrock2006,
       author = {{Osterbrock}, Donald E. and {Ferland}, Gary J.},
        title = "{Astrophysics of gaseous nebulae and active galactic nuclei}",
         year = 2006,
       adsurl = {https://ui.adsabs.harvard.edu/abs/2006agna.book.....O}
}

@article{Shapley2025,
  title = {The {{AURORA Survey}}: {{A New Era}} of {{Emission-line Diagrams}} with {{JWST}}/{{NIRSpec}}},
  shorttitle = {The {{AURORA Survey}}},
  author = {Shapley, Alice E. and Sanders, Ryan L. and Topping, Michael W. and Reddy, Naveen A. and Berg, Danielle A. and Bouwens, Rychard J. and Brammer, Gabriel and Carnall, Adam C. and Cullen, Fergus and Dav{\'e}, Romeel and Dunlop, James S. and Ellis, Richard S. and F{\"o}rster Schreiber, N. M. and Furlanetto, Steven R. and Glazebrook, Karl and Illingworth, Garth D. and Jones, Tucker and Kriek, Mariska and McLeod, Derek J. and McLure, Ross J. and Narayanan, Desika and Oesch, Pascal and Pahl, Anthony J. and Pettini, Max and Schaerer, Daniel and Stark, Daniel P. and Steidel, Charles C. and Tang, Mengtao and Clarke, Leonardo and Donnan, Callum T. and Kehoe, Emily},
  year = {2025},
  month = feb,
  journal = {The Astrophysical Journal},
  volume = {980},
  number = {2},
  pages = {242},
  issn = {0004-637X, 1538-4357},
  doi = {10.3847/1538-4357/adad68},
  urldate = {2025-06-25}
}

@article{Sankrit2000,
  title = {Modeling the {{Photoionized Interface}} in {{Blister H}} {\textsc{Ii}} {{Regions}}},
  shorttitle = {Modeling the {{Photoionized Interface}} in {{Blister H}}},
  author = {Sankrit, Ravi and Hester, J. Jeff},
  year = {2000},
  month = jun,
  journal = {The Astrophysical Journal},
  volume = {535},
  number = {2},
  pages = {847--856},
  issn = {0004-637X, 1538-4357},
  doi = {10.1086/308872},
  urldate = {2025-06-25},
  langid = {english}
}

@ARTICLE{McClymont2025,
       author = {{McClymont}, William and {Tacchella}, Sandro and {Smith}, Aaron and {Kannan}, Rahul and {Puchwein}, Ewald and {Borrow}, Josh and {Garaldi}, Enrico and {Keating}, Laura and {Vogelsberger}, Mark and {Zier}, Oliver and {Shen}, Xuejian and {Popovic}, Filip},
        title = "{The THESAN-ZOOM project: central starbursts and inside-out quenching govern galaxy sizes in the early Universe}",
      journal = {arXiv e-prints},
         year = 2025,
        month = mar,
          eid = {arXiv:2503.04894},
        pages = {arXiv:2503.04894},
          doi = {10.48550/arXiv.2503.04894},
archivePrefix = {arXiv},
       eprint = {2503.04894},
 primaryClass = {astro-ph.GA},
       adsurl = {https://ui.adsabs.harvard.edu/abs/2025arXiv250304894M}
}

@article{Walch2015,
  title = {The {{SILCC}} ({{SImulating}} the {{LifeCycle}} of Molecular {{Clouds}}) Project -- {{I}}. {{Chemical}} Evolution of the Supernova-Driven {{ISM}}},
  author = {Walch, S. and Girichidis, P. and Naab, T. and Gatto, A. and Glover, S. C. O. and W{\"u}nsch, R. and Klessen, R. S. and Clark, P. C. and Peters, T. and Derigs, D. and Baczynski, C.},
  year = {2015},
  month = nov,
  journal = {Monthly Notices of the Royal Astronomical Society},
  volume = {454},
  number = {1},
  pages = {246--276},
  issn = {0035-8711, 1365-2966},
  doi = {10.1093/mnras/stv1975},
  urldate = {2025-06-25},
  langid = {english}
}

@BOOK{Rybicki1979,
       author = {{Rybicki}, George B. and {Lightman}, Alan P.},
        title = "{Radiative processes in astrophysics}",
         year = 1979,
       adsurl = {https://ui.adsabs.harvard.edu/abs/1979rpa..book.....R}
}

@article{Pellegrini2012,
  title = {{{THE OPTICAL DEPTH OF H II REGIONS IN THE MAGELLANIC CLOUDS}}},
  author = {Pellegrini, E. W. and Oey, M. S. and Winkler, P. F. and Points, S. D. and Smith, R. C. and Jaskot, A. E. and Zastrow, J.},
  year = {2012},
  month = aug,
  journal = {The Astrophysical Journal},
  volume = {755},
  number = {1},
  pages = {40},
  issn = {0004-637X, 1538-4357},
  doi = {10.1088/0004-637X/755/1/40},
  urldate = {2023-07-11}
}

@article{Labzowsky2006,
  title = {Two-Photon {{E1M1}} and {{E1E2}} Transitions between 2p and 1s Levels in Hydrogen},
  author = {Labzowsky, L. and Solovyev, D. and Plunien, G. and Soff, G.},
  year = {2006},
  month = mar,
  journal = {The European Physical Journal D},
  volume = {37},
  number = {3},
  pages = {335--343},
  issn = {1434-6060, 1434-6079},
  doi = {10.1140/epjd/e2006-00022-6},
  urldate = {2024-07-29},
  copyright = {http://www.springer.com/tdm},
  langid = {english}
}

@article{Vrinceanu2001,
  title = {Classical and Quantal Collisional {{Stark}} Mixing at Ultralow Energies},
  author = {Vrinceanu, D. and Flannery, M. R.},
  year = {2001},
  month = feb,
  journal = {Physical Review A},
  volume = {63},
  number = {3},
  pages = {032701},
  issn = {1050-2947, 1094-1622},
  doi = {10.1103/PhysRevA.63.032701},
  urldate = {2024-10-07},
  copyright = {http://link.aps.org/licenses/aps-default-license},
  langid = {english}
}

@article{Pengelly1964b,
  title = {Recombination {{Spectra}}: {{II}}. {{Collisional Transitions Between States}} of {{Degenerate Energy Levels}}},
  shorttitle = {Recombination {{Spectra}}},
  author = {Pengelly, R. M. and Seaton, M. J.},
  year = {1964},
  month = jan,
  journal = {Monthly Notices of the Royal Astronomical Society},
  volume = {127},
  number = {2},
  pages = {165--175},
  issn = {0035-8711, 1365-2966},
  doi = {10.1093/mnras/127.2.165},
  urldate = {2025-07-01},
  langid = {english}
}

@article{Vrinceanu2019,
  title = {Efficient {{Computation}} of {{Collisional}} {$\ell$}-Mixing {{Rate Coefficients}} in {{Astrophysical Plasmas}}},
  author = {Vrinceanu, D. and Onofrio, R. and Oonk, J. B. R. and Salas, P. and Sadeghpour, H. R.},
  year = {2019},
  month = jul,
  journal = {The Astrophysical Journal},
  volume = {879},
  number = {2},
  pages = {115},
  issn = {0004-637X, 1538-4357},
  doi = {10.3847/1538-4357/ab218c},
  urldate = {2025-07-01}
}

@article{Sutherland2018,
  title = {{{MAPPINGS V}}: {{Astrophysical}} Plasma Modeling Code},
  shorttitle = {{{MAPPINGS V}}},
  author = {Sutherland, Ralph and Dopita, Mike and Binette, Luc and Groves, Brent},
  year = {2018},
  month = jul,
  journal = {Astrophysics Source Code Library},
  pages = {ascl:1807.005},
  urldate = {2025-07-10}
}

@misc{Cloudy2017,
  title = {The 2017 {{Release Cloudy}}},
  author = {Ferland, G. J. and Chatzikos, M. and Guzm{\'a}n, F. and Lykins, M. L. and {van Hoof}, P. A. M. and Williams, R. J. R. and Abel, N. P. and Badnell, N. R. and Keenan, F. P. and Porter, R. L. and Stancil, P. C.},
  year = {2017},
  month = oct,
  volume = {53},
  publisher = {arXiv},
  issn = {0185-1101},
  doi = {10.48550/arXiv.1705.10877},
  urldate = {2025-07-10}
}

@article{Cloudy1999,
  title = {Cloudy: {{Numerical}} Simulation of Plasmas and Their Spectra},
  shorttitle = {Cloudy},
  author = {Ferland, Gary J. and {van Hoof}, Peter A. M. and Chatzikos, Marios and Gunasekera, Chamani M. and Chakraborty, Priyanka and Shaw, Gargi and Dehghanian, Maryam},
  year = {1999},
  month = oct,
  journal = {Astrophysics Source Code Library},
  pages = {ascl:9910.001},
  urldate = {2025-07-10}
}

@article{Guzman2025,
  title = {H-, {{He-like}} Recombination Spectra -- {{V}}. {{On}} the Dependence of the Simulated Line Intensities on the Number of Electronic Levels of the Atoms},
  author = {Guzm{\'a}n, F and Chatzikos, M and Ferland, G J},
  year = {2025},
  month = may,
  journal = {Monthly Notices of the Royal Astronomical Society},
  volume = {539},
  number = {4},
  pages = {2939--2956},
  issn = {0035-8711, 1365-2966},
  doi = {10.1093/mnras/staf605},
  urldate = {2025-07-03},
  copyright = {https://creativecommons.org/licenses/by/4.0/},
  langid = {english}
}

@article{Rubin1989,
  title = {The Effect of Density Variations on Elemental Abundance Ratios in Gaseous Nebulae},
  author = {Rubin, R. H.},
  year = {1989},
  month = apr,
  journal = {The Astrophysical Journal Supplement Series},
  volume = {69},
  pages = {897},
  publisher = {American Astronomical Society},
  issn = {0067-0049, 1538-4365},
  doi = {10.1086/191330},
  urldate = {2025-07-22},
  langid = {english}
}

@article{Seaton1957,
  title = {Relative [o {{II}}] {{Intensities}} in {{Gaseous Nebulae}}.},
  author = {Seaton, M. J. and Osterbrock, D. E.},
  year = {1957},
  month = jan,
  journal = {The Astrophysical Journal},
  volume = {125},
  pages = {66},
  publisher = {American Astronomical Society},
  issn = {0004-637X, 1538-4357},
  doi = {10.1086/146282},
  urldate = {2025-07-22},
  langid = {english}
}

@article{Peimbert1967,
  title = {Temperature {{Determinations}} of {{H II Regions}}},
  author = {Peimbert, Manuel},
  year = {1967},
  month = dec,
  journal = {The Astrophysical Journal},
  volume = {150},
  pages = {825},
  publisher = {American Astronomical Society},
  issn = {0004-637X, 1538-4357},
  doi = {10.1086/149385},
  urldate = {2025-07-22},
  langid = {english}
}

@article{Silva2018,
  title = {Tomographic Intensity Mapping versus Galaxy Surveys: Observing the {{Universe}} in {{H}}\,{$\alpha$} Emission with New Generation Instruments},
  shorttitle = {Tomographic Intensity Mapping versus Galaxy Surveys},
  author = {Silva, B. Marta and Zaroubi, Saleem and Kooistra, Robin and Cooray, Asantha},
  year = {2018},
  month = apr,
  journal = {Monthly Notices of the Royal Astronomical Society},
  volume = {475},
  number = {2},
  pages = {1587--1608},
  publisher = {Oxford University Press (OUP)},
  issn = {0035-8711, 1365-2966},
  doi = {10.1093/mnras/stx3265},
  urldate = {2025-07-22},
  langid = {english}
}

@article{Rodriguez-gonzalez2023,
  title = {Numerical {{Models}} of {{Planetary Nebulae}} with {{Different Episodes}} of {{Mass Ejection}}: {{The Particular Case}} of {{HuBi}} 1},
  shorttitle = {Numerical {{Models}} of {{Planetary Nebulae}} with {{Different Episodes}} of {{Mass Ejection}}},
  author = {{Rodr{\'i}guez-Gonz{\'a}lez}, Ary and Pe{\~n}a, Miriam and {Hern{\'a}ndez-Mart{\'i}nez}, Liliana and {Ruiz-Escobedo}, Francisco and Raga, Alejandro and Stasi{\'n}ska, Grazyna and Castorena, Jorge Ivan},
  year = {2023},
  month = oct,
  journal = {The Astrophysical Journal},
  volume = {955},
  number = {2},
  pages = {151},
  publisher = {American Astronomical Society},
  issn = {0004-637X, 1538-4357},
  doi = {10.3847/1538-4357/acf0bc},
  urldate = {2025-07-22},
  copyright = {http://creativecommons.org/licenses/by/4.0/}
}

@article{Hummer1992,
  title = {Recombination Line Intensities for Hydrogenic Ions -- {{III}}. {{Effects}} of Finite Optical Depth and Dust},
  author = {Hummer, D. G. and Storey, P. J.},
  year = {1992},
  month = jan,
  journal = {Monthly Notices of the Royal Astronomical Society},
  volume = {254},
  number = {2},
  pages = {277--290},
  publisher = {Oxford University Press (OUP)},
  issn = {0035-8711, 1365-2966},
  doi = {10.1093/mnras/254.2.277},
  urldate = {2025-07-23},
  langid = {english}
}

@article{Schaller2024,
  title = {{\textsc{ {\textbf{Swift}} }} : A Modern Highly Parallel Gravity and Smoothed Particle Hydrodynamics Solver for Astrophysical and Cosmological Applications},
  author = {Schaller, Matthieu and Borrow, Josh and Draper, Peter W and Ivkovic, Mladen and McAlpine, Stuart and Vandenbroucke, Bert and Bah{\'e}, Yannick and Chaikin, Evgenii and Chalk, Aidan B G and Chan, Tsang Keung and Correa, Camila and {van~Daalen}, Marcel and Elbers, Willem and Gonnet, Pedro and Hausammann, Lo{\"i}c and Helly, John and Hu{\v s}ko, Filip and Kegerreis, Jacob A and Nobels, Folkert S J and Ploeckinger, Sylvia and Revaz, Yves and Roper, William J and {Ruiz-Bonilla}, Sergio and Sandnes, Thomas D and Uyttenhove, Yolan and Willis, James S and Xiang, Zhen},
  year = {2024},
  month = apr,
  journal = {Monthly Notices of the Royal Astronomical Society},
  volume = {530},
  number = {2},
  pages = {2378--2419},
  issn = {0035-8711, 1365-2966},
  doi = {10.1093/mnras/stae922},
  urldate = {2024-09-04},
  copyright = {https://creativecommons.org/licenses/by/4.0/},
  langid = {english}
}

@Article{Hunter2007,
  Author    = {Hunter, J. D.},
  Title     = {Matplotlib: A 2D graphics environment},
  Journal   = {Computing in Science \& Engineering},
  Volume    = {9},
  Number    = {3},
  Pages     = {90--95},
  publisher = {IEEE COMPUTER SOC},
  doi       = {10.1109/MCSE.2007.55},
  year      = 2007
}

@article{Harris2020,
  title = {Array Programming with {{NumPy}}},
  author = {Harris, Charles R. and Millman, K. Jarrod and Van Der Walt, St{\'e}fan J. and Gommers, Ralf and Virtanen, Pauli and Cournapeau, David and Wieser, Eric and Taylor, Julian and Berg, Sebastian and Smith, Nathaniel J. and Kern, Robert and Picus, Matti and Hoyer, Stephan and Van Kerkwijk, Marten H. and Brett, Matthew and Haldane, Allan and Del R{\'i}o, Jaime Fern{\'a}ndez and Wiebe, Mark and Peterson, Pearu and {G{\'e}rard-Marchant}, Pierre and Sheppard, Kevin and Reddy, Tyler and Weckesser, Warren and Abbasi, Hameer and Gohlke, Christoph and Oliphant, Travis E.},
  year = {2020},
  month = sep,
  journal = {Nature},
  volume = {585},
  number = {7825},
  pages = {357--362},
  issn = {0028-0836, 1476-4687},
  doi = {10.1038/s41586-020-2649-2},
  urldate = {2025-08-19},
  langid = {english}
}

@Misc{Jones2001,
  author =    {Eric Jones and Travis Oliphant and Pearu Peterson and others},
  title =     {{SciPy}: Open source scientific tools for {Python}},
  year =      {2001},
  url = "http://www.scipy.org/"
}

@ARTICLE{Virtanen2020,
  author  = {Virtanen, Pauli and Gommers, Ralf and Oliphant, Travis E. and
            Haberland, Matt and Reddy, Tyler and Cournapeau, David and
            Burovski, Evgeni and Peterson, Pearu and Weckesser, Warren and
            Bright, Jonathan and {van der Walt}, St{\'e}fan J. and
            Brett, Matthew and Wilson, Joshua and Millman, K. Jarrod and
            Mayorov, Nikolay and Nelson, Andrew R. J. and Jones, Eric and
            Kern, Robert and Larson, Eric and Carey, C J and
            Polat, {\.I}lhan and Feng, Yu and Moore, Eric W. and
            {VanderPlas}, Jake and Laxalde, Denis and Perktold, Josef and
            Cimrman, Robert and Henriksen, Ian and Quintero, E. A. and
            Harris, Charles R. and Archibald, Anne M. and
            Ribeiro, Ant{\^o}nio H. and Pedregosa, Fabian and
            {van Mulbregt}, Paul and {SciPy 1.0 Contributors}},
  title   = {{{SciPy} 1.0: Fundamental Algorithms for Scientific
            Computing in Python}},
  journal = {Nature Methods},
  year    = {2020},
  volume  = {17},
  pages   = {261--272},
  adsurl  = {https://rdcu.be/b08Wh},
  doi     = {10.1038/s41592-019-0686-2},
}

@article{Borrow2020,
  title = {Swiftsimio: {{A Python}} Library for Reading {{SWIFT}} Data},
  shorttitle = {Swiftsimio},
  author = {Borrow, Josh and Borrisov, Alexei},
  year = {2020},
  month = aug,
  journal = {Journal of Open Source Software},
  volume = {5},
  number = {52},
  pages = {2430},
  issn = {2475-9066},
  doi = {10.21105/joss.02430},
  urldate = {2025-08-19},
  copyright = {http://creativecommons.org/licenses/by/4.0/}
}

@article{Goldbaum2018,
  doi = {10.21105/joss.00809},
  url = {https://doi.org/10.21105/joss.00809},
  year  = {2018},
  month = {aug},
  publisher = {The Open Journal},
  volume = {3},
  number = {28},
  pages = {809},
  author = {Nathan J. Goldbaum and John A. ZuHone and Matthew J. Turk and Kacper Kowalik and Anna L. Rosen},
  title = {unyt: Handle,  manipulate,  and convert data with units in Python},
  journal = {Journal of Open Source Software}
}

@software{Collette2022,
  author       = {Andrew Collette and
                  Thomas Kluyver and
                  Thomas A Caswell and
                  James Tocknell and
                  Jerome Kieffer and
                  Aleksandar Jelenak and
                  Anthony Scopatz and
                  Darren Dale and
                  Chen and
                  Thomas VINCENT and
                  Matt Einhorn and
                  payno and
                  juliagarriga and
                  Pierlauro Sciarelli and
                  Valentin Valls and
                  Satrajit Ghosh and
                  Ulrik Kofoed Pedersen and
                  jakirkham and
                  Martin Raspaud and
                  Cyril Danilevski and
                  Hameer Abbasi and
                  John Readey and
                  Kai Mühlbauer and
                  Andrey Paramonov and
                  Lawrence Chan and
                  V. Armando Solé and
                  jialin and
                  Daniel Hay Guest and
                  Yu Feng and
                  Mark Kittisopikul},
  title        = {h5py/h5py: 3.7.0},
  month        = may,
  year         = 2022,
  publisher    = {Zenodo},
  version      = {3.7.0},
  doi          = {10.5281/zenodo.6575970},
  url          = {https://doi.org/10.5281/zenodo.6575970},
}

@software{Reback2020,
    author       = {{The pandas development team}},
    title        = {pandas-dev/pandas: Pandas},
    month        = feb,
    year         = 2020,
    publisher    = {Zenodo},
    version      = {2.2.2},
    doi          = {10.5281/zenodo.3509134},
    url          = {https://doi.org/10.5281/zenodo.3509134}
}

@inproceedings{McKinney2010,
  title = {Data {{Structures}} for {{Statistical Computing}} in {{Python}}},
  booktitle = {Python in {{Science Conference}}},
  author = {McKinney, Wes},
  year = {2010},
  pages = {56--61},
  address = {Austin, Texas},
  doi = {10.25080/Majora-92bf1922-00a},
  urldate = {2025-02-13}
}

@article{Chan2026,
  title = {{\textsc{Sparcs}} -- Combining Radiation Hydrodynamics with Non-Equilibrium Metal Chemistry in the {{{\textsc{Swift}}}} Astrophysical Code},
  author = {Chan, Tsang Keung and Richings, Alexander J and Theuns, Tom and Liu, Yuankang and Schaller, Matthieu and Ivkovic, Mladen},
  year = 2026,
  month = jan,
  journal = {Monthly Notices of the Royal Astronomical Society},
  volume = {546},
  number = {2},
  pages = {stag004},
  issn = {0035-8711, 1365-2966},
  doi = {10.1093/mnras/stag004},
  urldate = {2026-02-23},
  copyright = {https://creativecommons.org/licenses/by/4.0/},
  langid = {english}
}

@article{Mcleod2021,
  title = {The Impact of Pre-Supernova Feedback and Its Dependence on Environment},
  author = {McLeod, Anna F and Ali, Ahmad A and Chevance, M{\'e}lanie and Della~Bruna, Lorenza and Schruba, Andreas and Stevance, Heloise F and Adamo, Angela and Kruijssen, J M Diederik and Longmore, Steven N and Weisz, Daniel R and Zeidler, Peter},
  year = {2021},
  month = oct,
  journal = {Monthly Notices of the Royal Astronomical Society},
  volume = {508},
  number = {4},
  pages = {5425--5448},
  issn = {0035-8711, 1365-2966},
  doi = {10.1093/mnras/stab2726},
  urldate = {2025-08-21},
  copyright = {https://creativecommons.org/licenses/by/4.0/},
  langid = {english}
}

@article{Groves2023,
  title = {The {{PHANGS}}--{{MUSE}} Nebular Catalogue},
  author = {Groves, B and Kreckel, K and Santoro, F and Belfiore, F and Zavodnik, E and Congiu, E and Egorov, O V and Emsellem, E and Grasha, K and Leroy, A and Scheuermann, F and Schinnerer, E and Watkins, E J and Barnes, A T and Bigiel, F and Dale, D A and Glover, S C O and Pessa, I and {Sanchez-Blazquez}, P and Williams, T G},
  year = {2023},
  month = feb,
  journal = {Monthly Notices of the Royal Astronomical Society},
  volume = {520},
  number = {4},
  pages = {4902--4952},
  issn = {0035-8711, 1365-2966},
  doi = {10.1093/mnras/stad114},
  urldate = {2025-08-21},
  copyright = {https://creativecommons.org/licenses/by/4.0/},
  langid = {english}
}

@article{Drory2024,
  title = {The {{SDSS-V Local Volume Mapper}} ({{LVM}}): {{Scientific Motivation}} and {{Project Overview}}},
  shorttitle = {The {{SDSS-V Local Volume Mapper}} ({{LVM}})},
  author = {Drory, Niv and Blanc, Guillermo A. and Kreckel, Kathryn and S{\'a}nchez, Sebasti{\'a}n F. and {Mej{\'i}a-Narv{\'a}ez}, Alfredo and Johnston, Evelyn J. and Jones, Amy M. and Pellegrini, Eric W. and Konidaris, Nicholas P. and Herbst, Tom and {S{\'a}nchez-Gallego}, Jos{\'e} and Kollmeier, Juna A. and De Almeida, Florence and {Barrera-Ballesteros}, Jorge K. and Bizyaev, Dmitry and Brownstein, Joel R. and I Saguer, Mar Canal and Cherinka, Brian and Cioni, Maria-Rosa L. and Congiu, Enrico and Cosens, Maren and Dias, Bruno and Donor, John and Egorov, Oleg and Egorova, Evgeniia and Froning, Cynthia S. and Garc{\'i}a, Pablo and Glover, Simon C. O. and Greve, Hannah and H{\"a}berle, Maximilian and Hoy, Kevin and Ibarra, Hector and Li, Jing and Klessen, Ralf S. and Krishnarao, Dhanesh and Kumari, Nimisha and Long, Knox S. and {M{\'e}ndez-Delgado}, Jos{\'e} Eduardo and Popa, Silvia Anastasia and Ramirez, Solange and Rix, Hans-Walter and S{\'a}nchez, Aurora Mata and Sankrit, Ravi and Sattler, Natascha and Sayres, Conor and Singh, Amrita and Stringfellow, Guy and Wachter, Stefanie and Watkins, Elizabeth Jayne and Wong, Tony and Wofford, Aida},
  year = {2024},
  month = nov,
  journal = {The Astronomical Journal},
  volume = {168},
  number = {5},
  pages = {198},
  issn = {0004-6256, 1538-3881},
  doi = {10.3847/1538-3881/ad6de9},
  urldate = {2025-08-22}
}

@article{Rousseau-nepton2019,
  title = {{{SIGNALS}}: {{I}}. {{Survey}} Description},
  shorttitle = {{{SIGNALS}}},
  author = {{Rousseau-Nepton}, L and Martin, R P and Robert, C and Drissen, L and Amram, P and Prunet, S and Martin, T and Moumen, I and Adamo, A and Alarie, A and Barmby, P and Boselli, A and Bresolin, F and Bureau, M and Chemin, L and Fernandes, R C and Combes, F and Crowder, C and Della~Bruna, L and Duarte~Puertas, S and Egusa, F and Epinat, B and Ksoll, V F and Girard, M and G{\'o}mez~Llanos, V and Gouliermis, D and Grasha, K and Higgs, C and {Hlavacek-Larrondo}, J and Ho, I-T and {Iglesias-P{\'a}ramo}, J and Joncas, G and Kam, Z S and Karera, P and Kennicutt, R C and Klessen, R S and Lianou, S and Liu, L and Liu, Q and {de~Amorim}, A Luiz and Lyman, J D and Martel, H and {Mazzilli-Ciraulo}, B and McLeod, A F and Melchior, A-L and Millan, I and Moll{\'a}, M and Momose, R and Morisset, C and Pan, H-A and Pati, A K and Pellerin, A and Pellegrini, E and P{\'e}rez, I and Petric, A and Plana, H and Rahner, D and Ruiz~Lara, T and {S{\'a}nchez-Menguiano}, L and Spekkens, K and Stasi{\'n}ska, G and Takamiya, M and Vale~Asari, N and V{\'i}lchez, J M},
  year = {2019},
  month = nov,
  journal = {Monthly Notices of the Royal Astronomical Society},
  volume = {489},
  number = {4},
  pages = {5530--5546},
  issn = {0035-8711, 1365-2966},
  doi = {10.1093/mnras/stz2455},
  urldate = {2025-08-22},
  copyright = {https://academic.oup.com/journals/pages/open\_access/funder\_policies/chorus/standard\_publication\_model},
  langid = {english}
}

@article{Chluba2006,
  title = {Induced Two-Photon Decay of the 2s Level and the Rate of Cosmological Hydrogen Recombination},
  author = {Chluba, J. and Sunyaev, R. A.},
  year = {2006},
  month = jan,
  journal = {Astronomy \& Astrophysics},
  volume = {446},
  number = {1},
  pages = {39--42},
  issn = {0004-6361, 1432-0746},
  doi = {10.1051/0004-6361:20053988},
  urldate = {2024-02-23}
}

@article{Bottorff2006,
  title = {Two-{{Photon Transitions}} and {{Continuous Emission}} from {{Hydrogenic Species}}},
  author = {Bottorff, Mark~C. and Ferland, Gary~J. and Straley, Joseph~P.},
  year = {2006},
  month = aug,
  journal = {Publications of the Astronomical Society of the Pacific},
  volume = {118},
  number = {846},
  pages = {1176--1179},
  issn = {0004-6280, 1538-3873},
  doi = {10.1086/506974},
  urldate = {2023-09-20},
  langid = {english}
}

@article{Grinin2020,
  title = {Two-Photon Frequency Comb Spectroscopy of Atomic Hydrogen},
  author = {Grinin, Alexey and Matveev, Arthur and Yost, Dylan C. and Maisenbacher, Lothar and Wirthl, Vitaly and Pohl, Randolf and H{\"a}nsch, Theodor W. and Udem, Thomas},
  year = {2020},
  month = nov,
  journal = {Science},
  volume = {370},
  number = {6520},
  pages = {1061--1066},
  issn = {0036-8075, 1095-9203},
  doi = {10.1126/science.abc7776},
  urldate = {2025-08-27},
  langid = {english}
}

@software{Dullemond2012,
       author = {{Dullemond}, C.~P. and {Juhasz}, A. and {Pohl}, A. and {Sereshti}, F. and {Shetty}, R. and {Peters}, T. and {Commercon}, B. and {Flock}, M.},
        title = "{RADMC-3D: A multi-purpose radiative transfer tool}",
 howpublished = {Astrophysics Source Code Library, record ascl:1202.015},
         year = 2012,
        month = feb,
          eid = {ascl:1202.015},
       adsurl = {https://ui.adsabs.harvard.edu/abs/2012ascl.soft02015D}
}

@article{Anderson2000,
	title = {An \textit{{R}} -matrix with pseudostates approach to the electron-impact excitation of {H} {I} for diagnostic applications in fusion plasmas},
	volume = {33},
	issn = {0953-4075, 1361-6455},
	url = {https://iopscience.iop.org/article/10.1088/0953-4075/33/6/311},
	doi = {10.1088/0953-4075/33/6/311},
	number = {6},
	urldate = {2023-11-13},
	journal = {Journal of Physics B: Atomic, Molecular and Optical Physics},
	author = {Anderson, H and Ballance, C P and Badnell, N R and Summers, H P},
	month = mar,
	year = {2000},
	pages = {1255--1262},
}

@article{Anderson2002,
	title = {An {R}-matrix with pseudo-states approach to the electron-impact excitation of {H} {I} for diagnostic applications in fusion plasmas},
	volume = {35},
	issn = {09534075},
	url = {https://iopscience.iop.org/article/10.1088/0953-4075/35/6/701},
	doi = {10.1088/0953-4075/35/6/701},
	number = {6},
	urldate = {2023-11-13},
	journal = {Journal of Physics B: Atomic, Molecular and Optical Physics},
	author = {Anderson, H and Ballance, C P and Badnell, N R and Summers, H P},
	month = mar,
	year = {2002},
	pages = {1613--1615},
}

@article{VanRegemorter1962,
  title = {Rate of {{Collisional Excitation}} in {{Stellar Atmospheres}}.},
  author = {Van Regemorter, Henri},
  year = {1962},
  month = nov,
  journal = {The Astrophysical Journal},
  volume = {136},
  pages = {906},
  issn = {0004-637X, 1538-4357},
  doi = {10.1086/147445},
  urldate = {2024-11-26},
  langid = {english}
}

@article{Lykins2015,
	title = {{STOUT}: {CLOUDY}’{S} {ATOMIC} {AND} {MOLECULAR} {DATABASE}},
	volume = {807},
	issn = {1538-4357},
	shorttitle = {{STOUT}},
	url = {https://iopscience.iop.org/article/10.1088/0004-637X/807/2/118},
	doi = {10.1088/0004-637X/807/2/118},
	number = {2},
	urldate = {2023-10-24},
	journal = {The Astrophysical Journal},
	author = {Lykins, M. L. and Ferland, G. J. and Kisielius, R. and Chatzikos, M. and Porter, R. L. and Hoof, P. A. M. Van and Williams, R. J. R. and Keenan, F. P. and Stancil, P. C.},
	month = jul,
	year = {2015},
	pages = {118},
}

@article{Minerbo1978,
  title = {Maximum Entropy {{Eddington}} Factors},
  author = {Minerbo, Gerald N.},
  year = {1978},
  month = dec,
  journal = {Journal of Quantitative Spectroscopy and Radiative Transfer},
  volume = {20},
  number = {6},
  pages = {541--545},
  issn = {00224073},
  doi = {10.1016/0022-4073(78)90024-9},
  urldate = {2025-08-27},
  copyright = {https://www.elsevier.com/tdm/userlicense/1.0/},
  langid = {english}
}

@article{Levermore1984,
  title = {Relating {{Eddington}} Factors to Flux Limiters},
  author = {Levermore, C.D.},
  year = {1984},
  month = feb,
  journal = {Journal of Quantitative Spectroscopy and Radiative Transfer},
  volume = {31},
  number = {2},
  pages = {149--160},
  issn = {00224073},
  doi = {10.1016/0022-4073(84)90112-2},
  urldate = {2025-08-27},
  copyright = {https://www.elsevier.com/tdm/userlicense/1.0/},
  langid = {english}
}

@article{Isobe2025,
  title = {{{JADES}}: Nitrogen Enhancement in High-Redshift Broad-Line Active Galactic Nuclei},
  shorttitle = {{{JADES}}},
  author = {Isobe, Yuki and Maiolino, Roberto and D'Eugenio, Francesco and Curti, Mirko and Ji, Xihan and Juod{\v z}balis, Ignas and Scholtz, Jan and Feltre, Anne and Charlot, St{\'e}phane and {\"U}bler, Hannah and J.~Bunker, Andrew and Carniani, Stefano and {Curtis-Lake}, Emma and Ji, Zhiyuan and Kumari, Nimisha and Rinaldi, Pierluigi and Robertson, Brant and Willott, Chris and Witstok, Joris},
  year = {2025},
  month = may,
  journal = {Monthly Notices of the Royal Astronomical Society: Letters},
  volume = {541},
  number = {1},
  pages = {L71-L79},
  issn = {1745-3925, 1745-3933},
  doi = {10.1093/mnrasl/slaf056},
  urldate = {2025-08-27},
  copyright = {https://creativecommons.org/licenses/by/4.0/},
  langid = {english}
}

@article{Wiersma2009,
  title = {Chemical Enrichment in Cosmological, Smoothed Particle Hydrodynamics Simulations},
  author = {Wiersma, Robert P. C. and Schaye, Joop and Theuns, Tom and Dalla Vecchia, Claudio and Tornatore, Luca},
  year = {2009},
  month = oct,
  journal = {Monthly Notices of the Royal Astronomical Society},
  volume = {399},
  number = {2},
  pages = {574--600},
  issn = {00358711, 13652966},
  doi = {10.1111/j.1365-2966.2009.15331.x},
  urldate = {2025-08-27},
  langid = {english}
}

@article{Price2018,
  title = {{\textsc{Phantom}} : {{A Smoothed Particle Hydrodynamics}} and {{Magnetohydrodynamics Code}} for {{Astrophysics}}},
  author = {Price, Daniel J. and Wurster, James and Tricco, Terrence S. and Nixon, Chris and Toupin, St{\'e}ven and Pettitt, Alex and Chan, Conrad and Mentiplay, Daniel and Laibe, Guillaume and Glover, Simon and Dobbs, Clare and Nealon, Rebecca and Liptai, David and Worpel, Hauke and Bonnerot, Cl{\'e}ment and Dipierro, Giovanni and Ballabio, Giulia and Ragusa, Enrico and Federrath, Christoph and Iaconi, Roberto and Reichardt, Thomas and Forgan, Duncan and Hutchison, Mark and Constantino, Thomas and Ayliffe, Ben and Hirsh, Kieran and Lodato, Giuseppe},
  year = {2018},
  journal = {Publications of the Astronomical Society of Australia},
  volume = {35},
  pages = {e031},
  issn = {1323-3580, 1448-6083},
  doi = {10.1017/pasa.2018.25},
  urldate = {2025-08-27},
  copyright = {https://www.cambridge.org/core/terms},
  langid = {english}
}

@article{Hirashita2003,
  title = {Star Formation Rate in Galaxies from {{UV}}, {{IR}}, and {{H}} {\emph{{$\alpha$}}} Estimators},
  author = {Hirashita, H. and Buat, V. and Inoue, A. K.},
  year = {2003},
  month = oct,
  journal = {Astronomy \& Astrophysics},
  volume = {410},
  number = {1},
  pages = {83--100},
  issn = {0004-6361, 1432-0746},
  doi = {10.1051/0004-6361:20031144},
  urldate = {2025-04-13}
}

@article{Grudic2021,
  title = {{{STARFORGE}}: {{Towards}} a Comprehensive Numerical Model of Star Cluster Formation and Feedback},
  shorttitle = {{{STARFORGE}}},
  author = {Grudi{\'c}, Michael Y and Guszejnov, D{\'a}vid and Hopkins, Philip F and Offner, Stella S R and {Faucher-Gigu{\`e}re}, Claude-Andr{\'e}},
  year = {2021},
  month = jul,
  journal = {Monthly Notices of the Royal Astronomical Society},
  volume = {506},
  number = {2},
  pages = {2199--2231},
  issn = {0035-8711, 1365-2966},
  doi = {10.1093/mnras/stab1347},
  urldate = {2025-08-27},
  copyright = {https://academic.oup.com/journals/pages/open\_access/funder\_policies/chorus/standard\_publication\_model},
  langid = {english}
}

@article{Cohen1996,
  title = {{{CVODE}}, {{A Stiff}}/{{Nonstiff ODE Solver}} in {{C}}},
  author = {Cohen, Scott D. and Hindmarsh, Alan C. and Dubois, Paul F.},
  year = {1996},
  month = mar,
  journal = {Computers in Physics},
  volume = {10},
  number = {2},
  pages = {138--143},
  issn = {0894-1866},
  doi = {10.1063/1.4822377},
  urldate = {2025-08-27},
  langid = {english}
}

@article{Hoang-Binh1990,
  title = {An Exact Calculation of Hydrogenic Radial Integrals and Oscillator Strengths, for Principal Quantum Numbers up to {{N}} Equal to about 1000},
  author = {{Hoang-Binh}, D.},
  year = {1990},
  month = nov,
  journal = {Astronomy and Astrophysics},
  volume = {238},
  pages = {449--451},
  issn = {0004-6361},
  urldate = {2023-11-01}
}

@article{Hoang-Binh2005,
  title = {A Program to Compute Exact Hydrogenic Radial Integrals, Oscillator Strengths, and {{Einstein}} Coefficients, for Principal Quantum Numbers up To},
  author = {{Hoang-Binh}, D.},
  year = {2005},
  month = mar,
  journal = {Computer Physics Communications},
  volume = {166},
  number = {3},
  pages = {191--196},
  issn = {00104655},
  doi = {10.1016/j.cpc.2004.11.005},
  urldate = {2023-11-01},
  langid = {english}
}

@article{Hirschmann2023b,
  title = {High-Redshift Metallicity Calibrations for {{{\emph{JWST}}}} Spectra: Insights from Line Emission in Cosmological Simulations},
  shorttitle = {High-Redshift Metallicity Calibrations for {{{\emph{JWST}}}} Spectra},
  author = {Hirschmann, Michaela and Charlot, Stephane and Somerville, Rachel S},
  year = {2023},
  month = oct,
  journal = {Monthly Notices of the Royal Astronomical Society},
  volume = {526},
  number = {3},
  pages = {3504--3518},
  issn = {0035-8711, 1365-2966},
  doi = {10.1093/mnras/stad2745},
  urldate = {2025-08-27},
  copyright = {https://creativecommons.org/licenses/by/4.0/},
  langid = {english}
}

@article{Ferland1992,
  title = {Anisotropic Line Emission and the Geometry of the Broad-Line Region in Active Galactic Nuclei},
  author = {Ferland, G. J. and Peterson, B. M. and Horne, K. and Welsh, W. F. and Nahar, S. N.},
  year = {1992},
  month = mar,
  journal = {The Astrophysical Journal},
  volume = {387},
  pages = {95},
  issn = {0004-637X, 1538-4357},
  doi = {10.1086/171063},
  urldate = {2025-08-27},
  langid = {english}
}

@book{Pradhan2011, place={Cambridge}, title={Atomic Astrophysics and Spectroscopy}, publisher={Cambridge University Press}, author={Pradhan, Anil K. and Nahar, Sultana N.}, year={2011}}

@article{Nahar2021,
  title = {Photoionization and {{Electron-Ion Recombination}} of n = 1 to {{Very High}} n-{{Values}} of {{Hydrogenic Ions}}},
  author = {Nahar, Sultana},
  year = 2021,
  month = oct,
  journal = {Atoms},
  volume = {9},
  number = {4},
  pages = {73},
  issn = {2218-2004},
  doi = {10.3390/atoms9040073},
  urldate = {2025-10-27},
  langid = {english}
}

@article{Monaghan1988,
  title = {An Introduction to {{SPH}}},
  author = {Monaghan, J.J.},
  year = 1988,
  month = jan,
  journal = {Computer Physics Communications},
  volume = {48},
  number = {1},
  pages = {89--96},
  issn = {00104655},
  doi = {10.1016/0010-4655(88)90026-4},
  urldate = {2025-11-19},
  copyright = {https://www.elsevier.com/tdm/userlicense/1.0/},
  langid = {english}
}

@article{Howatson2025,
  title = {Emission Line Tracers of Galactic Outflows Driven by Stellar Feedback in Simulations of Isolated Disc Galaxies},
  author = {Howatson, Elliot L and Richings, Alexander J and Roediger, Elke and {Faucher-Gigu{\`e}re}, Claude-Andr{\'e} and Theuns, Tom and Liu, Yuankang and Chan, Tsang Keung and Thompson, Oliver and Carr, Cody and {Angl{\'e}s-Alc{\'a}zar}, Daniel},
  year = 2025,
  month = oct,
  journal = {Monthly Notices of the Royal Astronomical Society},
  volume = {543},
  number = {4},
  pages = {3428--3446},
  issn = {0035-8711, 1365-2966},
  doi = {10.1093/mnras/staf1641},
  urldate = {2026-01-15},
  copyright = {https://creativecommons.org/licenses/by/4.0/},
  langid = {english}
}

@article{Verner1996,
  title = {Atomic {{Data}} for {{Astrophysics}}. {{II}}. {{New Analytic FITS}} for {{Photoionization Cross Sections}} of {{Atoms}} and {{Ions}}},
  author = {Verner, D. A. and Ferland, G. J. and Korista, K. T. and Yakovlev, D. G.},
  year = 1996,
  month = jul,
  journal = {The Astrophysical Journal},
  volume = {465},
  pages = {487},
  issn = {0004-637X, 1538-4357},
  doi = {10.1086/177435},
  urldate = {2025-03-27},
  langid = {english}
}

@article{Bauman2005,
  title = {{\emph{J}} -{{Resolved He}} {\textsc{i}} {{Emission Predictions}} in the {{Low}}-{{Density Limit}}},
  shorttitle = {{\emph{J}} -{{Resolved He}}},
  author = {Bauman, R. P. and Porter, R. L. and Ferland, G. J. and MacAdam, K. B.},
  year = 2005,
  month = jul,
  journal = {The Astrophysical Journal},
  volume = {628},
  number = {1},
  pages = {541--554},
  issn = {0004-637X, 1538-4357},
  doi = {10.1086/430665},
  urldate = {2023-10-24},
  langid = {english}
}
